\documentclass[a4paper,10pt]{article}
\usepackage[utf8x]{inputenc}

\usepackage{latexsym, amssymb, enumerate, amsmath}
\usepackage{graphicx,subfigure}
\usepackage{fancyhdr}
\usepackage{longtable}
\usepackage{url}
\usepackage{color}
\usepackage{makecell} %for \gape command in table environments
\usepackage{esint} %for mean value integral sign
\usepackage{bm} % for bold greek letters
\usepackage{placeins}
\newcommand{\N}{\mathbb{N}}

\newcommand{\R}{\mathbb{R}}

\newcommand{\pd}{\partial}
\newcommand{\abs}[1]{\left| #1 \right|}

\newcommand{\eps}{\varepsilon}
\newcommand{\Laplace}{\Delta}
\newcommand{\surf}{\nabla_{\Gamma}}
\newcommand{\surfND}{\nabla_{\Gamma_{*}}}

\newcommand{\LB}{\Delta_{\Gamma}}

\newcommand{\md}{\pd^{\bullet}_{t}}
\newcommand{\mdND}{\pd^{\bullet}_{t_{*}}}
\newcommand{\nd}{\pd^{\circ}_{t}}

\def\eps{\varepsilon}

\def\bbb{\boldsymbol}

\def\rrr{\textrm}

\def\mathref#1{\ifmmode\mathrm{(\ref{#1})}\else{\rm(\ref{#1})}\fi} 
\def\nref#1{\ifmmode\mathrm{\ref{#1}}\else{\rm\ref{#1}}\fi} 
\def\mathcite#1{{\rm\cite{#1}}} 

\newtheorem{thm}{Theorem}[section]

\numberwithin{equation}{section}
\begin{document}

\title{Diffuse interface modelling of soluble surfactants in two-phase flow
\thanks{%{Received date / Revised version date}
          % The correct dates will be entered by the CMS editor}}
}}
          %For each author, make a block with the following four macros:
\author{Harald Garcke
\thanks {address: Fakult\"at f\"ur Mathematik, Universit\"at Regensburg, 93040 Regensburg, Germany, (email: harald.garcke@mathematik.uni-regensburg.de).}
\and Kei Fong Lam \thanks {address: Mathematics Institute, Zeeman Building, University of Warwick, Coventry, CV4 7AL, UK, (email: a.k.f.lam@warwick.ac.uk).} 
\and Bj\"orn Stinner \thanks{address: Mathematics Institute and Centre for Scientific Computing, University of Warwick, Coventry, CV4 7AL, UK, (email: bjorn.stinner@warwick.ac.uk).}}
          %{Put the URL for your home page here if you have one}

          %Use \thanks statements for acknowledgements of grants and
          %support. They will appear below all the authors' addresses, so be
          %specific about which author is thanking whom:

          %\thanks{}
\date{ }
\pagestyle{myheadings} 
% \markboth{Diffuse interface modelling of soluble surfactants in two-phase flow}{Harald Garcke, Kei Fong Lam and Bj\"orn Stinner}
\maketitle

\begin{abstract}
% % % 
% % % Insert your REQUIRED abstract here. If possible, do not use any
% % % math symbols or references to the bibliography to facilitate
% % % putting the abstract online in an .html
% % %           format.
Phase field models for two-phase flow with a surfactant soluble in possibly both fluids are derived from balance equations and an energy inequality so that thermodynamic consistency is guaranteed. Via a formal asymptotic analysis, they are related to sharp interface models. Both cases of dynamic as well as instantaneous adsorption are covered. Flexibility with respect to the choice of bulk and surface free energies allows to realise various isotherms and relations of state between surface tension and surfactant. Some numerical simulations display the effectiveness of the presented approach.
\end{abstract}

% \begin{keywords}
% two-phase flow, surfactant, phase field model, adsorption isotherm
% 
% \smallskip
% 
% {\bf subject classifications.}
% 35R35, %FBPs
% 35R01, %PDEs on manifolds
% 76T99, %two phase flow
% 76D45, %capillarity (surface tension)
% %35Q35, %PDEs in fluids
% 35C20, %asymptotic expansions in PDEs
% 35Q35 %PDEs in connection with fluid mechanics
% \end{keywords}

\section{Introduction}\label{intro}
Surface active agents (surfactants) reduce the surface tension of fluid interfaces and, via surface tension gradients, can lead to tangential forces resulting in the Marangoni effect. Biological systems take advantage of their impact on fluids with interfaces, but surfactants are also important for industrial applications such as processes of emulsification or mixing. While often much experience and knowledge is available on how surfactants influence the rheology of multi-phase fluids, the goal is to understand how exactly the presence of a surfactant influences coalescence and segregation of droplets.

Surfactants can be soluble in at least one of the fluid phases and the exchange of surfactants between the bulk phases and the fluid interfaces is governed by the process of adsorption and desorption.  Ward and Tordai \cite{article:WardTordai46} derived a time-dependent relation for the surfactant density at the interface and the surfactant density at the adjacent bulk phase (known as the sub-layer or sub-surface).  To compute the interfacial density, a closure relation between the two quantities has been proposed in the
form of several different equilibrium isotherms \cite{article:EastoeDalton2000,incoll:Kralchevsky,article:KralchevskyDanovBrozeMehreteab}, where the underlying assumption is that the interface is in equilibrium with the sub-layer at all times.  This corresponds to the case of diffusion-limited adsorption studied in Diamant and Andelman \cite{article:DiamantAndelman96}, where the process of adsorption to the interface is fast compared to the kinetics in the bulk phases.  However, instantaneous adsorption is not valid in the context of ionic surfactant systems \cite{article:DiamantAndelman96} or when the diffusion is not limited to a thin layer \cite{article:Coutelieris02,article:CoutelierisKainourgiakisStubos03,article:CoutelierisKainourgiakisStubos05}.  Therefore, we would like to be able to account for non-instantaneous adsorption in our models.

Two-phase flow with surfactant is classically modelled with moving hypersurfaces describing the interfaces separating the two fluids.  We will derive the following sharp interface model for a domain $\Omega$ containing two fluids of different mass densities.  We denote by $\Omega^{(1)}(t)$, $\Omega^{(2)}(t)$ the domains of the fluids which are separated by an interface $\Gamma(t)$: 
\begin{align}
\nabla \cdot \bm{v} =  0 &\quad& \text{ in } \Omega^{(i)}(t), \label{SIM:incompress} \\
\pd_{t}(\overline{\rho}^{(i)} \bm{v}) + \nabla \cdot (\overline{\rho}^{(i)} \bm{v} \otimes \bm{v}) =  \nabla \cdot \left ( - p \bm{I} + 2\eta^{(i)} D(\bm{v}) \right ) &\quad& \text{ in } \Omega^{(i)}(t), \label{SIM:momentum}\\
\md c^{(i)} =  \nabla \cdot (M_{c}^{(i)} \nabla G_{i}'(c^{(i)})) &\quad& \text{ in } \Omega^{(i)}(t), \label{SIM:bulk} \\
[\bm{v}]_{1}^{2} =  0, \quad \bm{v} \cdot \bm{\nu} = u_{\Gamma}  &\quad& \text{ on } \Gamma(t), \label{SIM:velocityjump}\\
[p \bm{I} - 2\eta^{(i)} D(\bm{v})]_{1}^{2} \bm{\nu} =  \sigma(c^{\Gamma})\kappa \bm{\nu} + \surf \sigma(c^{\Gamma}) &\quad& \text{ on } \Gamma(t), \label{SIM:stressjump} \\
\md c^{\Gamma} + c^{\Gamma} \surf \cdot \bm{v} - \surf \cdot (M_{\Gamma}\surf \gamma'(c^{\Gamma})) =  [M_{c}^{(i)} \nabla G'_{i}(c^{(i)})]_{1}^{2}\bm{\nu} &\quad& \text{ on } \Gamma(t), \label{SIM:interface} \\ 
\alpha^{(i)}(-1)^{i}M_{c}^{(i)} \nabla G'_{i}(c^{(i)}) \cdot \bm{\nu} =  -(\gamma'(c^{\Gamma}) - G_{i}'(c^{(i)})) &\quad& \text{ on } \Gamma(t). \label{SIM:dynamicAdsorp}
\end{align}
Here $\bm{v}$ denotes the fluid velocity, $\overline{\rho}^{(i)}$ is the constant mass density for fluid $i$, $\eta^{(i)}$ is the viscosity of fluid $i$, $D(\bm{v}) = \tfrac{1}{2}(\nabla \bm{v} + (\nabla \bm{v})^{\perp})$ is the rate of deformation tensor, $p$ is the pressure, $\bm{I}$ is the identity tensor, $\md(\cdot) = \pd_{t}(\cdot) + \bm{v} \cdot \nabla (\cdot)$ is the material derivative, $c^{(i)}$ is the bulk density of surfactant in fluid $i$, $M_{c}^{(i)}$ is the mobility of surfactants in fluid $i$, $G_{i}(c^{(i)})$ is the bulk free energy density associated to the bulk surfactant in fluid $i$.  On the interface, $u_{\Gamma}$ is the normal velocity, $\bm{\nu}$ is the unit normal on $\Gamma$ pointing into $\Omega^{(2)}$, $c^{\Gamma}$ is the interfacial surfactant density, $\sigma(c^{\Gamma})$ is the density dependent surface tension, $\kappa$ is the mean curvature of $\Gamma$, $\surf$ is the surface gradient operator, $\surf \cdot$ is the surface divergence, $M_{\Gamma}$ is the mobility of the interfacial surfactants, $\gamma(c^{\Gamma})$ is the free energy density associated to the interfacial surfactant, and $\alpha^{(i)} \geq 0$ is a kinetic factor that relates to the speed of adsorption. The above model satisfies the second law of thermodynamics in an isothermal situation in the form of an energy dissipation inequality.

Equations $(\ref{SIM:incompress})$ and $(\ref{SIM:momentum})$ are the classical incompressibility condition and momentum equation, respectively.  The mass balance equation for bulk surfactants is given by $(\ref{SIM:bulk})$.  Equation $(\ref{SIM:velocityjump})$ states that the interface is transported with the flow and that not only the normal components but also the tangential components of the velocity field match up. The force balance on the interface $(\ref{SIM:stressjump})$ relates the jump in the stress tensor across the interface to the surface tension force and the Marangoni force at the interface.  The mass balance of the interfacial surfactants is given by $(\ref{SIM:interface})$, and the closure condition $(\ref{SIM:dynamicAdsorp})$ tells us whether adsorption is instantaneous ($\alpha^{(i)} = 0$, an isotherm is obtained) or dynamic ($\alpha^{(i)} > 0$, the mass flux into the interface is proportional to the difference in chemical potentials).

The model studied in \cite{article:BothePruss10,incoll:BothePrussSimonett} bears the most resemblance to the above model, where the setting of these papers is the diffusion-limited regime with a surfactant which is soluble in one phase only and $(\ref{SIM:dynamicAdsorp})$ is replaced by the relation
\begin{align}\label{SIM:InstAdsorp}
 \gamma'(c^{\Gamma}) = G'(c) \quad \Longleftrightarrow \quad c^{\Gamma} = g(c) := (\gamma')^{-1}(G'(c)),
\end{align}
in which $g$ plays the role of the equilibrium isotherm and where $G$ is the bulk free energy of the phase in which the surfactant is soluble.  Our approach is based on a free energy formulation, originated from \cite{article:DiamantAndelman96, article:DiamantArielAndelman01}, where we gain access to equilibrium isotherms by setting $\alpha^{(i)} = 0$ and choosing suitable functions for $\gamma$ and $G_{i}$.  Furthermore, for positive values of $\alpha^{(i)}$ we are able to include the dynamics of non-equilibrium adsorption.

The governing equations $(\ref{SIM:incompress})-(\ref{SIM:dynamicAdsorp})$ form a free boundary problem.  The phase boundary $\Gamma(t)$ is unknown a priori and hence must be computed as part of the solution.  Much previous work have been dedicated to explicitly tracking and capturing
the interface using various numerical methods \cite{article:YonPozrikidis98,article:JamesLowengrub04,article:XuLiLowengrubZhao06,article:LaiTsengHuang08,article:MuradogluTryggvason08,article:KhatriTornberg11}.
However, the sharp interface description breaks down when topological changes occur.  Phenomena such as breakup of fluid droplets, reconnection of fluid interfaces and tip-streaming driven by Marangoni forces \cite{article:FernandezHomsy04,article:KrechetnikovHomsy04,article:KrechetnikovHomsy04B} involve changes in the topology of the interface.  Numerically, complications also arise when the shape of the interface becomes complicated or exhibits self-intersections.  These difficulties have led to the development of diffuse interface or phase field models to provide an alternative description of fluid/fluid interfaces.

At the core of these models, the sharp interface is replaced by an interfacial layer of finite width and an order parameter is used to distinguish between the bulk fluids and interfacial layer.  The order parameter takes distinct constant values in each of the bulk fluids and varies smoothly across the narrow interfacial layer.  The original sharp interface can then be represented as the zero level set of the order parameter, thus allowing different level sets to exhibit different topologies.

The width of the interfacial layer is characterised by the length scale over which the order parameter varies from its values at the bulk regions.  The phase field model can be related to the sharp interface model in the asymptotic limit in which this width is small compared to the length scales associated to the bulk regions.  Hence one can also view the phase field methodology purely as a tool for approximating the sharp interface equations.  If the objective is to ensure that, in the limit of vanishing interfacial thickness, certain sharp interface models are recovered then there is a lot of freedom in constructing phase field models to meet one's needs (see e.g. \cite{article:LiLowengrubRatzVoigt09}).
 
The review \cite{article:AndersonMcFaddenWheeler98} provides an overview on diffuse interface methods in the context of fluid flows. In \cite{article:GurtinPolignoneVinals96,article:HohenbergHalperin77} it was already proposed to combine a Cahn-Hilliard equation for distinguishing the two phases with a Navier-Stokes system. An additional term was included in the momentum equation to model the surface contributions to forces. In the case of different densities, Lowengrub and Truskinovsky \cite{article:LowengrubTruskinovsky98} derived quasi-incompressible models, where the fluid velocity is not divergence free.  On the other hand, Abels, Garcke and Gr\"un \cite{article:AbelsGarckeGrun} derived a thermodynamically consistent diffuse interface model for two-phase flow with different densities and with solenoidal fluid velocities.  Following the derivation in \cite{article:AbelsGarckeGrun}, we will derive three diffuse interface models, which approximate the sharp interface models in the diffuse-limited regime.

More precisely, for the case of non-instantaneous adsorption ($\alpha^{(i)} > 0$), we will derive the following model (denoted Model A)
\begin{align}
\nabla \cdot \bm{v} & = 0, \label{PFMA:incompress} \\
\pd_{t}(\rho \bm{v})  + \nabla \cdot ( \rho \bm{v} \otimes \bm{v} ) & = \nabla \cdot \Big{(} - p \bm{I} + 2 \eta(\varphi)D(\bm{v}) + \bm{v} \otimes \tfrac{\overline{\rho}^{(2)} - \overline{\rho}^{(1)}}{2} m(\varphi) \nabla \mu \Big{)} \label{PFMA:momentum} \\
\nonumber & \quad + \nabla \cdot  \big{(} K\sigma(c^{\Gamma}) (\delta(\varphi, \nabla \varphi) \bm{I}  - \eps \nabla \varphi \otimes \nabla \varphi) \big{)}, \\
\md \varphi & =  \nabla \cdot (m(\varphi) \nabla \mu),  \label{PFMA:phase} \\
\mu + \nabla \cdot ( K\eps \sigma(c^{\Gamma}) \nabla \varphi) & = \frac{K}{\eps}\sigma(c^{\Gamma}) W'(\varphi) + \sum_{i=1,2}\xi'_{i}(\varphi)(G_{i}(c^{(i)}) - G_{i}'(c^{(i)}) c^{(i)}), \label{PFMA:chem}  \\ 
\md (\xi_{i}(\varphi) c^{(i)}) & = \nabla \cdot (M_{c}^{(i)}(c^{(i)}) \xi_{i}(\varphi)\nabla G_{i}'(c^{(i)})) \label{PFMA:bulk} \\
\nonumber & \quad + \beta^{(i)} \delta(\varphi, \nabla \varphi)(\gamma'(c^{\Gamma}) - G'_{i}(c^{(i)})), \quad i = 1, 2,  \\
\md (K\delta(\varphi, \nabla \varphi) c^{\Gamma}) & = \nabla \cdot \Big{(}M_{\Gamma}(c^{\Gamma}) K\delta(\varphi, \nabla \varphi) \nabla \gamma'(c^{\Gamma}) \Big{)}  \label{PFMA:interface}  \\
\nonumber & \quad - \delta(\varphi, \nabla \varphi) \sum_{i=1,2} \beta^{(i)}(\gamma'(c^{\Gamma}) - G_{i}'(c^{(i)})).  \displaybreak[0]
\end{align}
Here $\eps$ is a length scale associated with the interfacial width, $\varphi$ is the order parameter that distinguishes the two bulk phases. In fact $\varphi$ takes values close to $\pm 1$ in the two phases and rapidly changes from $-1$ to $1$ in an interfacial layer. The functions $\xi_{i}(\varphi)$ and $\delta(\varphi, \nabla \varphi)$ act as regularisation to the indicator functions of $\Omega^{(i)}$ and $\Gamma$, respectively.  The quantity $\beta^{(i)} = K / \alpha^{(i)}$ is related to the adsorption kinetics and $K$ is a constant.  Equations $(\ref{PFMA:incompress})$ and $(\ref{PFMA:momentum})$ are the incompressibility condition and the phase field momentum equations, respectively.  Equation $(\ref{PFMA:phase})$ together with $(\ref{PFMA:chem})$ governs how the order parameter evolves and equations $(\ref{PFMA:bulk})$ and $(\ref{PFMA:interface})$ are the
bulk and interfacial surfactant equations, respectively.

We derive two additional models for instantaneous adsorption ($\alpha^{(i)} = 0$):
Model B models the case where the surfactant is soluble in only one of the bulk phases.  It consists of $(\ref{PFMA:incompress}) - (\ref{PFMA:chem})$ and replaces the bulk and interface surfactant equations  (\ref{PFMA:bulk}), (\ref{PFMA:interface}) with
\begin{align}\label{PFMC}
\md (\xi(\varphi)c + K \delta g(c)) - \nabla \cdot (M(c)\xi(\varphi)\nabla G'(c))  - \nabla \cdot (M_{\Gamma}(g(c)) K \delta \nabla G'(c)) = 0,
\end{align}
where $g(c)$ is the adsorption relation between interface and bulk densities as in (\ref{SIM:InstAdsorp}).

The case where the surfactant is soluble in both bulk phases is covered by Model C, which consists of $(\ref{PFMA:incompress}) - (\ref{PFMA:chem})$ and
\begin{align}\label{PFMD}
\md (\xi_{1}(\varphi)c^{(1)}(q) + \xi_{2}(\varphi) c^{(2)}(q) + K \delta c^{\Gamma}(q))  &- \sum_{i=1,2} \nabla \cdot (M_{i}(c^{(i)}(q))\xi_{i}(\varphi)\nabla q) \\
\nonumber & - \nabla \cdot (M_{\Gamma}(c^{\Gamma}(q)) K\delta \nabla q) = 0.
\end{align}
Here, $q$ denotes a chemical potential where, as will be discussed in Section 3,  we can express the surfactant densities as functions of $q$.
 
The  Model A is related to the approach in \cite{article:TeigenLiLowengrubWangVoigt09}. We modify the approach of \cite{article:TeigenLiLowengrubWangVoigt09} in such a way that an energy inequality is valid and such that we recover the isotherm relations for adsorption phenomena in the limit of instantaneous adsorption. We deepen the asymptotic analysis in that it works with the original equation for the surface quantity and does not require the assumption of extending the surface quantity continuously in normal direction.  Phase field models of surfactant adsorption that utilise the free energy approach of \cite{article:DiamantAndelman96,article:DiamantArielAndelman01} can be traced back to the models of \cite{article:TheissenGompper99,article:TeramotoYonezawa01,article:vanderSmanvanderGraaf06}, where the latter is extended in \cite{article:LiuZhang10} and solved
using lattice Boltzmann methods.  The issue of ill-posedness of the model is discussed in \cite{preprint:EngblomDoQuangAmbergTornberg} and three alternatives have been suggested.  Phase field models that look into the behaviour of equilibrium configurations of fluid-surfactant systems can be found in 
\cite{article:FonsecaMoriniSlastikov07,article:TengChernLai12} and a detailed comparison of previous phase field models can be found in \cite{article:LiKim12}.

The structure of this article is as follows: In Section 2 we will derive the sharp interface model $(\ref{SIM:incompress}) - (\ref{SIM:dynamicAdsorp})$ from basic conservation laws. We show that the sharp interface model satisfies a local energy inequality and present the functional forms for $\gamma$ and $G$ that lead to five of the popular adsorption isotherms when $\alpha^{(i)} = 0$, namely those of Henry, Langmuir, Volmer, Frumkin and Freundlich.  In Section 3, we present the derivation of phase field models based on the Lagrange multiplier method presented in \cite{article:AbelsGarckeGrun} and show all of them satisfy a local dissipation inequality.  In Section 4 we show, via formally matched asymptotics, that we recover $(\ref{SIM:incompress}) - (\ref{SIM:dynamicAdsorp})$ from Model A and $(\ref{SIM:InstAdsorp})$ from Models B and C in the limit $\eps \to
0$. In addition, Model A can be shown to converge to the sharp interface problem with instantaneous adsorption when the kinetic term is chosen appropriately. In Section 5, we present 1D and 2D numerics to support the asymptotic analysis.

% % % % % \section{Put title of the next section here}\label{an apprpriate
% % % % % label}
% % % % % 
% % % % %           %If you have subsections use:
% % % % % \subsection{Subsection title}\label{another label}
% % % % % 
% % % % % Don't forget to give each section, subsection, equation, theorem,
% % % % % corollary, etc. a unique label, and when you refer to the results
% % % % % later in the text use \ref{<labelname>} instead of explicitly
% % % % % writing the number of the environment in question.
% % % % % 
% % % % % This use of \ label and \ ref is REQUIRED for  papers.
% % % % % 
% % % % % Similarly, always use \cite{biblabelname} to refer to
% % % % % bibliographic references, which would then be entered in the
% % % % % bibliography via
% % % % %           %\bibitem{biblabelname}.
% % % % % 
% % % % %           %
% % % % %           % For figures use
% % % % % 
% % % % %           %\begin{figure}
% % % % % 
% % % % %           %The use of .eps files is encouraged, in which case you should
% % % % %           %un-comment the \uspackage{graphics} command above, and use the
% % % % %           %command
% % % % %           %\include{figure.eps}
% % % % %           % to insert the figure file.
% % % % % 
% % % % %           %\end{figure}
% % % % % 
% % % % % 
% % % % %           % BibTeX users please use
% % % % % 
% % % % %           % \bibliographystyle{}
% % % % % 
% % % % %           % \bibliography{}
% % % % % 
% % % % %           %
% % % % % 
% % % % %           % Non-BibTeX users please use
\section{Sharp interface model}\label{SIM}
\subsection{Balance equations}
We consider a domain $\Omega \subset \R^{d}$, $d = 1,2,3$, containing two immiscible, incompressible Newtonian fluids with possibly different constant mass densities $\overline{\rho}^{(i)}, i = 1, 2$. The domain occupied by the fluid with density $\overline{\rho}^{(i)}$ is labelled as $\Omega^{(i)} \subset \R \times \R^{d}$ where we set $\Omega^{(i)}(t) := \{ x \in \Omega ; (t,x) \in \Omega^{(i)}\}$.  The two domains are separated by an interface $\Gamma$ which is a hypersurface in $\R \times \R^{d}$ such that $\Gamma(t) \cap \pd \Omega = \emptyset$ where $\Gamma(t) := \{ x \in \Omega; (t, x) \in \Gamma\}$.  A surfactant is present which alters the surface tension by adsorbing to the fluid interface and, provided it is soluble in the corresponding fluid, it is subject to diffusion in the phases $\Omega^{(i)}$.  We denote the fluid velocity field by $\bm{v}$, the pressure by $p$, the bulk surfactant densities by $c^{(i)}, i = 1, 2$, and the interface surfactant density by $c^{\Gamma}$.  

Balance of mass and linear momentum inside the phases lead to the following equations
\begin{align*}
 \nabla \cdot \bm{v} = 0, \quad \pd_{t}(\overline{\rho}^{(i)} \bm{v}) + \nabla \cdot (\overline{\rho}^{(i)} \bm{v} \otimes \bm{v}) = \md(\overline{\rho}^{(i)} \bm{v})  =  \nabla \cdot \bm{T}^{(i)},
\end{align*}
where $\md$ denotes the material derivative and $\bm{T}^{(i)}$, $i=1, 2$, is the symmetric stress tensor (due to conservation of angular momentum). These equation hold in $\Omega^{(1)}(t) \cup \Omega^{(2)}(t)$.  We assume that the two fluids do not undergo phase transitions and the phase boundary $\Gamma(t)$ is purely transported with the flow where we also assume that there is no-slip at the interface, hence the tangential velocities match:
\begin{align*}
 [\bm{v}]_{1}^{2} & = 0,  \quad \bm{v} \cdot \bm{\nu} = u_{\Gamma}.
\end{align*}
Here $[\cdot]_{1}^{2}$ denotes the jump of the quantity in brackets across $\Gamma$ from $\Omega^{(1)}$ to $\Omega^{(2)}$, $\bm{\nu}$ is the unit outward normal of $\Gamma(t)$ pointing into $\Omega^{(2)}(t)$, and $u_{\Gamma}$ is the normal velocity of the interface.  

Let $V(t)$ be an arbitrary material test volume in $\Omega$ with external unit normal $\bm{\nu}_{ext}$ of $V(t) \cap \Omega$. If $V(t) \cap \Gamma(t)$ is non-empty then we denote its external unit co-normal by $\bm{\mu}$ and write $\bm{\nu}_{ext}^{(i)}$ for the external unit normal of $V(t) \cap \Omega^{(i)}(t)$, $i=1,2$.  In the bulk fluid regions, surfactants will be subjected to transport mechanisms consisting of only diffusion and convection.  Hence, mass balance for bulk surfactants in a material test volume $V(t)$ away from the interface $\Gamma(t)$ yields
\begin{align*}
 \frac{d}{dt} \int_{V(t)} c^{(i)} = - \int_{\pd V(t)} \bm{J}^{(i)}_{c} \cdot \bm{\nu}_{ext} 
\end{align*}
where $\bm{J}^{(i)}_{c}$ is the molecular flux.  By Reynold's transport theorem and using that $\nabla \cdot \bm{v} = 0$, this leads to the pointwise law
\begin{align}\label{eq:generalC}
 \md c^{(i)} + \nabla \cdot \bm{J}^{(i)}_{c} = 0, \quad i = 1, 2.
\end{align}
For a test volume $V(t)$ intersecting $\Gamma(t)$, we postulate
\begin{align}\label{eq:changeoftotalmass}
& \frac{d}{dt} \left ( \sum_{i=1,2} \int_{V(t) \cap \Omega^{(i)}(t)} c^{(i)} + \int_{\Gamma(t) \cap V(t)} c^{\Gamma} \right )  \\
\nonumber = & \sum_{i=1,2} \int_{\pd(V(t) \cap \Omega^{(i)}(t))\setminus \Gamma(t)} - \bm{J}_{c}^{(i)} \cdot \bm{\nu}_{ext} + \int_{\pd(V(t) \cap \Gamma(t))} -\bm{J}_{\Gamma} \cdot \bm{\mu}, \displaybreak[0]
\end{align}
where $\bm{J}_{\Gamma}$ is the interfacial molecular flux, tangential to $\Gamma$.  Using Reynold's transport theorem, the surface transport theorem and the surface divergence theorem (see \cite{article:Betounes86}) we obtain
\begin{align*}
& \frac{d}{dt} \left ( \sum_{i=1,2} \int_{V(t) \cap \Omega^{(i)}(t)} c^{(i)} + \int_{\Gamma(t) \cap V(t)} c^{\Gamma} \right )  \\
= & \sum_{i=1}^{2} \int_{V(t) \cap \Omega^{(i)}(t)} \md c^{(i)} + \int_{V(t) \cap \Gamma(t)} \left ( \md c^{\Gamma} + c^{\Gamma} \surf \cdot \bm{v} \right )
\end{align*}
for the left hand side and 
\begin{align*}
& \sum_{i=1,2} -\int_{\pd(V(t) \cap \Omega^{(i)}(t))\setminus \Gamma(t)} \bm{J}_{c}^{(i)} \cdot \bm{\nu}_{ext} - \int_{\pd(V(t) \cap \Gamma(t))} \bm{J}_{\Gamma} \cdot \bm{\mu} \\
= & \sum_{i=1,2} -\int_{\pd(V(t) \cap \Omega^{(i)}(t))} \bm{J}_{c}^{(i)} \cdot \bm{\nu}_{ext}^{(i)} - \int_{V(t) \cap \Gamma(t)} ([\bm{J}_{c}^{(i)}]_{1}^{2} \bm{\nu} + \surf \cdot \bm{J}_{\Gamma})
\end{align*}
for the right hand side. Hence, using $(\ref{eq:generalC})$ the mass balance $(\ref{eq:changeoftotalmass})$ yields the following pointwise law for the interfacial surfactant:
\begin{align*}
 \md c^{\Gamma} + c^{\Gamma} \surf \cdot \bm{v} = - \surf \cdot \bm{J}_{\Gamma} + q_{AD},
\end{align*}
where $q_{AD} = -[\bm{J}_{c}^{(i)}]_{1}^{2} \bm{\nu}$ is the mass flux for the transfer of surfactant to the interface from the adjacent sub-layers.  When the mass flux $q_{AD}$ is zero and the interfacial molecular flux is modelled by Fick's law, $\bm{J}_{\Gamma} = - D_{s}\surf c^{\Gamma}$, we obtain the mass balance equation in \cite{article:WongRumschitzkiMaldarelli96}.

\subsection{Energy inequality}
We postulate a total energy of the form 
\begin{align}\label{eq:SIM_FreeEnergy}
 \int_{\Omega^{(1)}(t)}[\tfrac{\overline{\rho}^{(1)}}{2}\abs{\bm{v}}^{2} + G_{1}(c^{(1)}) ]+ \int_{\Omega^{(2)}(t)} [\tfrac{\overline{\rho}^{(2)}}{2} \abs{\bm{v}}^{2} + G_{2}(c^{(2)})] + \int_{\Gamma(t)} \gamma(c^{\Gamma}),
\end{align} 
where $G_{1}, G_{2}$ are the bulk free energy densities, and $\gamma$ is a surface free energy density.  We assume that $\gamma'' > 0$ and $G_{i}'' > 0$. The Legendre transform of the surface energy density then is well defined, and the density dependent surface tension $\sigma(c^{\Gamma})$ is defined as
\begin{align}\label{surfacetension}
 \sigma(c^{\Gamma}) := \gamma(c^{\Gamma}) - c^{\Gamma} \gamma'(c^{\Gamma}).
\end{align}
Let $V(t)$ be an arbitrary material test volume. Then
\begin{align*}
& \frac{d}{dt} \left ( \sum_{i=1}^{2} \int_{V(t) \cap \Omega^{(i)}(t)}(\tfrac{\overline{\rho}^{(i)}}{2}\abs{\bm{v}}^{2} + G_{i}(c^{(i)}) ) + \int_{V (t)\cap \Gamma(t)} \gamma(c^{\Gamma}) \right ) \\
=& \sum_{i=1}^{2} \int_{V \cap \Omega^{(i)}} \left ( \overline{\rho}^{(i)} \bm{v} \cdot \md \bm{v} + G'_{i}(c^{(i)}) \md c^{(i)} \right ) + \int_{V \cap \Gamma} \left ( \gamma'(c^{\Gamma}) \md c^{\Gamma} + \gamma(c^{\Gamma}) \surf \cdot \bm{v} \right ) \\
=& \sum_{i=1}^{2} \int_{V(t) \cap \Omega^{(i)}(t)}\left ( \nabla \cdot ((\bm{T}^{(i)})^{\perp} \bm{v}- G'_{i}(c^{(i)}) \bm{J}_{c}^{(i)}) - \bm{T}^{(i)} \colon \nabla \bm{v} + \nabla G'_{i}(c^{(i)}) \cdot \bm{J}_{c}^{(i)} \right )\\
& + \int_{V(t) \cap \Gamma(t)} \gamma'(c^{\Gamma}) (- \surf \cdot \bm{J}_{\Gamma} + q_{AD}) + \sigma(c^{\Gamma}) \surf \cdot \bm{v}   \displaybreak[0]  \\ 
 =& \sum_{i=1}^{2} \int_{V(t) \cap \Omega^{(i)}(t)} -\bm{T}^{(i)} \colon \nabla \bm{v} + \nabla G'_{i}(c^{(i)}) \cdot \bm{J}_{c}^{(i)} + \int_{\pd(V(t) \cap \Gamma(t))} - \gamma'(c^{\Gamma}) \bm{J}_{\Gamma}\cdot \bm{\mu}\\
& + \sum_{i=1}^{2} \int_{\pd(V(t) \cap \Omega^{(i)}(t))\setminus \Gamma(t)} ((\bm{T}^{(i)})^{\perp} \bm{v} - G'_{i}(c^{(i)}) \bm{J}_{c}^{(i)}) \cdot \bm{\nu}_{ext} \\
& + \int_{V(t) \cap \Gamma(t)} ((\bm{T}^{(1)})^{\perp} \bm{v} - G_{1}'(c^{(1)}) \bm{J}_{c}^{(1)}) \cdot \bm{\nu} + ((\bm{T}^{(2)})^{\perp} \bm{v} - G_{2}'(c^{(2)}) \bm{J}_{c}^{(2)}) \cdot (-\bm{\nu}) \\
& + \int_{V(t) \cap \Gamma(t)} \bm{J}_{\Gamma} \cdot \surf \gamma'(c^{\Gamma}) + \gamma'(c^{\Gamma})(\bm{J}_{c}^{(1)} \cdot \bm{\nu} - \bm{J}_{c}^{(2)} \cdot \bm{\nu}) + \sigma(c^{\Gamma}) \surf \cdot \bm{v}. 
\end{align*}
Decomposing the velocity field $\bm{v}$ on $\Gamma(t)$ into its normal and tangential components
\begin{align*}
\bm{v} = u_{\Gamma} \bm{\nu} + \bm{v}_{\tau},
\end{align*}
then gives
\begin{align*}
&\int_{V(t) \cap \Gamma(t)} \sigma(c^{\Gamma}) \surf \cdot (u_{\Gamma} \bm{\nu} + \bm{v}_{\tau}) = \int_{V(t) \cap \Gamma(t)} \sigma(c^{\Gamma}) (\underbrace{\surf u_{\Gamma} \cdot \bm{\nu}}_{=0} + \underbrace{u_{\Gamma} \surf \cdot \bm{\nu}}_{- \kappa u_{\Gamma}} + \surf \cdot \bm{v}_{\tau}) \\
& = \int_{V(t) \cap \Gamma(t)} - \sigma(c^{\Gamma}) \kappa u_{\Gamma} - \surf \sigma(c^{\Gamma}) \cdot \bm{v} + \int_{\pd(V(t) \cap \Gamma(t))} \sigma(c^{\Gamma}) \bm{v}_{\tau} \cdot \bm{\mu},
\end{align*}
where $\kappa = - \surf \cdot \bm{\nu}$ is the mean curvature and we have used integration by parts to obtain the last equality.  Altogether we have 
\begin{align*}
& \frac{d}{dt} \left ( \sum_{i=1}^{2} \int_{V(t) \cap \Omega^{(i)}(t)}[\tfrac{\overline{\rho}^{(i)}}{2}\abs{\bm{v}}^{2} + G_{i}(c^{(i)}) ] + \int_{V(t) \cap \Gamma(t)} \gamma(c^{\Gamma}) \right ) \\
=& \sum_{i=1}^{2} \int_{\pd(V(t) \cap \Omega^{(i)}(t))\setminus \Gamma(t)} ((\bm{T}^{(i)})^{\perp} \bm{v} - G'_{i}(c^{(i)}) \bm{J}_{c}^{(i)}) \cdot \bm{\nu}_{ext} \\
& + \int_{\pd(V(t) \cap \Gamma(t))} \left ( - \gamma'(c^{\Gamma}) \bm{J}_{\Gamma}\cdot \bm{\mu} + \sigma(c^{\Gamma}) \bm{v}_{\tau} \cdot \bm{\mu} \right ) \\
& + \sum_{i=1}^{2} \int_{V(t) \cap \Omega^{(i)}(t)} \left ( -\bm{T}^{(i)} \colon \nabla \bm{v} + \nabla G'_{i}(c^{(i)}) \cdot \bm{J}_{c}^{(i)} \right ) \\
& + \int_{V(t) \cap \Gamma(t)} \bm{J}_{\Gamma} \cdot \surf \gamma'(c^{\Gamma}) \displaybreak[0] \\ 
& + \int_{V(t) \cap \Gamma(t)}\left ( (\gamma'(c^{\Gamma}) - G_{1}'(c^{(1)})) \bm{J}_{c}^{(1)} \cdot \bm{\nu} - (\gamma'(c^{\Gamma}) - G_{2}'(c^{(2)})) \bm{J}_{c}^{(2)} \cdot \bm{\nu} \right ) \\
& + \int_{V(t) \cap \Gamma(t)} \left ( \bm{T}^{(1)} \bm{\nu} \cdot \bm{v} - \bm{T}^{(2)} \bm{\nu} \cdot \bm{v} - \sigma(c^{\Gamma}) \kappa \bm{v} \cdot \bm{\nu} - \surf \sigma(c^{\Gamma}) \cdot \bm{v} \right ). 
\end{align*}

Hence, if
\begin{align*}
\bm{J}_{c}^{(i)} \cdot \nabla G_{i}'(c^{(i)}) & \leq 0, &\quad& \text{ in } \Omega^{(i)}(t), \quad i = 1, 2, \\
\bm{T}^{(i)} \colon \nabla \bm{v} & \geq 0, &\quad& \text{ in } \Omega^{(i)}(t), \quad i = 1, 2,  \displaybreak[0] \\
\bm{J}_{\Gamma} \cdot \surf \gamma'(c^{\Gamma}) & \leq 0, &\quad& \text{ on } \Gamma(t), \\
(\bm{J}_{c}^{(1)} \cdot \bm{\nu})(\gamma'(c^{\Gamma}) - G_{1}'(c^{(1)})) & \leq 0, &\quad& \text{ on } \Gamma(t), \\
(-\bm{J}_{c}^{(2)} \cdot \bm{\nu})(\gamma'(c^{\Gamma}) - G_{2}'(c^{(2)})) & \leq 0, &\quad& \text{ on } \Gamma(t), \\
(-[\bm{T}]_{1}^{2}\bm{\nu} - \sigma(c^{\Gamma})\kappa \bm{\nu} - \surf \sigma(c^{\Gamma}))\cdot \bm{v} & \leq 0, &\quad& \text{ on } \Gamma(t),
\end{align*}
then we obtain the following energy inequality:
\begin{align*}
& \frac{d}{dt} \left ( \sum_{i=1}^{2} \int_{V(t) \cap \Omega^{(i)}(t)}(\tfrac{\overline{\rho}^{(i)}}{2}\abs{\bm{v}}^{2} + G_{i}(c^{(i)}) ) + \int_{V(t) \cap \Gamma(t)} \gamma(c^{\Gamma}) \right ) \\
& \leq \sum_{i=1}^{2} \left ( \int_{\pd(V(t) \cap \Omega^{(i)}(t))\setminus \Gamma(t)} ((\bm{T}^{(i)})^{\perp} \bm{v} - G'_{i}(c^{(i)}) \bm{J}_{c}^{(i)}) \cdot \bm{\nu}_{ext} \right ) \\
& + \int_{\pd(V(t) \cap \Gamma(t))} \left ( - \gamma'(c^{\Gamma}) \bm{J}_{\Gamma}\cdot \bm{\mu} + \sigma(c^{\Gamma}) \bm{v}_{\tau} \cdot \bm{\mu} \right ),
\end{align*}
where the right hand side represents the working on the arbitrary material test volume $V(t)$ and the inequality indicates that the dissipation is non-positive, thus guaranteeing thermodynamic consistency \cite{article:FriedGurtin93,article:GurtinPolignoneVinals96}.

\subsection{General model}
We make the following constitutive assumptions:
\begin{align}
\nonumber \bm{J}_{c}^{(i)} & = -M_{c}^{(i)}(c^{(i)}) \nabla G'_{i}(c^{(i)}), \\
\nonumber \bm{J}_{\Gamma} & = - M_{\Gamma}(c^{\Gamma}) \surf \gamma'(c^{\Gamma}), \\
\alpha^{(i)}(c^{\Gamma},c^{(i)})(-1)^{i+1}\bm{J}^{(i)}_{c} \cdot \bm{\nu} & = -(\gamma'(c^{\Gamma}) - G_{i}'(c^{(i)})), \label{eq:alpha}\\
\nonumber \bm{T}^{(i)} & = - p \bm{I} + 2\eta^{(i)} D(\bm{v}), \\
\nonumber -[\bm{T}]_{1}^{2} \bm{\nu} & = \sigma(c^{\Gamma}) \kappa \bm{\nu} + \surf \sigma(c^{\Gamma}),
\end{align}
where $M_{c}^{(i)}(c^{(i)}) > 0$, $M_{\Gamma}(c^{\Gamma}) > 0$, and $\alpha^{(i)}(c^{\Gamma},c^{(i)}) \geq 0$.

The formulation presented in $(\ref{eq:alpha})$ utilises a free energy approach, first applied to the kinetics of surfactant adsorption in \cite{article:DiamantAndelman96, article:DiamantArielAndelman01}, to model instantaneous adsorption kinetics.  At adsorption/desorption equilibrium, the chemical potentials $\gamma'(c^{\Gamma})$ and $G'(c)$ must be equal \cite{article:Zhdanov01,article:LiuZhang10,article:vanderSmanvanderGraaf06} and thus this approach allows us to cover the adsorption isotherms often used in the literature by selecting suitable functional forms for $\gamma$ and $G$.  Hence, $\alpha^{(i)} > 0$ can be seen as a kinetic factor which relates the speed of adsorption to the interface or desorption from the interface to the deviation from local thermodynamical equilibrium. Let us summarise the governing equations of the general model for two-phase flow with soluble surfactant:

\textbf{Balance equations in $\Omega^{(i)}(t), \; i= 1, 2$ :}
\begin{align}
\nabla \cdot \bm{v} & = 0, \label{SIM:eq1} \\
\pd_{t}(\overline{\rho}^{(i)} \bm{v}) + \nabla \cdot (p \bm{I} - 2\eta^{(i)}D(\bm{v}) + \overline{\rho}^{(i)} \bm{v} \otimes \bm{v}) & = 0, \\
\md c^{(i)} - \nabla \cdot (M_{c}^{(i)} \nabla G_{i}'(c^{(i)})) & = 0. \label{SIM:eq3}
\end{align}
\textbf{Free boundary conditions on $\Gamma(t)$:}
\begin{align}
[\bm{v}]_{1}^{2} & = 0, \quad \bm{v} \cdot \bm{\nu} = u_{\Gamma}, \\
[p]_{1}^{2} \bm{\nu} - 2[\eta^{(i)} D(\bm{v})]_{1}^{2} \bm{\nu} & = \sigma(c^{\Gamma})\kappa \bm{\nu} + \surf \sigma(c^{\Gamma}), \\
\md c^{\Gamma} + c^{\Gamma} \surf \cdot \bm{v} & =  \surf \cdot (M_{\Gamma}\surf \gamma'(c^{\Gamma})) +[M_{c}^{(i)} \nabla G_{i}'(c^{(i)})]_{1}^{2}\bm{\nu} , \label{SIM:eq6} \\
\alpha^{(i)}(-1)^{i}M_{c}^{(i)}\nabla G_{i}'(c^{(i)}) \cdot \bm{\nu} & = -(\gamma'(c^{\Gamma}) - G_{i}'(c^{(i)})). \label{SIM:eq7}
\end{align}

In this model, the surface tension $\sigma : \R^{+} \to \R^{+}$ is a (usually decreasing) function of the surfactant density $c^{\Gamma}$. The phenomenon known as Marangoni effect, where tangential stress at the phase boundary leads to flows along the interface, is incorporated into the model via the surface gradient of $\sigma$ in the momentum jump free boundary condition.  

\subsection{Specific models}
\subsubsection{Fick's law for fluxes}\label{subsubsec:Ficks}
By appropriate choice of the mobilities we obtain Fick's law for the surfactant both in the bulk and on the surface. If we set
\begin{align*}
M_{c}^{(i)}(c^{(i)}) = D_{c}^{(i)} \frac{1}{G''_{i}(c^{(i)})}, \quad M_{\Gamma}(c^{\Gamma}) = D_{\Gamma} \frac{1}{\gamma''(c^{\Gamma})},
\end{align*}
for constant Fickian diffusivities $D_{c}^{(i)}, D_{\Gamma} > 0$.  Then
\begin{align*}
\bm{J}_{c}^{(i)}  = -D_{c}^{(i)}\nabla c^{(i)}, \quad \bm{J}_{\Gamma} = - D_{\Gamma}\surf c^{\Gamma}. 
\end{align*}

\subsubsection{Instantaneous adsorption and local equilibrium}\label{subsubsec:InstantaneousAdsorptionLocalEquil}
We may assume that the process of adsorption of surfactant at the interface is instantaneous, i.e. fast compared to the timescale of convective and diffusive transport. This local equilibrium corresponds to the case that the bulk chemical potential $G'(c)$ and the interface chemical potential $\gamma'(c^{\Gamma})$ are equal, i.e. we set $\alpha = 0$ in $(\ref{eq:alpha})$ (we here only consider one of the bulk phases adjacent to the interface and, for simplicity, drop the upper index ${(i)}$). We obtain the following relation (also see \cite{article:BothePruss10,incoll:BothePrussSimonett}):
\begin{align}\label{eq:instadsorption}
 \gamma'(c^{\Gamma})  = G'(c) \quad \Longleftrightarrow \quad  c^{\Gamma} = g(c) := (\gamma')^{-1}(G'(c)),
\end{align}
where $g: \R_{+} \to \R_{+}$ is strictly increasing.  This function $g$ plays the role of various adsorption isotherms which state the equilibrium relations between the two densities.

Table \ref{tbl:Isotherms} displays the functional forms for $\gamma$ and $G$ in order to obtain the adsorption isotherms of Henry, Langmuir, Freundlich, Volmer and Frumkin. The free energies are (variants of) ideal solutions. Here, $c^{\Gamma}_{M}$ is the maximum surfactant density on the interface, $K$ a constant relating the surface density to the bulk density in equilibrium, $\sigma_{0}$ denotes the surface tension of a clean interface, $B$ essentially is the sensitivity of the surface tension to surfactant, $A$ in the Frumkin isotherm is known as surface interaction parameter while, in the Freundlich isotherm, $A_{c}$ measures the adsorbent capacity and $N$ is the intensity of adsorption.

\begin{table}[p]
\begin{tabular}{|c|c|c|}
 \hline
Isotherm & Henry & Langmuir \\
\hline
Relation  & $Kc = \frac{c^{\Gamma}}{c^{\Gamma}_{M}}$ & \gape{$Kc = \frac{c^{\Gamma}}{c^{\Gamma}_{M} - c^{\Gamma}}$} \\ [2ex]
\hline
$\gamma(c^{\Gamma}) - \sigma_{0}$ & $ B c^{\Gamma}(\log \frac{c^{\Gamma}}{c^{\Gamma}_{M}} -1)$ & \gape{$B \left (c^{\Gamma} \log \frac{c^{\Gamma}}{c^{\Gamma}_{M} - c^{\Gamma}} + c^{\Gamma}_{M} \log ( 1 - \frac{c^{\Gamma}}{c^{\Gamma}_{M}} ) \right ) $} \\ [2ex]
\hline 
$G(c) $ & $Bc(\log(Kc)-1)$ & \gape{$B c (\log(Kc)-1)$}  \\ 
\hline
$\sigma - \sigma_{0} $ & $-B c^{\Gamma}$  & \gape{$Bc^{\Gamma}_{M} \log \left ( 1 - \frac{c^{\Gamma}}{c^{\Gamma}_{M}} \right )$} \\ [2ex]
\hline
\end{tabular}
\vskip3pt
\begin{tabular}{|c|c|c|}
\hline
 Isotherm & Freundlich & Volmer \\
\hline
Relation & \gape{$Kc = \frac{1}{A_{c}} \left (\frac{c^{\Gamma}}{c^{\Gamma}_{M}} \right )^{N}$} & \gape{$Kc = \frac{c^{\Gamma}}{c^{\Gamma}_{M} - c^{\Gamma}}\exp \left ( \frac{c^{\Gamma}}{c^{\Gamma}_{M} - c^{\Gamma}} \right )$} \\ [2ex]
\hline
$\gamma(c^{\Gamma}) - \sigma_{0}$  & $NB c^{\Gamma}(\log \frac{c^{\Gamma}}{c^{\Gamma}_{M}} -1 )$ &  \gape{$B c^{\Gamma}\log \frac{c^{\Gamma}}{c^{\Gamma}_{M}- c^{\Gamma}} $} \\ [2ex]
\hline 
$G(c) $ & $B c(\log (A_{c}^{N}Kc) - 1 )$ & \gape{$B c\log(Kc)$}  \\ 
\hline
$\sigma - \sigma_{0} $ & $-NB c^{\Gamma} $ & \gape{$-B \frac{c^{\Gamma} c^{\Gamma}_{M}}{c^{\Gamma}_{M} - c^{\Gamma}} $} \\ [2ex]
\hline
\end{tabular}
\vskip3pt
\begin{tabular}{|c|c|}
\hline
Isotherm & Frumkin  \\
\hline
Relation & \gape{$Kc = \frac{c^{\Gamma}}{c^{\Gamma}_{M} - c^{\Gamma}}\exp \left (- \frac{A c^{\Gamma}}{B} \right )$}   \\ [2ex]
\hline
$\gamma(c^{\Gamma}) - \sigma_{0}  $ & \gape{$B \left ( c^{\Gamma} \log \frac{c^{\Gamma}}{c^{\Gamma}_{M} - c^{\Gamma}} + c^{\Gamma}_{M} \log ( 1 - \frac{ c^{\Gamma}}{c^{\Gamma}_{M}} ) \right ) - \frac{A (c^{\Gamma})^{2}}{2 }$} \\ [2ex]
\hline
$G(c) $ & \gape{$Bc (\log(Kc)-1)$}  \\ 
\hline
$\sigma - \sigma_{0} $ & \gape{$\frac{A(c^{\Gamma})^{2}}{2} + Bc^{\Gamma}_{M} \log \left ( 1 - \frac{c^{\Gamma}}{c^{\Gamma}_{M}} \right ) $} \\ [2ex]
\hline
\end{tabular}
\caption{Possible functional forms for $\gamma$ and $G$ to obtain the most frequently used adsorption isotherms and equations of state.}
\label{tbl:Isotherms}
\end{table}

\subsubsection{Insoluble surfactants}\label{subsubsec:Insoluble}
Neglecting $(\ref{SIM:eq3})$, $(\ref{SIM:eq7})$ and the jump term in $(\ref{SIM:eq6})$ gives a two-phase flow model with insoluble surfactant.
\subsection{Reformulation of the surfactant equations}
The strong form of the surfactant equations $(\ref{SIM:eq3}),(\ref{SIM:eq6}),(\ref{SIM:eq7})$ can be reformulated into an equivalent distributional form using a result from Alt \cite{article:Alt09}.  Let $\chi_{\Omega^{(i)}}$ and $\delta_{\Gamma}$ denote the distributions given by the indicator functions on $\Omega^{(i)}$ and $\Gamma$ respectively, see the Appendix for a precise definition.  We now define
\begin{align*} 
j_{1} = \frac{1}{\alpha^{(1)}}(\gamma'(c^{\Gamma}) - G_{1}'(c^{(1)})), \quad j_{2} = \frac{1}{\alpha^{(2)}}(\gamma'(c^{\Gamma}) - G_{2}'(c^{(2)})).
\end{align*}
In the Appendix we show that 
\begin{align}
\pd_{t}(\chi_{\Omega^{(1)}}c^{(1)}) + \nabla \cdot (\chi_{\Omega^{(1)}} c^{(1)} \bm{v} - \chi_{\Omega^{(1)}}M_{c}^{(1)} \nabla G_{1}'(c^{(1)})) & = \delta_{\Gamma} j_{1}, \label{eq:bulk1distributional} \\
\pd_{t}(\chi_{\Omega^{(2)}}c^{(2)}) + \nabla \cdot (\chi_{\Omega^{(2)}} c^{(1)} \bm{v} - \chi_{\Omega^{(2)}}M_{c}^{(2)} \nabla G_{2}'(c^{(2)})) & = \delta_{\Gamma} j_{2}, \label{eq:bulk2distributional} \\
\pd_{t}(\delta_{\Gamma} c^{\Gamma}) + \nabla \cdot (\delta_{\Gamma} c^{\Gamma} \bm{v} - M_{\Gamma} \delta_{\Gamma} \nabla \gamma'(c^{\Gamma})) & = -\delta_{\Gamma} (j_{1} + j_{2}), \label{eq:interfacedistributional}
\end{align} 
interpreted in its distributional formulation are equivalent to
\begin{align*}
 \pd_{t}c^{(1)} + \nabla \cdot (c^{(1)} \bm{v} - M_{c}^{(1)}\nabla G_{1}'(c^{(1)})) & = 0, \text{ in } \Omega^{(1)}, \\
 M_{c}^{(1)} \nabla G_{1}'(c^{(1)})\cdot \bm{\nu} & = j_{1}, \text{ on } \Gamma, \\
 \pd_{t}c^{(2)} + \nabla \cdot (c^{(2)} \bm{v} - M_{c}^{(2)}\nabla G_{2}'(c^{(2)})) & = 0, \text{ in } \Omega^{(2)} ,\\
 -M_{c}^{(2)} \nabla G_{2}'(c^{(2)})\cdot \bm{\nu} & = j_{2}, \text{ on } \Gamma
\end{align*}
and $(\ref{SIM:eq6})$ respectively.

\subsection{Non-dimensional evolution equations}
To derive equations in a dimensionless form we pick a length scale $L$, a time scale $T$ (or, equivalently, a scale for the velocity $V = L/T$), a scale $\Sigma$ for the surface tension, and let $C^{\Gamma} = L^{-2}, C = L^{-3}$ denote scales for the surfactant densities in the interface and in the bulk, respectively.

The Reynolds number, as the ratio of advective to viscous forces, is defined as $\text{Re} := (\overline{\rho}^{(2)}L^2)/(\eta^{(2)}T)$. The capillary number, as the ratio of viscous to surface tension forces, is defined as $\text{Ca} = (\eta^{(2)}L)/(T \Sigma)$.  Scaling the pressure by $T^2/(\overline{\rho}^{(2)}L^2)$ we arrive at the following dimensionless fluid equations 
\begin{align}
 \nabla_{*} \cdot \bm{v}_{*} & = 0, \label{Nond:fluid}\\
\pd_{t_{*}}(\overline{\rho}^{\pm} \bm{v}_{*}) + \nabla_{*} \cdot \left ( p_{*} \bm{I} - \frac{2 \eta^{\pm}}{\text{Re}}D(\bm{v}_{*}) + \overline{\rho}^{\pm} \bm{v}_{*} \otimes \bm{v}_{*} \right ) & = 0, \\
[\bm{v}_{*}]_{1}^{2} = 0, \quad \bm{v}_{*} \cdot \bm{\nu} & = u_{\Gamma_{*}}, \\
\left [ p_{*}\bm{I} - \frac{2\eta^{\pm}}{\text{Re}} D(\bm{v}_{*}) \right ]_{1}^{2} \bm{\nu} = \frac{1}{\text{ReCa}}(\sigma_{*} \kappa \bm{\nu}  + & \surfND  \sigma_{*}),
\end{align}
where $\eta^{+} = 1$, $\eta^{-} = \eta^{(1)}/\eta^{(2)}$, $\overline{\rho}^{+} = 1$, $\overline{\rho}^{-} = \overline{\rho}^{(1)}/\overline{\rho}^{(2)}$.  Let 
\begin{align*}
 \gamma_{*} = \frac{\gamma}{\Sigma}, \quad G_{i,*} = \frac{G_{i} L}{\Sigma}, \quad M_{c,*}^{(i)} = M_{c}^{(i)}\Sigma T L^{3}, \quad M_{\Gamma,*} = M_{\Gamma}\Sigma T L^{2},
\end{align*}
where $\gamma_{*}, G_{i,*}$ denote the dimensionless free energies and $M_{c,*}^{(i)}, M_{\Gamma,*}$ denote the dimensionless mobilities.  The dimensionless surfactant equations are given by
\begin{align}
 \mdND c^{(i)}_{*} - \nabla_{*} \cdot \left ( M_{c,*}^{(i)} \nabla_{*} G'_{i,*}(c^{(i)}_{*}) \right ) & = 0, \\
 \mdND c^{\Gamma}_{*} + c^{\Gamma}_{*} \surfND \cdot \bm{v}_{*} - \surfND \cdot \left ( M_{\Gamma,*} \surfND \gamma'_{*}(c^{\Gamma}_{*}) \right ) & = \left [ M_{c,*}^{(i)} \nabla_{*} G'_{i,*}(c^{(i)}_{*}) \right ]_{1}^{2} \bm{\nu}, \\
 \alpha_{*}^{(i)}(-1)^{i} M_{c,*}^{(i)}\nabla_{*} G_{i}'(c^{(i)}_{*}) \cdot \bm{\nu} & = -(\gamma_{*}'(c^{\Gamma}_{*}) - G'_{*,i}(c^{(i)}_{*})), \label{Nond:adsorp}
\end{align}
where $\alpha_{*}^{(i)} = \alpha^{(i)}/(T \Sigma L^{4})$ is the dimensionless kinetic factor.  If we consider the mobilities in Section \ref{subsubsec:Ficks}, then we have the relation
\begin{align*}
 M_{c,*}^{(i)} = \frac{1}{\text{Pe}_{c,i}} \frac{1}{G_{i,*}''(c^{(i)}_{*})}, \quad M_{\Gamma,*} = \frac{1}{\text{Pe}_{\Gamma}} \frac{1}{\gamma_{*}''(c^{\Gamma}_{*})},
\end{align*}
where $\text{Pe}_{c,i} = L^2/(T D_{c}^{(i)})$, as the ratio of advection to diffusion of bulk surfactants, is the bulk Peclet number and $\text{Pe}_{\Gamma} = L^2/(T D_{\Gamma})$ is the corresponding interface Peclet number.  The dimensionless surfactant equations with Fickian diffusion read as
\begin{align}
 \mdND c^{(i)}_{*} - \nabla_{*} \cdot \left ( \frac{1}{\text{Pe}_{c,i}} \nabla_{*} c^{(i)}_{*} \right ) & = 0, \\
 \mdND c^{\Gamma}_{*} + c^{\Gamma}_{*} \surfND \cdot \bm{v}_{*} - \surfND \cdot \left ( \frac{1}{\text{Pe}_{\Gamma}} \surfND c^{\Gamma}_{*} \right ) & = \left [ \frac{1}{\text{Pe}_{c,i}}\nabla_{*} c^{(i)}_{*} \right ]_{1}^{2}  \bm{\nu}, \\
 \alpha_{*}^{(i)}\frac{(-1)^{i}}{\text{Pe}_{c,i}} \nabla_{*} c^{(i)}_{*} \cdot \bm{\nu} & = -(\gamma_{*}'(c^{\Gamma}_{*}) - G'_{*,i}(c^{(i)}_{*})).
\end{align}

\section{Phase field model}\label{PFM}
\subsection{Model for two-phase fluid flow}
In this section we will derive a phase field model for two-phase flow with surfactant generalizing the work by Abels, Garcke and Gr\"un on phase field modelling of two-phase flow \cite{article:AbelsGarckeGrun}. We start by recapitulating their essential assumptions and governing equations.

For a test volume $V \subset \Omega$, let $\rho$ denote the total mass density of the mixture in $V$ and, for $i = 1, 2$, denote by $\overline{\rho}^{(i)}, V_{i}$ the bulk density and the volume occupied by fluid $i$ in $V$, respectively.  Let $u_{i} = V_{i}/V$ denote the volume fraction occupied by fluid $i$ in $V$.  Assuming zero excess volume due to mixing, we have
\begin{align}\label{density:zeroexcess}
 u_{1} + u_{2} = 1.
\end{align}
Then the total density $\rho$ can be expressed as a function of the difference in volume fraction $\varphi = u_{2} - u_{1}$, which is a natural choice for the order parameter that distinguishes the two fluids,
\begin{align*}
 \rho = \rho(\varphi) = \frac{\overline{\rho}^{(2)}(1+\varphi)}{2} + \frac{\overline{\rho}^{(1)}(1-\varphi)}{2} = \frac{\overline{\rho}^{(2)}-\overline{\rho}^{(1)}}{2}\varphi + \frac{\overline{\rho}^{(2)} + \overline{\rho}^{(1)}}{2}.
\end{align*}
As in \cite{article:AbelsGarckeGrun,article:GurtinPolignoneVinals96}, we assume that the inertia and kinetic energy due to the motion of the fluid relative to the gross motion is negligible.  Therefore we consider the mixture as a single fluid with velocity $\bm{v}$.  If one chooses $\bm{v}$ to be the volume averaged velocity then the prototype diffuse interface model for incompressible two-phase flow with different densities is:
\begin{align}
 \nabla \cdot \bm{v} & = 0, \label{proto:massbalance} \\
 \pd_{t}(\rho \bm{v}) + \nabla \cdot (\rho \bm{v} \otimes \bm{v}) & = \nabla \cdot \bm{T}, \label{proto:momentumbalance} \\
\pd_{t}\varphi + \nabla \cdot (\varphi \bm{v}) & = - \nabla \cdot \bm{J}_{\varphi}, \label{proto:phase}
\end{align}
where $\bm{T}$ is a tensor yet to be specified, $\bm{J}_{\varphi}$ is a flux related to the mass flux $\overline{\bm{J}}$ by
\begin{align}\label{PFM:flux}
(\overline{\rho}^{(2)} - \overline{\rho}^{(1)}) \bm{J}_{\varphi} = 2\overline{\bm{J}}.
\end{align}
As a consequence of (\ref{proto:phase}) we obtain the mass balance law
\begin{align}
\pd_{t}\rho +  \nabla \cdot (\rho \bm{v})  = - \nabla \cdot \overline{\bm{J}} . \label{proto:density} 
\end{align}

Our goal is now to extend this model to the case where surfactants are present, distinguishing the cases of dynamic and instantaneous adsorption. We proceed as in the sharp interface setting by postulating appropriate mass balance equation(s) for the surfactant and deriving models from constitutive assumptions such that thermodynamic consistency is guaranteed.

\subsection{Dynamic adsorption (Model A)}
\subsubsection{Mass balance equations}
We will use the distributional forms for the bulk and interfacial surfactant equations to derive the phase field surfactant equations.  Since the sharp interface is replaced by an interfacial layer, we consider regularisations of $\chi_{\Omega^{(i)}}$ and $\delta_{\Gamma}$ that appear in $(\ref{eq:bulk1distributional}),(\ref{eq:bulk2distributional}),(\ref{eq:interfacedistributional})$.  In the context of phase field models, many regularisations of the delta function are available from the literature \cite{article:TeigenSongLowengrubVoigt11, article:ElliottStinnerStylesWelford11, article:RatzVoigt06}, but it will turn out that the Ginzburg--Landau free energy density $$\delta(\varphi, \nabla \varphi) = \frac{\eps}{2} \abs{\nabla \varphi}^{2} + \frac{1}{\eps}W(\varphi)$$ is a suitable regularisation for a multiple of $\delta_{\Gamma}$, where $\eps$ is a measure of interfacial thickness and $W(\varphi)$ is a potential of double-well or double-obstacle type \cite{incoll:BloweyElliott93} with equal minima at $\varphi = \pm 1$ and symmetric about $\varphi = 0$.  For example, one can choose $W(\varphi) = \frac{1}{4}(1-\varphi^{2})^{2}$ for a potential of double-well type or
\begin{align*}
 W(\varphi) = \frac{1}{2}(1-\varphi^2) + I_{[-1,1]}(\varphi), \quad I_{[-1,1]}(\varphi) = 
\begin{cases}
0, & \text{ if } \abs{\varphi} \leq 1, \\
\infty, & \text{ else}
\end{cases}
\end{align*}
for a potential of double-obstacle type. However, in the following derivation we assume a smooth potential for convenience.  The potential term $W(\varphi)$ in $\delta(\varphi, \nabla \varphi)$ prefers the order parameter $\varphi$ in its minima at $\pm 1$ and the gradient term $\abs{\nabla \varphi}^2$ penalises large jumps in gradient.  This leads to the development of regions where $\varphi$ is close to $\pm 1$ which are separated by a narrow interfacial layer.  For the regularisation of $\chi_{\Omega^{(2)}}$, we consider $\xi_{2}(\varphi)$ to be a non-negative cut-off function such that $\xi_{2}(1) = 1$, $\xi_{2}(-1) = 0$ and $\xi_{2}$ varies smoothly across $\abs{\varphi} < 1$.  For example, in some of the subsequent numerical experiments we used
\begin{align*}
  \xi_{2}(\varphi) = \begin{cases}
           1, & \varphi \geq 1, \\
\frac{1}{2} (1+\frac{1}{2}\varphi(3-\varphi^2)), & \abs{\varphi} < 1, \\
0, & \varphi \leq -1.
          \end{cases}
\end{align*}
Similarly, $\xi_{1}(\varphi) = 1 - \xi_{2}(\varphi)$ will be the regularisation of $\chi_{\Omega^{(1)}}$.

Our ansatz for the case of dynamic adsorption of the surfactant to the interface is motivated by the distributional formulation in (\ref{eq:bulk1distributional})-(\ref{eq:interfacedistributional})
\begin{align}
& \pd_{t} (\xi_{i}c^{(i)}) + \nabla \cdot (\xi_{i}c^{(i)}\bm{v}) + \nabla \cdot (\xi_{i}\bm{J}_{c}^{(i)}) = \delta j_{i}, \quad i = 1, 2, \label{proto:bulk} \\
& \pd_{t}  ( K\delta c^{\Gamma} ) + \nabla \cdot ( K\delta c^{\Gamma}\bm{v}) + \nabla \cdot \Big{(}K\delta \bm{J}_{\Gamma} \Big{)} = -\delta(j_{1} + j_{2}), \label{proto:interface}
\end{align}
where $K \neq 0$ is a calibration constant which depends on $W$, chosen such that $K\delta(\varphi, \nabla \varphi)$ regularises $\delta_{\Gamma}$, see \cite{article:ModicaMortola77}.  In particular we set
\begin{align*}
 \frac{1}{K} = \mathcal{W} =  
\begin{cases} 
 \displaystyle  \int_{-\infty}^{\infty} 2W(\tanh(z/\sqrt{2})) dz = 2\sqrt{2}/3, & \text{ for the double-well,} \\
 \displaystyle  \int_{-\pi/2}^{\pi/2} 2W(\sin(z)) dz = \pi/2, & \text{ for the double-obstacle.}
\end{cases}
\end{align*}
Furthermore, $\bm{J}_{c}^{(i)}$ is the bulk surfactant flux, $\bm{J}_{\Gamma}$ is the interfacial surfactant flux and $j_{i}, i = 1, 2,$ denote the mass exchange between the bulk and the interfacial regions.  In the above prototype model we allow the situation where there are surfactants present either in both bulk phases or in just one bulk phase.  We denote the former as the two-sided model and the latter as the one-sided model.  In the one-sided model, we set $c^{(1)} \equiv 0, \xi_{1}(\varphi) \equiv 0, j_{1} \equiv 0, \bm{J}_{c,1} \equiv \bm{0}$ and we drop the subscripts so that equations $(\ref{proto:bulk}), (\ref{proto:interface})$ are written as
\begin{align*}
\pd_{t} (\xi(\varphi)c) + \nabla \cdot (\xi(\varphi) c \bm{v}) + \nabla \cdot (\xi(\varphi)\bm{J}_{c}) & = \delta(\varphi, \nabla \varphi)j, \\
\pd_{t} \Big{(} K\delta(\varphi, \nabla \varphi)c^{\Gamma} \Big{)} + \nabla \cdot (K \delta(\varphi, \nabla \varphi) c^{\Gamma} \bm{v}) + \nabla \cdot \Big{(}K\delta(\varphi, \nabla \varphi)\bm{J}_{\Gamma} \Big{)} & = -\delta(\varphi, \nabla \varphi)j.
\end{align*}

Observe that, for a test volume $V(t)$ with external normal $\bm{\nu}$, we have
\begin{align*}
\frac{d}{dt} \Big{(} \sum_{i=1,2} \int_{V(t)} \xi_{i}c^{(i)} + \int_{V(t)} K\delta c^{\Gamma} \Big{)} =- \int_{\pd V(t)} (\xi_{1} \bm{J}_{c}^{(1)} + \xi_{2} \bm{J}_{c}^{(2)} + K\delta  \bm{J}_{\Gamma}) \cdot \bm{\nu},                                                                                                                                                                                                                
\end{align*}
which is analogous to $(\ref{eq:changeoftotalmass})$.

\subsubsection{Energy inequality}
We introduce a Helmholtz free energy density $a(\varphi, \nabla \varphi, c^{(i)}, c^{\Gamma})$ which will play the role of the bulk and interfacial free energy density for the diffuse interface model.  As in the sharp interface setting and in analogy to $(\ref{eq:SIM_FreeEnergy})$ the total energy in a test volume $V$ is the sum of the kinetic and free energy:
\begin{align}\label{totalenergy}
\int_{V} e(\bm{v}, \varphi, \nabla \varphi, c^{(i)}, c^{\Gamma}) = \int_{V}  \rho  \frac{\abs{\bm{v}}^{2}}{2} + \int_{V} a(\varphi, \nabla \varphi, c^{(i)}, c^{\Gamma}) 
\end{align}
where 
\begin{align*}
a(\varphi, \nabla \varphi, c, c^{\Gamma}) = K\delta(\varphi, \nabla \varphi)\gamma(c^{\Gamma})+ \xi_{1}(\varphi)G_{1}(c^{(1)}) + \xi_{2}(\varphi)G_{2}(c^{(2)}).
\end{align*}
Since $K\delta(\varphi, \nabla \varphi)$ approximates $\delta_{\Gamma}$ we can consider the first term as an approximation of the surface free energy density.  We assume that the free energy densities satisfy $\gamma'' > 0, G_{i}'' > 0$ and that the following dissipation law holds pointwise in $V$:
\begin{align}\label{eq:dissipation}
 - \mathcal{D} := \pd_{t} e + \nabla \cdot (\bm{v} e) + \nabla \cdot \bm{J}_{e} \leq 0
\end{align}
where $\bm{J}_{e}$ is an energy flux that we will determine later.  

From $(\ref{proto:density})$ and $(\ref{proto:momentumbalance})$ we have 
\begin{align*}
\pd_{t} \Big{(}\tfrac{\rho \abs{\bm{v}}^{2}}{2} \Big{)} + \nabla \cdot \Big{(} \tfrac{\rho \abs{\bm{v}}^{2}}{2} \bm{v}	 \Big{)} & = -\tfrac{\abs{\bm{v}}^{2}}{2} \nabla \cdot \overline{\bm{J}} + (\nabla \cdot \bm{T})\cdot \bm{v} + [(\nabla \cdot \overline{\bm{J}}) \bm{v}] \cdot \bm{v}\\
& = -\tfrac{\abs{\bm{v}}^{2}}{2} \nabla \cdot \overline{\bm{J}} +  (\nabla \cdot \bm{T}) \cdot \bm{v} + [\nabla \cdot (\bm{v} \otimes \overline{\bm{J}})] \cdot \bm{v} - [(\overline{\bm{J}} \cdot \nabla \bm{v})] \cdot \bm{v}\\
& = \nabla \cdot \Big{(}-\tfrac{\abs{\bm{v}}^{2}}{2} \overline{\bm{J}} + \bm{T}^{\perp} \bm{v} \Big{)} - \bm{T} \colon \nabla \bm{v} + [\nabla \cdot (\bm{v} \otimes \overline{\bm{J}})] \cdot \bm{v} \\
& = \nabla \cdot \Big{(}-\tfrac{\abs{\bm{v}}^{2}}{2} \overline{\bm{J}} + (\bm{T}^{\perp} + [\bm{v} \otimes \overline{\bm{J}}]^{\perp})\bm{v} \Big{)} - (\bm{T} + (\bm{v} \otimes \overline{\bm{J}})) \colon \nabla \bm{v}.
\end{align*}
We use the identities 
\begin{align*}
 \md \nabla \varphi & = \nabla \md \varphi - (\nabla \bm{v})^{\perp}\nabla \varphi, \quad \md(ab) = a\md b + b\md a, \\
\md (K \delta(\varphi, \nabla \varphi) \gamma(c^{\Gamma})) &= \md (K \delta) \gamma(c^{\Gamma}) + \gamma'(c^{\Gamma}) \md c^{\Gamma} K \delta \\
& = \md (K \delta) \gamma(c^{\Gamma}) + \gamma'(c^{\Gamma}) \md (K \delta c^{\Gamma}) - \gamma'(c^{\Gamma})c^{\Gamma} \md (K \delta), \\
\md (\xi_{i}(\varphi) G_{i}(c^{(i)})) &= \md (\xi_{i}(\varphi) c^{(i)})G'_{i}(c^{(i)}) + \md (\xi_{i}(\varphi))(G_{i}(c^{(i)}) - c^{(i)} G'_{i}(c^{(i)}))
\end{align*}
to obtain after some lengthy calculations that
\begin{align*}
 -\mathcal{D} & = \nabla \cdot \Big{(} \bm{J}_{e} - \overline{\bm{J}} \tfrac{\abs{\bm{v}}^{2}}{2} + \bm{T}^{\perp} \bm{v} + (\bm{v} \otimes \overline{\bm{J}})\bm{v} \Big{)} \\
& + \nabla \cdot \Big{(} - K \delta \gamma'(c^{(\Gamma}) \bm{J}_{\Gamma} - \sum_{i=1,2}\xi_{i} G'_{i}(c^{(i)}) \bm{J}_{c}^{(i)} + K \eps \sigma \nabla \varphi \md \varphi \Big{)} \\
& + \nabla \cdot \Big{(} \bm{J}_{\varphi} \Big{(} \sum_{i=1,2} \xi'_{i}(\varphi)(G_{i}(c^{(i)}) - G'_{i}(c^{(i)})c^{(i)}) - \nabla \cdot ( K \eps \sigma \nabla \varphi) + \frac{K}{\eps}\sigma W'(\varphi) \Big{)} \Big{)} \\
& + K \delta \bm{J}_{\Gamma} \cdot \nabla \gamma'(c^{\Gamma}) + \xi_{1} \bm{J}_{c}^{(1)} \cdot \nabla G'_{1}(c^{(1)}) + \xi_{2} \bm{J}_{c}^{(2)} \cdot \nabla G'_{2}(c^{(2)}) \\
& - \delta j_{1}(\gamma'(c^{\Gamma}) - G_{1}'(c^{(1)})) - \delta j_{2}(\gamma'(c^{\Gamma})-G'_{2}(c^{(2)}))\\
& +  \bm{J}_{\varphi} \cdot \nabla \Big{(} \sum_{i=1,2} \xi'_{i}(\varphi)(G_{i}(c^{(i)}) - G'_{i}(c^{(i)})c^{(i)}) - \nabla \cdot ( K \eps \sigma \nabla \varphi) + \frac{K}{\eps}\sigma W'(\varphi) \Big{)}  \\
& + (\nabla \cdot \bm{v}) \Big{(} -\varphi \Big{(} \sum_{i=1,2} \xi'_{i}(\varphi)(G_{i}(c^{(i)}) - G'_{i}(c^{(i)})c^{(i)}) - \nabla \cdot ( K \eps \sigma \nabla \varphi) + \frac{K}{\eps}\sigma W'(\varphi) \Big{)} \Big{)}\\
& + (\nabla \cdot \bm{v}) \Big{(} K \delta \sigma + \xi_{1}(G_{1}(c^{(1)}) - G'_{1}(c^{(1)}) c^{(1)}) + \xi_{2}(G_{2}(c^{(2)}) - G_{2}'(c^{(2)}) c^{(2)})\Big{)} \\
& - \nabla \bm{v} \colon (\bm{T} + \bm{v} \otimes \overline{\bm{J}} + K \eps \sigma \nabla \varphi \otimes \nabla \varphi).
\end{align*}
In the case where the surfactant is present in only one of the bulk phases, a similar calculation shows that we obtain the above form for $-\mathcal{D}$ without any terms involving the subscript $1$. 

In any case, we choose $\bm{J}_{e}$ so that the divergence term cancels.

\subsubsection{Constitutive assumptions}
We set
\begin{align*}
 \mu =  -\nabla \cdot \big{(} K\eps \sigma(c^{\Gamma}) \nabla \varphi \big{)} + \frac{K}{\eps} \sigma(c^{\Gamma}) W'(\varphi) + \sum_{i=1,2} \xi_{i}'(\varphi)(G_{i}(c^{(i)}) - G_{i}'(c^{(i)})c^{(i)})
\end{align*}
and make the following constitutive assumptions:
\begin{align*}
\bm{J}_{\Gamma} & = - M_{\Gamma}(c^{\Gamma}) \nabla \gamma'(c^{\Gamma}), \\
\bm{J}_{c}^{(i)} & = -M_{c}^{(i)}(c^{(i)}) \nabla G_{i}'(c^{(i)}), \\
j_{i} & = \beta^{(i)} \big{(} \gamma'(c^{\Gamma}) - G_{i}'(c^{(i)}) \big{)}, \\
\bm{J}_{\varphi} & = - m(\varphi) \nabla \mu
\end{align*}
for some non-negative function $m(\varphi)$, and the $\beta^{(i)}, i = 1, 2$ are given by   
\begin{align*}
 \beta^{(i)} = \frac{K}{\alpha^{(i)}}.
\end{align*}
We choose the tensor $\bm{T}$ to be
\begin{multline*}
\bm{T} = \Big{(} K \sigma \delta + \sum_{i=1,2} \xi_{i}(G_{i}(c^{(i)}) - G'_{i}(c^{(i)})c^{(i)}) - \varphi \mu \Big{)} \bm{I} \\
- \bm{v} \otimes \overline{\bm{J}} - K \eps \sigma \nabla \varphi \otimes \nabla \varphi + 2 \eta(\varphi)D(\bm{v}) - p \bm{I}
\end{multline*}
where $p$ denotes the unknown pressure, $\eta(\varphi)> 0$ denotes the viscosity and from (\ref{PFM:flux}) the volume diffuse flux $\overline{\bm{J}}$ is given by 
\begin{align*}
 \overline{\bm{J}} = - \tfrac{\overline{\rho}^{(2)} - \overline{\rho}^{(1)}}{2}m(\varphi) \nabla \mu .
\end{align*}
Since the interface thickness will be of order $\eps$ it turns out that the term
\begin{align*}
 \nabla \cdot ( K \sigma (\delta(\varphi, \nabla \varphi) \bm{I} - \eps \nabla \varphi \otimes \nabla \varphi))
\end{align*}
scales with $\eps^{-2}$, while the term
\begin{align*}
 \nabla \cdot (\xi_{1}(G_{1}(c^{(1)}) - G'_{1}(c^{(1)}) c^{(1)})\bm{I} + \xi_{2}(G_{2}(c^{(2)}) - G'_{2}(c^{(2)}) c^{(2)}) \bm{I} - \varphi \mu \bm{I})
\end{align*}
scales with $\eps^{-1}$, the same order as the pressure $p$.  Hence we absorb the latter term as part of the pressure and reuse the variable $p$ as the rescaled pressure, leading to
\begin{align*}
 \bm{T} = K\sigma(c^{\Gamma}) ( \delta(\varphi, \nabla \varphi)\bm{I} - \eps \nabla \varphi \otimes \nabla \varphi) - p\bm{I} + 2 \eta(\varphi) D(\bm{v}) + \bm{v} \otimes \overline{\bm{J}}.
\end{align*}
We remark that the term $\nabla \cdot (K \sigma \delta(\varphi, \nabla \varphi)\bm{I})$ in the momentum equation is required to recover the surface gradient of the surface tension in the asymptotic analysis. It is present also in other diffuse interface models with Marangoni effects \cite{article:SunLiuXu09, article:Kim05, article:LiuShenFengYue05}.  

With the above assumptions we obtain the energy inequality
\begin{align*}
- \mathcal{D} = & - m(\varphi) \abs{\nabla \mu}^{2} - \sum_{i=1,2} M_{c}^{(i)}(c^{(i)})\xi_{i}(\varphi) \abs{\nabla G'_{i}(c^{(i)})}^{2} - 2 \eta(\varphi) \abs{D(\bm{v})}^{2}  \\
& - \beta^{(i)} \delta(\varphi, \nabla \varphi)\abs{\gamma'(c^{\Gamma}) - G_{i}'(c^{(i)})}^{2} - KM_{\Gamma}(c^{\Gamma})\delta(\varphi, \nabla \varphi)\abs{\nabla \gamma'(c^{\Gamma})}^{2} \leq 0,
\end{align*}
and the diffuse interface model (denoted Model A) for the case of dynamic adsorption reads 
\begin{align}
\nabla \cdot \bm{v} & = 0, \label{surfactant:mass} \\
\pd_{t}(\rho \bm{v})  + \nabla \cdot ( \rho \bm{v} \otimes \bm{v} ) & = \nabla \cdot \Big{(} - p \bm{I} + 2 \eta(\varphi)D(\bm{v}) + \bm{v} \otimes \tfrac{\overline{\rho}^{(2)} - \overline{\rho}^{(1)}}{2} m(\varphi) \nabla \mu \Big{)} \label{surfactant:momentum} \\
\nonumber & \quad + \nabla \cdot  \big{(} K\sigma(c^{\Gamma}) (\delta(\varphi, \nabla \varphi) \bm{I}  - \eps \nabla \varphi \otimes \nabla \varphi) \big{)}, \\
\md \varphi & =  \nabla \cdot (m(\varphi) \nabla \mu),  \label{surfactant:phase} \\
\mu + \nabla \cdot ( K\eps \sigma(c^{\Gamma}) \nabla \varphi) & = \frac{K}{\eps}\sigma(c^{\Gamma}) W'(\varphi) + \sum_{i=1,2}\xi'_{i}(\varphi)(G_{i}(c^{(i)}) - G_{i}'(c^{(i)}) c^{(i)}), \label{surfactant:chem} \\
\md (\xi_{i}(\varphi) c^{(i)}) & = \nabla \cdot (M_{c}^{(i)}(c^{(i)}) \xi_{i}(\varphi)\nabla G_{i}'(c^{(i)})) \label{surfactant:bulk} \\
\nonumber & \quad + \beta^{(i)} \delta(\varphi, \nabla \varphi)(\gamma'(c^{\Gamma}) - G'_{i}(c^{(i)})), \quad i = 1, 2,   \\
\md (K\delta(\varphi, \nabla \varphi) c^{\Gamma}) & = \nabla \cdot \Big{(}M_{\Gamma}(c^{\Gamma}) K\delta(\varphi, \nabla \varphi) \nabla \gamma'(c^{\Gamma}) \Big{)}  \label{surfactant:interface}  \\
\nonumber & \quad - \delta(\varphi, \nabla \varphi) \sum_{i=1,2} \beta^{(i)}(\gamma'(c^{\Gamma}) - G_{i}'(c^{(i)})).
\end{align}

\subsection{Instantaneous adsorption, one-sided (Model B)}
To model instantaneous adsorption,  we assume that the surfactant is insoluble in one phase $\Omega^{(1)}$.  Similar as in Section \ref{subsubsec:InstantaneousAdsorptionLocalEquil} we assume that the bulk surfactant in $\Omega^{(2)}$ and the interface surfactant are in local thermodynamical equilibrium.  This means that the bulk chemical potential $G'(c^{(2)})$ and the interface chemical potential $\gamma'(c^{\Gamma})$ are equal.  Hence we impose the constraint
\[
 \gamma'(c^{\Gamma}) = G'_{2}(c^{(2)})
\]
in order to replace $c^{\Gamma}$. For this purpose, since $\gamma'$ is strictly monotone (recall that $\gamma$ is strictly convex) we may set 
\begin{align*}
 g(c^{(2)}) = (\gamma')^{-1}(G'_{2}(c^{(2)})) = c^\Gamma.
\end{align*}
We then consider one surfactant mass balance equation which we obtain by adding (\ref{proto:bulk}) for $i = 2$, (\ref{proto:interface}) and setting $j_{1} = 0$
\begin{align}\label{OneEqu}
 \md(\xi(\varphi)c + K \delta(\varphi, \nabla \varphi) g(c)) + \nabla \cdot (\xi(\varphi)\bm{J}_{c} + K \delta(\varphi, \nabla \varphi)\bm{J}_{\Gamma}) = 0,
\end{align}
in place of $(\ref{proto:bulk})$ and $(\ref{proto:interface})$ (for convenience, we drop the index 2 of $\xi_{2}$, $c^{(2)}$, $\bm{J}_{c}^{(2)}$ etc.). 

The energy of the system is given by
\begin{align*}
 e(\bm{v}, \varphi, \nabla \varphi, c) = \frac{1}{2} \rho \abs{\bm{v}}^{2} + K\delta(\varphi, \nabla \varphi)\gamma(g(c)) + \xi(\varphi) G(c),
\end{align*}
and we set
\begin{align*}
 \mu = - \nabla \cdot ( K\eps \sigma(g(c)) \nabla \varphi) + \frac{K}{\eps}\sigma(g(c)) W'(\varphi) + \xi'(\varphi)(G(c) - G'(c) c),
\end{align*}
where
\begin{align*}
\sigma(g(c)) = \gamma(g(c)) - \gamma'(g(c)) g(c) = \gamma(g(c)) - G'(c) g(c).
\end{align*}
Then, a similar computation as in the previous model yields the following 
\begin{align*}
  -\mathcal{D} & = \nabla \cdot (\bm{J}_{e} - \overline{\bm{J}} \tfrac{\abs{\bm{v}}^{2}}{2} + (\bm{v} \otimes \overline{\bm{J}})\bm{v} - K \delta \gamma'(g(c)) \bm{J}_{\Gamma} - \xi G'(c) \bm{J}_{c} + K \eps \sigma(g(c))\nabla \varphi \md \varphi) \\
& + \nabla \cdot \Big{(} \bm{T}^{\perp} \bm{v} + \bm{J}_{\varphi} \mu \Big{)} +  \bm{J}_{\varphi} \cdot \nabla \mu - \nabla \bm{v} \colon (\bm{T} + \bm{v} \otimes \overline{\bm{J}} + K \eps \sigma(g(c)) \nabla \varphi \otimes \nabla \varphi) \\% + (\nabla \cdot \bm{v}) \Big{(} -\varphi \mu + K \delta \sigma(g(c)) + \xi(G(c) - G'(c)c) \Big{)} \\
& + K \delta \bm{J}_{\Gamma} \cdot \nabla \gamma'(g(c)) + \xi \bm{J}_{c} \cdot \nabla G'(c) .
\end{align*}

We choose $\bm{J}_{e}, \bm{T}, \bm{J}_{\varphi}$ as in Model A. Furthermore, we assume that
\begin{align*}
 \bm{J}_{c} = -M(c) \nabla G'(c), \quad \bm{J}_{\Gamma} = - M_{\Gamma}(g(c)) \nabla \gamma'(g(c)) = -M_{\Gamma}(g(c)) \nabla G'(c).
\end{align*}
We then get the energy inequality
\begin{align*}
 -\mathcal{D} = - 2 \eta(\varphi)\abs{D(\bm{v})}^{2} - m(\varphi)\abs{\nabla \mu}^{2} - ( M(c)\xi(\varphi) + K \delta(\varphi, \nabla \varphi ) M_{\Gamma}(g(c)) )\abs{\nabla G'(c)}^{2} \leq 0.
\end{align*}

The diffuse interface model for this case (denoted Model B) is
\begin{align}
\nabla \cdot \bm{v} & = 0, \label{inst2:mass} \\
\pd_{t}(\rho \bm{v}) + \nabla \cdot ( \rho \bm{v} \otimes \bm{v} ) & = \nabla \cdot \Big{(} -p\bm{I} + 2 \eta(\varphi)D(\bm{v}) + \bm{v} \otimes \tfrac{\overline{\rho}^{(2)} - \overline{\rho}^{(1)}}{2}m(\varphi)\nabla \mu  \Big{)} \label{inst2:momentum} \\
\nonumber & \quad + \nabla \cdot \big{(} K \sigma(g(c)) (\delta(\varphi, \nabla \varphi) \bm{I} - \eps \nabla \varphi \otimes \nabla \varphi) \big{)} , \\
\md \varphi & =  \nabla \cdot (m(\varphi) \nabla \mu),  \label{inst2:phase} \\
\mu + \nabla \cdot ( K\eps \sigma(g(c)) \nabla \varphi) & = \frac{K}{\eps}\sigma(g(c)) W'(\varphi) + \xi'(\varphi)(G(c) - G'(c)c), \label{inst2:chem} \\
\md (\xi(\varphi)c + K\delta(\varphi, \nabla \varphi) g(c)) & = \nabla \cdot (M(c)\xi(\varphi)\nabla G'(c)) \label{inst2:surfactant} \\
\nonumber & \quad + \nabla \cdot (M_{\Gamma}(g(c)) K\delta(\varphi, \nabla \varphi)\nabla G'(c)).
\end{align}

\subsection{Instantaneous adsorption, two-sided (Model C)}
We now derive an alternative model for instantaneous adsorption that is two-sided.  Since we assume local thermodynamical equilibrium, the chemical potentials $G_{1}'(c^{(1)}), G_{2}'(c^{(2)})$ and $\gamma'(c^{\Gamma})$ are equal on the interface.  We hence introduce a chemical potential, denoted by $q$, and consider this as unknown field rather than the densities of the surfactants.  Since the free energies $G_{i}, \gamma$ are strictly convex, their derivatives are strictly monotone and we obtain a one-to-one correspondence between the $c^{(i)}$ and $q$, i.e.
\begin{align*}
c^{(1)} = (G_{1}')^{-1}(q), \quad c^{(2)} = (G_{2}')^{-1}(q), \quad c^{\Gamma} = (\gamma')^{-1}(q).
\end{align*}
We then also may write the surface tension as a function of $q$, 
\begin{align*}
\tilde{\sigma}(q) = \sigma(c^{\Gamma}(q)) = \gamma(c^{\Gamma}(q)) - c^{\Gamma}(q) q.
\end{align*}
Summing $(\ref{proto:bulk})$ for $i=1, 2$ and $(\ref{proto:interface})$ we obtain the conservation of surfactants as follows:
\begin{multline*}
 \md(\xi_{1}(\varphi) c^{(1)}(q) + \xi_{2}(\varphi) c^{(2)}(q) + K \delta(\varphi, \nabla \varphi) c^{\Gamma}(q)) \\ 
= - \nabla \cdot \big{(} \xi_{1}(\varphi) \bm{J}_{c}^{(1)} + \xi_{2}(\varphi) \bm{J}_{c}^{(2)} + K \delta(\varphi, \nabla \varphi) \bm{J}_{\Gamma} \big{)}.
\end{multline*}

The energy density of the system is given by
\begin{align*}
 e(\varphi, \nabla \varphi, \bm{v}, q) = \frac{1}{2}\rho \abs{\bm{v}}^{2} + \xi_{1}(\varphi) G_{1}(c^{(1)}(q)) + \xi_{2}(\varphi) G_{2}(c^{(2)}(q)) + K \delta(\varphi, \nabla \varphi) \gamma(c^{\Gamma}(q))
\end{align*}
and similar computations as in the previous models yield
\begin{align*}
-\mathcal{D} & = \nabla \cdot (\bm{J}_{e} - \overline{\bm{J}} \tfrac{\abs{\bm{v}}^{2}}{2}  + (\bm{v} \otimes \overline{\bm{J}})\bm{v} - K \delta q \bm{J}_{\Gamma} - \xi_{1} q\bm{J}_{c}^{(1)} - \xi_{2} q \bm{J}_{c}^{(2)} + K \eps \tilde{\sigma}(q)\nabla \varphi \md \varphi) \\
& + \nabla \cdot \Big{(} \bm{T}^{\perp} \bm{v} + \bm{J}_{\varphi} \mu \Big{)} +  \bm{J}_{\varphi} \cdot \nabla \mu - \nabla \bm{v} \colon (\bm{T} + \bm{v} \otimes \overline{\bm{J}} + K \eps \tilde{\sigma}(q) \nabla \varphi \otimes \nabla \varphi) \\% +(\nabla \cdot \bm{v}) \Big{(} -\varphi \mu + K \delta \tilde{\sigma}(q) + \sum_{i=1,2}\xi_{i}(G_{i}(c^{(i)}(q)) - qc^{(i)}(q)) \Big{)} \\
& + K \delta \bm{J}_{\Gamma} \cdot \nabla q + \xi_{1}(\varphi) \bm{J}_{c}^{(1)} \cdot \nabla q + \xi_{2}(\varphi) \bm{J}_{c}^{(2)} \cdot \nabla q,
\end{align*}
where
\begin{align*}
 \mu = \sum_{i=1,2} \xi_{i}'(\varphi)(G_{i}(c^{(i)}(q)) - qc^{(i)}(q)) - \nabla \cdot ( K \eps \tilde{\sigma}(q) \nabla \varphi) + \frac{K}{\eps}\tilde{\sigma}(q) W'(\varphi). 
\end{align*}

Choosing $\bm{J}_{e}, \bm{T}, \bm{J}_{\varphi}$ as before (but with the $c^{(i)}$ now as functions of $q$), and setting
\begin{align*}
 \bm{J}_{c}^{(i)} = - M_{c}^{(i)}(c^{(i)}(q)) \nabla q, \quad \bm{J}_{\Gamma} = - M_{\Gamma}(c^{\Gamma}(q)) \nabla q,
\end{align*}
leads to the following energy inequality:
\begin{multline*}
 -\mathcal{D} = -2 \eta(\varphi) \abs{D(\bm{v})}^{2} - m(\varphi) \abs{\nabla \mu}^{2} \\
- \Big{(} \sum_{i=1,2} M_{c}^{(i)}(c^{(i)}(q))\xi_{i}(\varphi) + M_{\Gamma}(c^{\Gamma}(q))K \delta(\varphi, \nabla \varphi) \Big{)} \abs{\nabla q}^{2} \leq 0.
\end{multline*}

The diffuse interface model for this case of instantaneous adsorption based on the chemical potential as a field (denoted Model C) is
\begin{align}
\nabla \cdot \bm{v} & = 0, \label{inst3:mass} \\
\pd_{t}(\rho \bm{v})  + \nabla \cdot ( \rho \bm{v} \otimes \bm{v} + p\bm{I} - 2 \eta(\varphi)D(\bm{v}) ) & = \nabla \cdot \Big{(}- \bm{v} \otimes \tfrac{\overline{\rho}^{(2)} - \overline{\rho}^{(1)}}{2}m(\varphi)\nabla \mu \Big{)} \label{inst3:momentum} \\
\nonumber & \quad + \nabla \cdot \big{(} K \tilde{\sigma}(q) (\delta \bm{I} - \eps \nabla \varphi \otimes \nabla \varphi) \big{)} , \\
\md \varphi & =  \nabla \cdot (m(\varphi) \nabla \mu),  \label{inst3:phase} \\
\mu + \nabla \cdot ( K\eps \tilde{\sigma}(q) \nabla \varphi) - \frac{K}{\eps}\tilde{\sigma}(q) W'(\varphi) & = \sum_{i=1,2} \xi_{i}'(\varphi)(G(c^{(i)})-qc^{(i)}), \label{inst3:chem} \\
\md \big{(} \xi_{1} c^{(1)}(q) + \xi_{2} c^{(2)}(q) + K \delta c^{\Gamma}(q) \big{)} &= \sum_{i=1,2} \nabla \cdot \big{(} M_{c}^{(i)}(c^{(i)}(q))\xi_{i} \nabla q) \label{inst3:surfactant} \\
\nonumber & \quad + \nabla \cdot \big{(} M_{\Gamma}(c^{\Gamma}(q)) K \delta \nabla q \big{)}.
\end{align}

\subsection{Specific models} 
\subsubsection{Insoluble surfactants}\label{subsubsec:PFInsoluble}
Similar as in Section \ref{subsubsec:Insoluble}, we can consider a phase field model for insoluble surfactants.  The resulting model is a system for the unknowns $\bm{v}, p, \varphi, \mu, c^{\Gamma}$ and is obtained by setting $\xi_{i} \equiv 0$ and $\beta^{(i)} = 0$ in $(\ref{surfactant:mass}) - (\ref{surfactant:interface})$.
\subsubsection{One-sided model with non-instantaneous adsorption}  It is also possible to consider a one-sided version of Model A by setting $\xi_{1} \equiv 0$ and neglecting the unknown $c^{(1)}$.
\subsubsection{Mobility for the phase field equation}
We will choose the functional form of the mobility to be
\begin{align*}
m(\varphi) = m_{1}(1-\varphi^{2})_{+},
\end{align*}
where $m_{1} > 0$ is a constant and $(\cdot)_{+}$ denotes the positive part of the quantity in the brackets.  This degenerate mobility switches off diffusion in the bulk phases away from the interfacial layer. In this case, the phase field equations (\ref{surfactant:phase}),(\ref{surfactant:chem}) lead to a pure advection of the interface, see \cite{article:AbelsGarckeGrun}.

\subsubsection{Diffusivities}\label{subsubsec:PFFicks}
If we set
\begin{align*}
 M_{c}^{(i)} = D_{c}^{(i)}\frac{1}{G''_{i}(c^{(i)})}, \quad M_{\Gamma}(c^{\Gamma}) = D_{\Gamma}\frac{1}{\gamma''(c^{\Gamma})},
\end{align*}
for constants $D_{c}^{(i)}$ and $D_{\Gamma}$, then we derive Fick's law for the surfactant
\begin{align*}
 \bm{J}_{c}^{(i)} = - D_{c}^{(i)}\nabla c^{(i)}, \quad \bm{J}_{\Gamma} = - D_{\Gamma}\nabla c^{\Gamma}.
\end{align*}

\subsubsection{Obstacle potential}
If $W$ is chosen to be a potential of double-obstacle type, then equation $(\ref{surfactant:chem})$ is formulated as the following variational inequality:  For all $\psi \in \mathcal{K} := \{ \eta \in H^1(\Omega) : \abs{\eta} \leq 1 \}$,
\begin{align}
 \nonumber & \int_{\Omega}-\mu(\psi - \varphi) + K\eps \sigma(c^{\Gamma}) \nabla \varphi \cdot (\nabla \psi - \nabla \varphi) + \frac{K}{\eps} \sigma(c^{\Gamma}) W'(\varphi)(\psi - \varphi) \\
 & + \int_{\Omega} \sum_{i=1,2}\xi'_{i}(\varphi)(G_{i}(c^{(i)}) - G_{i}'(c^{(i)}) c^{(i)})(\psi - \varphi) \geq 0. \label{ObstacleVariationalIneq}
\end{align}
\subsubsection{Reformulation of the momentum equation}\label{subsubsec:ReformulationMom}
A short computation shows that
\begin{align*}
 \mu \nabla \varphi & = \nabla \cdot (K \sigma (\delta(\varphi, \nabla \varphi) \bm{I} - \eps \nabla \varphi \otimes \nabla \varphi)) - K \delta(\varphi, \nabla \varphi) \nabla \sigma \\
& + \sum_{i=1,2} \xi_{i}'(\varphi)(G_{i}(c^{(i)}) - G'_{i}(c^{(i)})c^{(i)}) \nabla \varphi,
\end{align*}
hence the momentum equation $(\ref{surfactant:momentum})$ can be reformulated as
\begin{multline*}
\pd_{t}(\rho \bm{v}) + \nabla \cdot (\rho \bm{v} \otimes \bm{v}) = \nabla \cdot \Big{(} - p \bm{I} + 2 \eta(\varphi)D(\bm{v}) + \bm{v} \otimes \tfrac{\overline{\rho}^{(2)} - \overline{\rho}^{(1)}}{2} m(\varphi) \nabla \mu \Big{)} \\
 + \mu \nabla \varphi + K \delta(\varphi, \nabla \varphi) \nabla \sigma - \sum_{i=1,2} \xi'(\varphi)(G_{i}(c^{(i)}) -G'_{i}(c^{(i)})c^{(i)}) \nabla \varphi.
\end{multline*}

\subsubsection{Non-dimensional evolution equations}
We consider the following dimensionless variables
\begin{align*}
\delta_{*} = L \delta, \; \eps_{*} = \frac{\eps}{L}, \; m_{*} = \frac{m(\varphi) \Sigma}{V L^2}, \; \mu_{*} = \frac{\mu L}{\Sigma} %, \; \gamma_{*} = \frac{\gamma}{\Sigma}, \; G_{*} = \frac{G L}{\Sigma}.
\end{align*}
with the characteristic length $L$, the scale $\Sigma$ for the surface tension and a characteristic velocity $V$.  In addition we scale the bulk densities by $C$, the interfacial density by $C^{\Gamma}$ and similar to the density, the viscosity $\eta(\varphi)$ can be decomposed to $\eta = u_{1} \eta^{(1)} + u_{2} \eta^{(2)}$. The dimensionless density and viscosity are
\begin{align*}
 \rho_{*} = \rho/\overline{\rho}^{(2)} = u_{1}\lambda_{\rho} + u_{2}, \quad \eta_{*} = \eta/\eta^{(2)} = u_{1}\lambda_{\eta} + u_{2}
\end{align*}
where $\lambda_{\rho} = \overline{\rho}^{(1)}/\overline{\rho}^{(2)}, \lambda_{\eta} = \eta^{(1)}/\eta^{(2)}$ are the density and viscosity ratios.  Set $\text{Re} = (\overline{\rho}^{(2)} L^2)/(T \eta^{(2)})$, $\text{Ca} = (\eta^{(2)}L)/(T \Sigma)$ to be the Reynolds and capillary numbers respectively. Then the dimensionless fluid and phase field equations are
\begin{align}
 \nabla_{*} \cdot \bm{v}_{*} & = 0, \label{phaseNonD:mass} \\
\nonumber \pd_{t_{*}}(\rho_{*} \bm{v}_{*}) + \nabla_{*} \cdot  ( \rho_{*} \bm{v}_{*} \otimes \bm{v}_{*} ) & = \nabla_{*} \cdot \Big{(} - p_{*}\bm{I} + \frac{2 \eta_{*}}{\text{Re}}D(\bm{v}_{*}) + \bm{v}_{*} \otimes \frac{1 - \lambda_{\rho}}{2}m_{*}(\varphi)\nabla_{*} \mu_{*} \Big{)}  \\
& \quad + \frac{1}{\text{ReCa}}\nabla_{*} \cdot \big{(} K \sigma_{*}(\delta_{*} \bm{I}  - \eps_{*} \nabla_{*} \varphi \otimes \nabla_{*} \varphi) \big{)}, \label{phaseNonD:momentum}\\
\mdND \varphi & =  \nabla_{*} \cdot (m_{*}(\varphi) \nabla_{*} \mu_{*}) \label{phaseNonD:phase}, \\
\nabla_{*} \cdot (K \eps_{*} \sigma_{*} \nabla_{*} \varphi) - \frac{K}{\eps_{*}} \sigma_{*} W'(\varphi) & =  - \mu_{*} +  \sum_{i=1,2} \xi_{i}'(\varphi) (G_{*,i}(c^{(i)}_{*}) - G_{*,i}'(c^{(i)}_{*})c^{(i)}_{*}), \label{phaseNonD:chem}
\end{align}
where $p_{*} = (pT^2)/(L^2 \overline{\rho}^{(2)})$ is the rescaled pressure.  The reformulated momentum equation from Section \ref{subsubsec:ReformulationMom} has the dimensionless form 
\begin{align}
\nonumber \pd_{t_{*}}(\rho_{*} \bm{v}_{*}) + \nabla_{*} \cdot ( \rho_{*} \bm{v}_{*} \otimes \bm{v}_{*} ) & = \nabla_{*} \cdot \Big{(} - p_{*}\bm{I} + \frac{2 \eta_{*}}{\text{Re}}D(\bm{v}_{*}) + \bm{v}_{*} \otimes \frac{1 - \lambda_{\rho}}{2}m_{*}(\varphi)\nabla_{*} \mu_{*} \Big{)} \\
 & \quad + \frac{1}{\text{ReCa}} \Big{(} \mu_{*} \nabla_{*}\varphi + K \delta_{*} \nabla_{*} \sigma_{*} \Big{)} \label{phaseNonD:AlternateMomentum} \\
 \nonumber & \quad + \frac{1}{\text{ReCa}} \Big{(} \sum_{i=1,2} \xi'_{i}(\varphi)(G_{i,*}(c^{(i)}_{*}) - G'_{i,*} (c^{(i)}_{*})c^{(i)}_{*})\nabla_{*}\varphi_{*}) \Big{)}. 
\end{align}
The dimensionless surfactant equations for Model A are
\begin{align}
\mdND(\xi_{i}c^{(i)}_{*}) & - \nabla_{*} \cdot \Big{(} M_{c,*}^{(i)} \xi_{i} \nabla_{*} G_{i,*}'(c^{(i)}_{*}) \Big{)} = \beta_{*}^{(i)} \delta_{*}(\gamma'_{*}(c^{\Gamma}_{*}) - G_{i,*}'(c^{(i)}_{*})),  \label{phaseNonD:bulk} \\
\mdND (K\delta_{*}c^{\Gamma}_{*}) & - \nabla_{*} \cdot \Big{(} K M_{\Gamma,*} \delta_{*} \nabla_{*} \gamma_{*}'(c^{\Gamma}_{*}) \Big{)}  = -\delta_{*} \sum_{i=1,2} \beta_{*}^{(i)}(\gamma'_{*}(c^{\Gamma}_{*}) - G_{*,i}'(c^{(i)}_{*})), \label{phaseNonD:interface}
\end{align}
where $\beta_{*}^{(i)} = \beta \Sigma T L^{4}$.  For Model B, the dimensionless surfactant equation reads
\begin{align}\label{phaseNonDinst2:surfactant}
 \mdND \big{(} \xi c_{*} + K \delta_{*} g_{*} \big{)} - \nabla_{*} \cdot \Big{(} M_{c,*} \xi  \nabla_{*} G_{*}'(c_{*}) + K M_{\Gamma,*} \delta_{*}  \nabla_{*} G_{*}'(c_{*}) \Big{)} = 0,
\end{align}
and for Model C, it reads as
\begin{align}\label{phaseNonDinst3:surfactant}
\nonumber \mdND \big{(} \xi_{1} c^{(1)}_{*}(q_{*}) & + \xi_{2} c^{(2)}_{*}(q_{*}) + K \delta_{*} c^{\Gamma}_{*}(q_{*}) \big{)} \\
& - \nabla_{*} \cdot \Big{(} M_{c,*}^{(1)} \xi_{1}  \nabla_{*}q_{*} + M_{c,*}^{(2)} \xi_{2} \nabla_{*} q_{*} + K M_{\Gamma,*} \delta_{*}  \nabla_{*} q_{*} \Big{)} = 0.
\end{align}
If we consider the mobilities in Section \ref{subsubsec:PFFicks}, the dimensionless surfactant equations for Model A are
\begin{align}
\mdND(\xi_{i}c^{(i)}_{*}) & - \nabla_{*} \cdot \Big{(} \frac{1}{\text{Pe}_{c,i}} \xi_{i} \nabla_{*} c^{(i)}_{*} \Big{)} = \beta_{*}^{(i)} \delta_{*}(\gamma'_{*}(c^{\Gamma}_{*}) - G_{*,i}'(c^{(i)}_{*})),  \label{phaseNonDFick:bulk} \\
\mdND (K\delta_{*}c^{\Gamma}_{*}) & - \nabla_{*} \cdot \Big{(}\frac{K}{\text{Pe}_{\Gamma}} \delta_{*} \nabla_{*} c^{\Gamma}_{*} \Big{)}  = - \delta_{*} \sum_{i=1,2} \beta_{*}^{(i)}(\gamma'_{*}(c^{\Gamma}_{*}) - G_{*,i}'(c^{(i)}_{*})). \label{phaseNonDFick:interface}
\end{align}
For Model B, the dimensionless surfactant equation with Fickian diffusion reads
\begin{align}\label{phaseNonDinst2Fick:surfactant}
 \mdND \big{(} \xi c_{*} + K \delta_{*} g_{*} \big{)} - \nabla_{*} \cdot \Big{(} \frac{1}{\text{Pe}_{c}} \xi  \nabla_{*} c_{*} + \frac{K}{\text{Pe}_{\Gamma}} \delta_{*}  \nabla_{*} c_{*} \Big{)} = 0.
\end{align}

\section{Sharp interface asymptotics}\label{Asymptotics}
In this section we identify the sharp interface limit of the diffuse interface models introduced in the previous section by the method of matching formal asymptotic expansions.  The procedure is based on the assumption that there exist a family of solutions, sufficiently smooth and indexed by $\eps$, to the diffuse interface models.  For small $\eps$, we assume that the domain $\Omega$ can at each time $t$ be divided into two open subdomains $\Omega^{\pm}(t;\eps)$, separated by an interface $\Gamma(t;\eps)$.  Furthermore, we assume that the solutions have an asymptotic expansion in $\eps$ in the bulk regions (away from $\Gamma(t;\eps)$) and another expansion in the interfacial regions (close to $\Gamma(t;\eps)$).  The idea is to analyse these expansions in a suitable region where they should match up.  We will apply this method to Model A, where we distinguish two different scalings of $\alpha^{(i)}$, namely $\mathcal{O}(1)$ and $\mathcal{O}(\eps)$. In the last section we briefly outline the procedure for Models B and C.  Details of the method can be found in \cite{article:FifePenrose95,article:GarckeStinner06,article:AbelsGarckeGrun} for the smooth double-well potential and in \cite{incoll:BloweyElliott93,article:BhateBowerKumar02} for the double-obstacle potential.  We remark that for some specific models this procedure has been rigorously justified (see \cite{article:AlikakosBatesChen94,article:deMottoniSchatzman95,article:CaginalpChen98}).

\subsection{Outer expansions, equations and solutions}
We assume there exist the following asymptotic expansions in $\eps$ for $u_{\eps} = u(t,\bm{x};\eps) \in \{\bm{v}_{\eps}, p_{\eps}, \varphi_{\eps}, \mu_{\eps}, c^{(i)}_{\eps}, c^{\Gamma}_{\eps}\}$ in the bulk regions away from the interface
\begin{align}\label{outer}
 u_{\eps}(t,\bm{x}) = u(t,\bm{x};\eps) = u_{0}(t,\bm{x}) + \eps u_{1}(t,\bm{x}) + \mathcal{O}(\eps^{2}).
\end{align}
Substituting these expansions into Model A and $(\ref{surfactant:chem})$ to order $-1$ gives
\begin{align*}
 0 = \sigma(c^{\Gamma}_{0}) W'(\varphi_{0}).
\end{align*}
As $\sigma > 0$, we obtain the stable solutions $\varphi_{0} = \pm 1$.  We denote $\Omega^{(2)}$ and $\Omega^{(1)}$ to be the sets where $\varphi_{0} = 1$ and $\varphi_{0} = -1$ respectively.  

The zeroth order expansions of the fluid equations yield 
\begin{align*}
\nabla \cdot \bm{v}_{0} & = 0, \\
\pd_{t}(\overline{\rho}^{(i)} \bm{v}_{0}) + \nabla \cdot (\overline{\rho}^{(i)} \bm{v}_{0} \otimes \bm{v}_{0} - 2 \eta^{(i)} D(\bm{v}_{0}) + p_{0} \bm{I}) & = 0.
\end{align*}
The bulk surfactant equation gives to the zeroth order
\begin{align*}
 \pd_{t} c^{(i)}_{0} + \bm{v}_{0} \cdot \nabla c^{(i)}_{0} - \nabla \cdot (M_{i}(c^{(i)}_{0}) \nabla G'_{i}(c^{(i)}_{0})) = 0, \quad i = 1, 2.
\end{align*}
Observe that $\delta(\varphi_{0}, \nabla \varphi_{0}) = 0$ so that $(\ref{surfactant:interface})$ fully degenerates in both domains $\Omega^{(2)}$ and $\Omega^{(1)}$, whence $c^\Gamma_{0}$ remains undetermined in the bulk. Similarly, $\mu_{0}$ is undetermined in the bulk due to the degenerate nature of the mobility $m(\varphi_{0})$. 

For the double-obstacle potential, equation $(\ref{surfactant:chem})$ is replaced by $(\ref{ObstacleVariationalIneq})$ which, to order $-1$, is the variational inequality
\begin{align*}
 \int_{\Omega}\sigma(c^{\Gamma}_{0}) W'(\varphi_{0})(\psi_{0} - \varphi_{0}) \geq 0, \quad \forall \psi_{0} \in \mathcal{K}.
\end{align*}
Here, $W'(\varphi) = - \varphi + \pd I_{[-1,1]}(\varphi)$ where $\pd I$ is the sub-differential of $I_{[-1,1]}$. Then the above can be expressed as
\begin{align*}
 - \int_{\Omega} \sigma(c^{\Gamma}_{0}) \varphi_{0}(\psi_{0} - \varphi_{0}) \geq 0, \quad \forall \psi_{0} \in \mathcal{K}.
\end{align*}
Since $\sigma > 0$, this implies that $\varphi_{0}$ must take the values $\pm 1$ and we can define sets $\Omega^{(2)}, \Omega^{(1)}$ as in the case with the double-well potential. 

\subsection{Inner expansions and matching conditions}
Let us assume that the zero level sets of $\varphi_{\eps}$ converge to some hypersurface $\Gamma$ moving with a normal velocity denoted by $u_{\Gamma}$ as $\eps \to 0$. Close to $\Gamma$, we denote by $d(t, \bm{x})$ the signed distance function of a point $\bm{x} \in \Omega$ to $\Gamma$ with the convention $d(t, \bm{x}) > 0$ if $x \in \Omega^{(2)}(t)$, and set $z(t, \bm{x}) = d(t, \bm{x})/ \eps$.  We write each field $u(t, \bm{x})$ close to $\Gamma$ in new coordinates $U(t,s,z)$ where $s$ are tangential spatial coordinates on $\Gamma$. The upshot is 
\begin{align*}
\pd_{t} u & = - \frac{1}{\eps} u_{\Gamma} \pd_{z} U + \nd U + \text{ h.o.t.}, \\
\nabla_{\bm{x}} u & = \frac{1}{\eps}\pd_{z}U \bm{\nu} + \surf U + \text{ h.o.t.}, \\
\Laplace_{\bm{x}} u & = \frac{1}{\eps^{2}} \pd_{zz} U - \frac{1}{\eps} \kappa \pd_{z} U - z \abs{\mathcal{S}}^{2} \pd_{z} U + \LB U + \text{ h.o.t.},
\end{align*}
where $\bm{\nu} = \nabla_{\bm{x}} d$ is the unit normal pointing into $\Omega^{(2)}$, $\nd(\cdot) = \pd_{t}(\cdot) + u_{\Gamma}\bm{\nu} \cdot \nabla_{\bm{x}}(\cdot)$ is the normal time derivative, $\surf$ is the spatial surface gradient on $\Gamma$, $\kappa$ is the mean curvature, $\abs{\mathcal{S}}$ is the spectral norm of the Weingarten map $\mathcal{S}$, $\LB$ is the Laplace--Beltrami operator on $\Gamma$ and h.o.t. denotes higher order terms (see the appendix of \cite{article:AbelsGarckeGrun} for a proof).

We assume that the inner expansions of unknown fields $u \in \{\bm{v}_{\eps}, p_{\eps}, \varphi_{\eps}, \mu_{\eps}, c^{(i)}_{\eps}, c^{\Gamma}_{\eps}\}$ take the form
\begin{align*}
u(t, \bm{x};\eps) = U(t,s,z;\eps) = U_{0}(t, s, z) + \eps U_{1}(t,s,z) + \mathcal{O}(\eps^{2})
\end{align*}
with inner variables $U \in \{\bm{V}, P, \Phi, M, C^{(i)}, C^{\Gamma}\}$.  We assume that $\Phi$ satisfies
\begin{align*}
 \Phi(t,s,0;\eps) = 0.
\end{align*}
Regarding the double-obstacle potential, we further assume that $\Phi$ is monotone increasing with $z$ and the interfacial layer has finite thickness of $2l$, where the value of $l$ will come out of the asymptotic analysis (see  \cite{incoll:BloweyElliott93}).  For the double-well potential we take $l = \infty$.  Furthermore, we assume that
\begin{align}\label{match:phi}
 \Phi(t,s,l;\eps) = 1, \quad \Phi(t,s,-l;\eps) = -1.
\end{align}
In order to match the inner expansions valid in the interfacial layers to outer expansions we employ following matching conditions \cite{article:GarckeStinner06}: 
As $z \to \pm l$,
\begin{align}
U_{0} (t,s,z) & \sim u_{0}^{\pm} (t, \bm{x}), \label{MC0} \\
\pd_{z} U_{0}(t,s,z) & \sim 0, \label{MC1} \\
\pd_{z} U_{1} (t,s,z) & \sim \nabla u_{0}^{\pm} (t, \bm{x}) \cdot \bm{\nu}, \label{MC2} \\
\pd_{z} U_{2} (t,s,z) & \sim \nabla u_{1}^{\pm} (t, \bm{x}) \cdot \bm{\nu} + \big{(} (\bm{\nu} \cdot \nabla)(\bm{\nu} \cdot \nabla) u_{0}^{\pm} (t, \bm{x}) \big{)} z, \label{MC3}
\end{align}
where $u_{0}^{\pm}$ denotes the limit $\lim_{\delta \searrow 0} u_{0}(\bm{x} \pm \delta \bm{\nu})$ at a point $\bm{x} \in \Gamma$.   

If the bulk fields are not determined by any equation, i.e., if $u = c^{\Gamma}$ or $u = \mu$, then we assume that the derivatives of the inner expansion in $z$ remain bounded as $z \to \pm l$.  More precisely, we assume that
\begin{align*}
 \pd_{z}C^{\Gamma}_{0}, \; \pd_{z}C^{\Gamma}_{1}, \; \pd_{z}C^{\Gamma}_{2}, \; \pd_{z}M_{0}, \; \pd_{z}M_{1} \text{  are bounded as } z \to \pm l.
\end{align*}
Moreover, we assume
\begin{align*}
 \pd_{z}C^{(1)}_{0}, \; \pd_{z}C^{(1)}_{1} \text{ are bounded as } z \to +l, \quad \pd_{z}C^{(2)}_{0}, \; \pd_{z}C^{(2)}_{1} \text{ are bounded as } z \to -l,
\end{align*}
since $c^{(1)}$ is not defined in $\Omega^{(2)}$ and $c^{(2)}$ is not defined in $\Omega^{(1)}$.  Similar assumptions are made for the asymptotic analysis of Models B and C.
\subsection{Asymptotics for Model A}
We begin by stating a few expansions of the most complicated terms for later use. These can be obtained by some short calculations. 
First, 
\begin{align*}
\eps \nabla \cdot & (\sigma(c^{\Gamma})\nabla \varphi \otimes \nabla \varphi) = \frac{1}{\eps^{2}} \pd_{z}(\sigma(c^{\Gamma})(\pd_{z} \Phi)^{2} \bm{\nu}) + \frac{1}{\eps} \pd_{z}(\sigma(c^{\Gamma})\pd_{z}\Phi \surf \Phi) \\
& + \frac{1}{\eps} \surf \cdot (\sigma(c^{\Gamma})(\pd_{z} \Phi)^{2} \bm{\nu} \otimes \bm{\nu}) + \surf \cdot (\sigma(c^{\Gamma}) \pd_{z}\Phi(\bm{\nu} \otimes \surf \Phi + \surf \Phi \otimes \bm{\nu})) + \text{ h.o.t}.
\end{align*}
where $\surf \cdot$ of a 2-tensor is the surface divergence applied to each row.  Then, setting $\mathcal{E}(\bm{A}) = \frac{1}{2}(\bm{A} + \bm{A}^{\perp})$ for a tensor $\bm{A}$ one can show that 
\begin{multline*}
\nabla \cdot (\eta(\varphi) D(\bm{v})) 
= \frac{1}{\eps^{2}} \pd_{z}(\eta(\Phi) \mathcal{E}(\pd_{z} \bm{V} \otimes \bm{\nu}) \bm{\nu}) \\
+ \frac{1}{\eps} \pd_{z}( \eta(\Phi) \mathcal{E}(\surf \bm{V}) \bm{\nu}) + \frac{1}{\eps} \surf \cdot ( \eta(\Phi)\mathcal{E}(\pd_{z} \bm{V} \otimes \bm{\nu})) + \text{ h.o.t}.
\end{multline*}
Finally, observe that
\begin{align*}
 \delta(\varphi, \nabla \varphi) 
& = \frac{1}{2\eps} \abs{\pd_{z}\Phi}^{2} + \frac{1}{\eps} W(\Phi)  + \frac{\eps}{2}\abs{\surf \Phi}^{2}  + \text{ h.o.t}.
\end{align*}

\subsubsection{Inner equations and solutions to leading order}
The order $-3$ terms in $(\ref{surfactant:interface})$ give
\begin{align*}
K\pd_{z}(M_{\Gamma}(C^{\Gamma}_{0})(\tfrac{1}{2}\abs{\pd_{z}\Phi_{0}}^{2} + W(\Phi_{0}) )\pd_{z}\gamma'(C^{\Gamma}_{0})) = 0.
\end{align*}
Integrating from $-l$ to $z$ and matching conditions $(\ref{MC0})$ and $(\ref{MC1})$ applied to $\Phi_{0}$ yields
\begin{align*}
M_{\Gamma}(C^{\Gamma}_{0})(\tfrac{1}{2}\abs{\pd_{z}\Phi_{0}}^{2} +W(\Phi_{0}(z))) \pd_{z}\gamma'(C^{\Gamma}_{0}(z)) = 0.
\end{align*}
We conclude that 
\begin{align*}
\pd_{z} \gamma'(C^{\Gamma}_{0}(z)) = 0 \text{ whenever } \abs{\Phi_{0}} < 1.
\end{align*}
Since $\gamma'' > 0$, we obtain that
\begin{align*}
\pd_{z} C^{\Gamma}_{0}(z) = 0 \text{ whenever } \abs{\Phi_{0}} < 1
\end{align*}
which means that $C^{\Gamma}_0$ is constant across the interfacial layer. Since the surface tension is given by $\sigma(C^{\Gamma}_{0}) = \gamma(C^{\Gamma}_{0}) - C^{\Gamma}_{0}\gamma'(C^{\Gamma}_{0})$, we also obtain
\begin{align*}
\pd_{z} \sigma(C^{\Gamma}_{0}(z)) = 0 \text{ whenever } \abs{\Phi_{0}} < 1.
\end{align*}

To order $-1$ in $(\ref{surfactant:chem})$ we have
\begin{align*}
 K\sigma(C^{\Gamma}_{0})(-\pd_{zz}\Phi_{0} + W'(\Phi_{0})) = 0.
\end{align*}
We can choose $\Phi_{0}$ such that it is independent of $s$ and solves
\begin{align}\label{inner:profile}
 -\pd_{zz} \Phi_{0} + W'(\Phi_{0}) = 0, 
\end{align}
with $\Phi_{0}(0) = 0$ and $\Phi_{0}(\pm l) = \pm 1$.  With the double-well potential $W(\varphi) = \frac{1}{4}(1-\varphi^2)^2$ we have the unique solution
\begin{align*}
 \Phi_{0}(z) = \tanh(z/\sqrt{2}),
\end{align*}
while for the double-obstacle potential, a unique solution to
\begin{align*}
 -\pd_{zz} \Phi_{0} - \Phi_{0} = 0, \quad \abs{\Phi_{0}} \leq 1, \quad \Phi_{0}(t,s,0) = 0
\end{align*}
is
\begin{align*}
 \Phi_{0}(z) = 
\begin{cases}
 +1, & \text{ for } z \geq \frac{\pi}{2}, \\
 \sin(z), & \text{ for } \abs{z} < \frac{\pi}{2}, \\
 -1, & \text{ for } z \leq -\frac{\pi}{2},
\end{cases}
\end{align*}
so that $l = \frac{\pi}{2}$ and from $(\ref{match:phi})$ we deduce that
\begin{align}\label{inner:Phi1}
\Phi_{1}(t,s,\pm \tfrac{\pi}{2}) = 0.
\end{align}  
Multiplying $(\ref{inner:profile})$ by $\pd_{z}\Phi_{0}$, integrating from $-l$ to $z$ and applying matching to $\Phi_{0}$ yield the equipartition of energy
\begin{align}\label{inner:equipartition}
 \frac{1}{2} \abs{\pd_{z}\Phi_{0}(z)}^{2} = W(\Phi_{0}(z)).
\end{align}

The order $-1$ term in the mass balance $(\ref{surfactant:mass})$ gives
\begin{align}\label{inner:normalvelocity}
 (\pd_{z} \bm{V}_{0}) \cdot \bm{\nu} = \pd_{z} (\bm{V}_{0} \cdot \bm{\nu}) = 0.
\end{align}
Integrating from $-l$ to $l$ and matching $(\ref{MC0})$ applied to $\bm{V}_{0}$ imply that $\bm{V}_{0} \cdot \bm{\nu}$ is constant in $z$ and 
\begin{align}\label{inner:velocitylimit}
\bm{v}_{0}^{(2)} \cdot \bm{\nu} = \lim_{z \to +\infty} \bm{V}_{0}(z) \cdot \bm{\nu} = \lim_{z \to -\infty} \bm{V}_{0}(z) \cdot \bm{\nu} = \bm{v}_{0}^{(1)} \cdot \bm{\nu},
\end{align}
i.e., the normal velocity is continuous across the interface.

Equation $(\ref{surfactant:bulk})$ gives to order $-2$
\begin{align*}
\pd_{z}(M_{i}(C^{(i)}_{0}) \xi_{i}(\Phi_{0}) G_{i}''(C^{(i)}_{0}) \pd_{z}C^{(i)}_{0}) = 0.
\end{align*}
In the two-sided model, for $i = 2$ we integrate from $-l$ to $z$ to obtain
\begin{align*}
 M_{2}(C^{(2)}_{0}) \xi_{2}(\Phi_{0}(z)) G_{2}''(C^{(2)}_{0}(z)) \pd_{z}C^{(2)}_{0}(z) = 0
\end{align*}
as $\xi_{2}(-1) = 0$.   Since $G''_{2}>0$ we have that $\pd_{z}C^{(2)}_{0} = 0$.  Similarly for $C^{(1)}_0$ where we integrate from $z$ to $+l$ to obtain
\begin{align*}
 M_{1}(C^{(1)}_{0}) \xi_{1}(\Phi_{0}(z)) G_{1}''(C^{(1)}_{0}(z)) \pd_{z}C^{(1)}_{0}(z) = 0
\end{align*}
as $\xi_{1}(+1) = 0$.  Thus $\pd_{z}C^{(1)}_{0} = 0$ follows from the same argument.  In the case of the one-sided model, we argue as above to obtain $\pd_{z}C_{0} = 0$.

Equation $(\ref{surfactant:phase})$ gives to order $-2$
\begin{align*}
 0 =\pd_{z}( m_{1}(1-\Phi_{0}^2)_{+} \pd_{z}M_{0}).
\end{align*}
Integrating from $-l$ to $z$ and matching $(\ref{MC0})$ applied to $\Phi_{0}$ gives
\begin{align*}
 0 = m_{1}(1-\Phi_{0}^{2}(z))_{+} \pd_{z}M_{0}(z).
\end{align*}
For $\abs{\Phi_{0}} <1$ we have $\pd_{z}M_{0} = 0$, hence the term $\nabla \cdot (\bm{v} \otimes \frac{\overline{\rho}^{(2)} - \overline{\rho}^{(1)}}{2} m(\varphi) \nabla \mu)$ plays no part in the order $-2$ expansion of the momentum equation $(\ref{surfactant:momentum})$.  To leading order the momentum equation gives
\begin{align}\label{inner:veloODE}
 \bm{0} =  2 \pd_{z}(\eta(\Phi_{0})\pd_{z}\bm{V}_{0}).
\end{align}
With the usual trick of integrating with respect to $z$ from $-l$ to a limit denoted by $z$ again and applying $(\ref{MC1})$ to $\bm{V}_{0}$ we obtain that $\eta(\Phi_{0}) \pd_z \bm{V}_{0} = 0$. Since $\eta > 0$ we conclude that $\pd_z \bm{V}_{0} = 0$ so that, using $(\ref{MC0})$, the tangential velocity is continuous across the interface:
\begin{align*}
 [\bm{v}_{0}]_{1}^{2} = 0.
\end{align*}

\subsubsection{Inner equations and solutions to first order}
Equation $(\ref{surfactant:mass})$ of the mass balance yields to zeroth order 
\begin{align}\label{inner:mass}
 \pd_{z}\bm{V}_{1} \cdot \bm{\nu} + \surf \cdot \bm{V}_{0} = 0,
\end{align}
while equation $(\ref{surfactant:phase})$ gives to order $-1$
\begin{align*}
 (-u_{\Gamma} + \bm{V}_{0} \cdot \bm{\nu})\pd_{z}\Phi_{0} 
& = \pd_{z}(m_{1}(1-\Phi_{0}^2)_{+} \pd_{z}M_{1} ), % + \surf \cdot(m_{1}(1-\Phi_{0}^2)_{+} \pd_{z}M_{0})
\end{align*}
where we used that $\pd_{z}M_{0} = 0$. Integrating from $-l$ to $+l$ and applying $(\ref{MC0})$ to $\Phi_{0}$ and $(\ref{MC2})$ to $M_0$ (see also the remark after further down after $(\ref{match:phi})$) then imply that
\begin{align*}
 2(u_{\Gamma} - \bm{v}_{0}\cdot \bm{\nu}) = [m_{1}(1-\Phi_{0}^2)_{+} \pd_{z}M_{1}]_{-l}^{+l} = 0,
\end{align*}
and we obtain
\begin{align}\label{inner:normalvelocity2}
u_{\Gamma} = \bm{v}_{0} \cdot \bm{\nu}.
\end{align}

Using equipartition of energy $(\ref{inner:equipartition})$, $\pd_{z} C^{(i)}_{0} = 0$ and $u_{\Gamma} = \bm{v}_{0} \cdot \bm{\nu}$, we obtain from $(\ref{surfactant:bulk})$ at order $-1$
\begin{align}\label{inner:bulkequ}
& 2\beta^{(i)}(\gamma'(C^{\Gamma}_{0}) - G'_{i}(C^{(i)}_{0}))W(\Phi_{0}) = -\pd_{z}(M_{i}(C^{(i)}_{0})\xi_{i}(\Phi_{0}) \pd_{z}(G_{i}''(C^{(i)}_{0})C^{(i)}_{1})). 
\end{align}
In the two-sided model, for $i=2$, integrating $(\ref{inner:bulkequ})$ from $-l$ to $+l$ and using $(\ref{MC2})$ leads to
\begin{align*}
0 & = [M_{2}(C^{(2)}_{0})\xi_{2}(\Phi_{0})G''_{2}(C^{(2)}_{0})\pd_{z}C^{(2)}_{1}]_{-l}^{+l} + \int_{-l}^{+l} 2\beta^{(2)} (\gamma'(C^{\Gamma}_{0}) - G_{2}'(C^{(2)}_{0})) W(\Phi_{0}) dz \\
& = M_{c}^{(2)}(c^{(2)}_{0})\nabla G'_{2}(c^{(2)}_{0})\cdot \bm{\nu} + \beta^{(2)} (\gamma'(c^{\Gamma}_{0}) - G_{2}'(c^{(2)}_{0}))\mathcal{W}.
\end{align*}
Proceeding similarly for $i=1$ and we recover the following free boundary conditions
\begin{align}\label{inner:freebdy}
\begin{array}{cc}
-M_{c}^{(2)}(c^{(2)})\nabla G'_{2}(c^{(2)}_{0})\cdot \bm{\nu} & = \bm{J}^{(2)}_{c,0} \cdot \bm{\nu} = \frac{1}{\alpha^{(2)}} (\gamma'(C^{\Gamma}_{0}) - G_{2}'(c^{(2)}_{0})), \\
M_{c}^{(1)}(c^{(1)})\nabla G'_{1}(c^{(1)}_{0})\cdot \bm{\nu} & = - \bm{J}^{(1)}_{c,0} \cdot \bm{\nu} =  \frac{1}{\alpha^{(1)}}(\gamma'(C^{\Gamma}_{0}) - G_{1}'(c^{(1)}_{0})).
\end{array}
\end{align}
The argument for the one-sided model is similar to the above case with $i = 2$.  

Using $\pd_{z}C^{\Gamma}_{0} = 0$, $u_{\Gamma} = \bm{v_{0}} \cdot \bm{\nu}$, and the equipartition of energy, after integrating from $-l$ to $z$ and matching, equation $(\ref{surfactant:interface})$ gives to order $-2$
\begin{align*}
2M_{\Gamma}(C^{\Gamma}_{0}) W(\Phi_{0}(z))\gamma''(C^{\Gamma}_{0}(z)) \pd_{z}C^{\Gamma}_{1}(z) = 0.
\end{align*}
Since $\gamma'' > 0$ we have that 
\begin{align*}
\pd_{z}C^{\Gamma}_{1} = 0 \mbox{ whenever } \abs{\Phi_{0}} < 1.
\end{align*}

Equation $(\ref{surfactant:chem})$ for the chemical potential gives to zeroth order 
\begin{align*}
M_{0} & = K\sigma(C^{\Gamma}_{0})(-\pd_{zz}\Phi_{1} + W''(\Phi_{0})\Phi_{1}) + K\sigma'(C^{\Gamma}_{0})C^{\Gamma}_{1}\underbrace{(- \pd_{zz}\Phi_{0} + W'(\Phi_{0})}_{=0}) \\
& - K\surf \cdot ( \sigma(C^{\Gamma}_{0}) \bm{\nu})\pd_{z}\Phi_{0} + \sum_{i= 1,2} \xi_{i}'(\Phi_{0})(G_{i}(C^{(i)}_{0}) - G_{i}'(C^{(i)}_{0})C^{(i)}_{0}).
\end{align*}
To obtain a solution $\Phi_{1}$, a solvability condition has to hold.  Multiply the above by $\pd_{z}\Phi_{0}$ and integrate from $-l$ to $+l$ gives
\begin{align*}
& \int_{-l}^{+l}M_{0}\pd_{z}\Phi_{0} dz  = K\int_{-l}^{+l} \sigma(C^{\Gamma}_{0})(-\pd_{zz}\Phi_{1} \pd_{z}\Phi_{0} +  W''(\Phi_{0})\Phi_{1} \pd_{z}\Phi_{0} ) dz  \\
& - K\int_{-l}^{+l} \surf \cdot (\sigma(C^{\Gamma}_{0}) \bm{\nu}) (\pd_{z}\Phi_{0})^{2} dz+ \int_{-l}^{+l} \sum_{i= 1,2} \xi_{i}'(\Phi_{0})(G_{i}(C^{(i)}_{0}) - G_{i}'(C^{(i)}_{0})C^{(i)}_{0})\pd_{z}\Phi_{0} dz.
\end{align*}
Integrating by parts, using $\pd_{z}C^{(i)}_{0} = 0, \pd_{z}C^{\Gamma}_{0} = 0$ and matching lead to
\begin{align*}
2 \mu_{0} & = K\int_{-l}^{+l} \sigma(C^{\Gamma}_{0})\underbrace{(\pd_{zz}\Phi_{0} - W'(\Phi_{0}))}_{=0}\pd_{z}\Phi_{1} dz - K[\sigma(C^{\Gamma}_{0})(\pd_{z}\Phi_{0} \pd_{z}\Phi_{1}  - W'(\Phi_{0}) \Phi_{1})]_{-l}^{+l} \\
& - K\surf \cdot (\sigma(C^{\Gamma}_{0}) \bm{\nu}) \int_{-l}^{+l} (\pd_{z}\Phi_{0})^{2} dz + \sum_{i= 1,2} [(G_{i}(C^{(i)}_{0}) - G_{i}'(C^{(i)}_{0})C^{(i)}_{0})\xi_{i}(\Phi_{0})]_{-l}^{+l}.
\end{align*}
We use the fact that $W'(\pm 1) = 0$ for the double-well potential and $(\ref{MC1})$ to cancel the first jump term. Furthermore
\begin{align*}
\surf \cdot (\sigma(C^{\Gamma}_{0}) \bm{\nu} ) = \sigma(C^{\Gamma}_{0})\surf \cdot \bm{\nu} + \underbrace{\surf \sigma(C^{\Gamma}_{0}) \cdot \bm{\nu}}_{=0} = - \kappa \sigma(C^{\Gamma}_{0})
\end{align*}
and by equipartition of energy $(\ref{inner:equipartition})$ we deduce that the solvability condition is
\begin{align}\label{inner:solvability}
 2\mu_{0} = \sigma(c^{\Gamma}_{0})\kappa + [G_{i}(c^{(i)}_{0}) - G_{i}'(c^{(i)}_{0})c^{(i)}_{0}]_{1}^{2}.
\end{align}
For the double-obstacle potential, the equation for $\Phi_{1}$ is expressed as a variational inequality: For all $\psi_{0} \in \mathcal{K}$
\begin{align*}
&K\Big{(} -\sigma(C^{\Gamma}_{0})(\pd_{zz}\Phi_{1} + \Phi_{1}) - \sigma'(C^{\Gamma}_{0})C^{\Gamma}_{1}(\pd_{zz}\Phi_{0} + \Phi_{0}) - \pd_{z}\Phi_{0} \surf \cdot (\sigma(C^{\Gamma}_{0}) \bm{\nu} ) , \psi_{0} - \Phi_{0} \Big{)} \\
& \geq \Big{(} M_{0} - \sum_{i= 1,2} \xi_{i}'(\Phi_{0})(G_{i}(C^{(i)}_{0}) - G_{i}'(C^{(i)}_{0})C^{(i)}_{0}), \psi_{0} - \Phi_{0} \Big{)}.
\end{align*}
Whenever $\abs{\Phi_{0}} < 1$, testing with $\psi_{0} = \Phi_{0} + \hat{\psi}_{0}$ with either a non-positive or a non-negative $\hat{\psi}_{0}$ we obtain the equality
\begin{align*}
& -M_{0} - K\sigma(C^{\Gamma}_{0})(\pd_{zz}\Phi_{1} + \Phi_{1}) - K\sigma'(C^{\Gamma}_{0})C^{\Gamma}_{1}(\pd_{zz}\Phi_{0} + \Phi_{1}) \\
& - K\pd_{z}\Phi_{0} \surf \cdot (\sigma(C^{\Gamma}_{0}) \bm{\nu} ) + \sum_{i= 1,2} \xi_{i}'(\Phi_{0})(G_{i}(C^{(i)}_{0}) - G_{i}'(C^{(i)}_{0})C^{(i)}_{0}) = 0.
\end{align*}
Multiplying by $\pd_{z}\Phi_{0}$ and integrate from $-l$ to $+l$ gives after matching
\begin{align*}
& 2\mu_{0} - \sigma(c^{\Gamma}_{0})\kappa -  \sum_{i=1,2} [\xi_{i}(\varphi_{0})(G_{i}(c^{(i)}_{0}) - G_{i}'(c^{(i)}_{0})c^{(i)}_{0}))]_{-l}^{+l} \\
& = K\int_{-l}^{+l} - \sigma(C^{\Gamma}_{0})(\pd_{zz}\Phi_{1} + \Phi_{1})\pd_{z}\Phi_{0} dz \\
& = -K[\sigma(C^{\Gamma}_{0})(\pd_{z}\Phi_{0}\pd_{z}\Phi_{1} + \Phi_{0}\Phi_{1})]_{-l}^{+l} + K\sigma(C^{\Gamma}_{0}) \int_{-l}^{+l} \pd_{z}\Phi_{1}(\pd_{zz}\Phi_{0} + \Phi_{0}) dz.
\end{align*}
The last integral term is zero due to $(\ref{inner:profile})$, and using $(\ref{MC1})$ for $\Phi_{0}$ and $(\ref{inner:Phi1})$ for $\Phi_{1}$ at $z = \pm l$ the jump term is also zero.  This leads to the same solvability condition as in $(\ref{inner:solvability})$.

Using $\pd_{z}M_{0} = 0$, $u_{\Gamma} = \bm{v}_{0} \cdot \bm{\nu}$, $\surf \Phi_{0} = 0$ and equipartition of energy, the momentum equation $(\ref{surfactant:momentum})$ gives to order $-1$
\begin{multline*}
 \pd_{z}P_{0} \bm{\nu} + \eta(\Phi_{0}) \mathcal{E}(\pd_{z}\bm{V}_{1} \otimes \bm{\nu})\bm{\nu} + \eta(\Phi_{0}) \mathcal{E}(\surf \bm{V}_{0}) \bm{\nu} - \pd_{z}(\bm{V}_{0}\otimes\tfrac{\overline{\rho}^{(2)} - \overline{\rho}^{(1)}}{2}m(\Phi_{0}) \pd_{z}M_{1}\bm{\nu}) \\
= K\abs{\pd_{z}\Phi_{0}}^{2}(\surf \cdot (\sigma(C^{\Gamma}_{0})\bm{I}) - \surf \cdot (\sigma(C^{\Gamma}_{0}) \bm{\nu} \otimes \bm{\nu}))
\end{multline*}
where we used that $\bm{V}_{0}$ is constant in $z$.  Matching $(\ref{MC2})$ requires that $\lim_{z \to \pm l} \pd_{z}\bm{V}_{1}(z) = \nabla \bm{v}_{0}^{\pm} \bm{\nu}$ and hence
\begin{align*}
 \pd_{z} \bm{V}_{1} \otimes \bm{\nu} + \surf \bm{V}_{0} \to \nabla \bm{v}_{0} \text{ for } z \to \pm l.
\end{align*}
Furthermore, a short calculation shows that 
\begin{align*}
 \surf \cdot (\sigma(C^{\Gamma}_{0}) \bm{I}) - \surf \cdot (\sigma(C^{\Gamma}_{0})\bm{\nu} \otimes \bm{\nu}) 
& = \surf \sigma(C^{\Gamma}_{0}) + \kappa \sigma(C^{\Gamma}_{0})\bm{\nu}.
\end{align*}
So upon integrating from $-l$ to $+l$, matching and using that $m(\pm 1) = 0$ we obtain 
\begin{align}\label{inner:stressjump}
[p_{0}]_{1}^{2} \bm{\nu} - 2\eta[D(\bm{v_{0}})]_{1}^{2} \bm{\nu} = \kappa \sigma(C^{\Gamma}_{0}) \bm{\nu} + \surf \sigma(C^{\Gamma}_{0}).
\end{align}

\subsubsection{Inner equations and solutions to second order}
\label{Sec:2ndOrder}
Using $u_{\Gamma} = \bm{v}_{0} \cdot \bm{\nu}$, $\pd_{z}C^{\Gamma}_{0} = \pd_{z}C^{\Gamma}_{1} = 0$ and equipartition of energy $(\ref{inner:equipartition})$, equation $(\ref{surfactant:interface})$ gives to order $-1$
\begin{align*}
& K \Big{(} \nd \left ( 2C^{\Gamma}_{0}W(\Phi_{0}) \right ) + \bm{V}_{0} \cdot \surf \left ( 2C^{\Gamma}_{0}W(\Phi_{0}) \right )  + (\bm{V}_{1} \cdot \bm{\nu}) \pd_{z}\left ( 2C^{\Gamma}_{0} W(\Phi_{0})  \right ) \Big{)} \\
& \quad = K\pd_{z} \big{(} 2M_{\Gamma}(C^{\Gamma}_{0}) W(\Phi_{0}) \gamma''(C^{\Gamma}_{0}) \pd_{z}C^{\Gamma}_{2} \big{)} 
+ K\surf \cdot \big{(} 2M_{\Gamma}(C^{\Gamma}_{0})W(\Phi_{0}) \surf \gamma'(C^{\Gamma}_{0}) \big{)} \\
& \quad \quad -2W(\Phi_{0})\sum_{i = 1,2} \beta^{(i)}(\gamma'(C^{\Gamma}_{0}) - G_{i}'(C^{(i)}_{0})).
\end{align*}
Integrating from $-l$ to $+l$, we obtain 
\begin{align*}
& \Big{(} \nd C^{\Gamma}_{0} + \bm{V}_{0} \cdot \surf C^{\Gamma}_{0} \Big{)} K \int_{-l}^{+l} 2W(\Phi_{0}) dz + K\int_{-l}^{+l} ( \bm{V}_{1} \cdot \bm{\nu} ) \pd_{z}(2W(\Phi_{0})C^{\Gamma}_{0}) dz\\
& = K [2M_{\Gamma}(C^{\Gamma}_{0}) W(\Phi_{0})\gamma''(C^{\Gamma}_{0})\pd_{z}C^{\Gamma}_{2} ]_{-l}^{+l} + \surf \cdot \big{(} M_{\Gamma}(C^{\Gamma}_{0}) \surf \gamma'(C^{\Gamma}_{0}) \big{)} K \int_{-l}^{+l} 2W(\Phi_{0}) dz \\
& \quad -\sum_{i = 1,2} \beta^{(i)}(\gamma'(C^{\Gamma}_{0}) - G_{i}'(C^{(i)}_{0}))\mathcal{W}.
\end{align*}
Applying the matching conditions $(\ref{MC0})$ to $\Phi_{0}$ and $C_{0}^{\Gamma}$ and $(\ref{MC3})$ to $C_{2}^{\Gamma}$ we see that
\begin{align*}
 [M_{\Gamma}(C^{\Gamma}_{0})\left (W(\Phi_{0})\gamma''(C^{\Gamma}_{0})\pd_{z}C^{\Gamma}_{2} \right )]_{-l}^{+l} = 0.
\end{align*}
By $(\ref{inner:mass})$ we have that
\begin{align*}
 \int_{-l}^{+l} (\bm{V}_{1} \cdot \bm{\nu})\pd_{z}(2C^{\Gamma}_{0}W(\Phi_{0})) dz & = [2(\bm{V}_{1} \cdot \bm{\nu}) C^{\Gamma}_{0}W(\Phi_{0})]_{-l}^{+l} - \int_{-l}^{+l} 2\pd_{z}(\bm{V}_{1} \cdot \bm{\nu}) W(\Phi_{0})C^{\Gamma}_{0} dz \\
& = 0 + c^{\Gamma}_{0} \int_{-l}^{+l} (\surf \cdot \bm{V}_{0}) 2W(\Phi_{0}) dz = \mathcal{W} c^{\Gamma}_{0} \surf \cdot \bm{v}_{0}, % = c^{\Gamma}_{0} \surf \cdot \bm{v}_{0},
\end{align*}
and by $(\ref{inner:freebdy})$ 
\begin{align*}
 \sum_{i = 1,2} \beta^{(i)}(\gamma'(c^{\Gamma}_{0}) - G_{i}'(c^{(i)}_{0}))\mathcal{W} = [\bm{J}_{c,0}^{(i)}]_{1}^{2} \bm{\nu}.
\end{align*}
Using $\md(\cdot) = \nd(\cdot) + \bm{v} \cdot \surf (\cdot)$, we finally obtain the desired surface surfactant equation
\begin{align}\label{inner:interfacialsurfactantbalance}
\md c^{\Gamma}_{0}  + c^{\Gamma}_{0} \surf \cdot \bm{v}_{0} - \surf \cdot \big{(} M_{\Gamma}(c^{\Gamma}_{0}) \surf \gamma'(c^{\Gamma}_{0}) \big{)} = [\bm{J}_{c,0}^{(i)}]_{2}^{1} \bm{\nu}.  
\end{align}

\subsection{Alternative asymptotic limit for Model A}
Let us now assume that $\beta^{(i)}$ scales with $\eps^{-1}$, i.e.
\begin{align*}
 \beta^{(i)} = \frac{1}{\mathcal{W} \eps}.
\end{align*}
Then we obtain instantaneous adsorption $(\ref{eq:instadsorption})$ instead of $(\ref{SIM:eq7})$ in the limit $\eps \to 0$, which will be demonstrated in what follows. 

\subsubsection{Inner equations and solutions to leading and first order}
We recover $[\bm{v}_{0} \cdot \bm{\nu}]_{1}^{2} = 0$ and obtain $\pd_{z}V_{1} \cdot \bm{\nu} + \surf \cdot \bm{V}_{0} = 0$ from equation $(\ref{surfactant:mass})$ to order $-1$ and to zeroth order respectively.  From equation $(\ref{surfactant:phase})$ we obtain $\pd_{z}M_{0} = 0$ and $u_{\Gamma} = \bm{v}_{0} \cdot \bm{\nu}$ to order $-2$ and to order $-1$ respectively.  To order $-2$ equation $(\ref{surfactant:momentum})$ gives $[\bm{v}_{0}]_{1}^{2} = 0$.

To order $-3$, the interfacial surfactant equation $(\ref{surfactant:interface})$ gives $\pd_{z} C^{\Gamma}_{0} = 0$.  This leads to the profile $\Phi_{0}$ and equipartition of energy $(\ref{inner:equipartition})$ from $(\ref{surfactant:chem})$.  Furthermore, we obtain the solvability condition $(\ref{inner:solvability})$ from $(\ref{surfactant:chem})$ at zeroth order and the jump in the stress tensor $(\ref{inner:stressjump})$ from $(\ref{surfactant:momentum})$ at order $-1$.  

To order $-2$ we obtain from $(\ref{surfactant:bulk})$ and $(\ref{surfactant:interface})$
\begin{align}
 -\mathcal{W} \pd_{z} \big{(} M_{c}^{(i)}(C^{(i)}_{0}) \xi_{i}(\Phi_{0}) \pd_{z}G'_{i}(C^{(i)}_{0}) \big{)} & = 2 W(\Phi_{0})(\gamma'(C^{\Gamma}_{0}) - G'_{i}(C^{(i)}_{0})), \label{AlternateLimit:bulk} \\
 \mathcal{W} \pd_{z} \big{(} M_{\Gamma}(C^{\Gamma}_{0}) K 2W(\Phi_{0}) \pd_{z} (\gamma''(C^{\Gamma}_{0})C^{\Gamma}_{1}) \big{)} & = \sum_{i=1,2} 2W(\Phi_{0})(\gamma'(C^{\Gamma}_{0}) - G'_{i}(C^{(i)}_{0})). \label{AlternateLimit:interface}
\end{align}
Now, multiplying $(\ref{AlternateLimit:bulk})$ by $G'_{i}(C^{(i)}_{0})$, $i=1,2$, and $(\ref{AlternateLimit:interface})$ by $\gamma'(C^{\Gamma}_{0})$ and subtracting gives
\begin{align*}
& - \mathcal{W} \sum_{i=1,2} \pd_{z} \big{(} M_{c}^{(i)}\xi_{i}(\Phi_{0}) \pd_{z}G'_{i}(C^{(i)}_{0}) \big{)} G'_{i}(C^{(i)}_{0}) + 2W(\Phi_{0})\sum_{i=1,2} \abs{\gamma'(C^{\Gamma}_{0}) - G'_{i}(C^{(i)}_{0})}^{2}\\
& - \mathcal{W} \pd_{z}\big{(} M_{\Gamma} 2K W(\Phi_{0}) \pd_{z}(\gamma''(C^{\Gamma}_{0})C^{\Gamma}_{1}) \big{)} \gamma'(C^{\Gamma}_{0}) = 0.
\end{align*}
Integrating from $-l$ to $+l$, integrating by parts and using that $\pd_{z}C^{\Gamma}_{0} = 0$ yields
\begin{multline*}
0 = \mathcal{W} \sum_{i=1,2} \int_{-l}^{+l} M_{c}^{(i)} \xi_{i}(\Phi_{0}) \abs{\pd_{z}G'_{i}(C^{(i)}_{0})}^{2} dz - \mathcal{W} \big{[} M_{c}^{(i)} \xi_{i}(\Phi_{0}) \pd_{z}G'_{i}(C^{(i)}_{0})  G'_{i}(C^{(i)}_{0}) \big{]}_{-l}^{+l} \\
- \big{[} M_{\Gamma} 2 K W(\Phi_{0}) \gamma''(C^{\Gamma}_{0}) \pd_{z} C^{\Gamma}_{1} \gamma'(C^{\Gamma}_{0}) \big{]}_{-l}^{+l} + \sum_{i=1,2} \int_{-l}^{+l} 2W(\Phi_{0}) \abs{\gamma'(C^{\Gamma}_{0}) - G'_{i}(C^{(i)}_{0})}^{2}.
\end{multline*}
The third term vanishes as $W(\pm 1) = 0$, and when applying $(\ref{MC1})$ to $C^{(i)}_{0}$ then the second term is zero, too. Hence we have
\begin{align*}
\mathcal{W} \sum_{i=1,2} \int_{-l}^{+l} M_{c}^{(i)} \xi_{i}(\Phi_{0}) \abs{\pd_{z}G'_{i}(C^{(i)}_{0})}^{2} + \sum_{i=1,2} \int_{-l}^{+l} 2W(\Phi_{0}) \abs{\gamma'(C^{\Gamma}_{0}) - G'_{i}(C^{(i)}_{0})}^{2} = 0.
\end{align*}
As all the terms are non-negative, this implies that
\begin{align*}
 \pd_{z}C^{(i)}_{0} = 0 \quad \text{and} \quad \gamma'(C^{\Gamma}_{0}) = G'_{i}(C^{(i)}_{0}).
\end{align*}

\subsubsection{Inner equations and solutions to second order}
Adding the surfactant equations $(\ref{surfactant:bulk})$ and $(\ref{surfactant:interface})$, the order $-1$ terms yield
\begin{align*}
& 2KW(\Phi_{0}) \big{(} \nd C^{\Gamma}_{0} + \bm{V}_{0} \cdot \surf C^{\Gamma}_{0} \big{)} + \bm{V}_{1} \cdot \bm{\nu} \pd_{z}(2 K W(\Phi_{0}) C^{\Gamma}_{0}) \\
& = \pd_{z} \big{(} M_{\Gamma} 2K W(\Phi_{0}) \gamma''(C^{\Gamma}_{0}) \pd_{z} C^{\Gamma}_{2} + M_{\Gamma} K (\pd_{z}\Phi_{0} \pd_{z} \Phi_{1} + W'(\Phi_{0}) \Phi_{1}) \gamma''(C^{\Gamma}_{0}) \pd_{z} C^{\Gamma}_{1} \big{)} \\
& \quad + \surf \cdot (M_{\Gamma} 2 K W(\Phi_{0}) \surf \gamma'(C^{\Gamma}_{0})) + \sum_{i=1,2} \pd_{z}(M_{c}^{(i)} \xi_{i}(\Phi_{0}) G''_{i}(C^{(i)}_{0}) \pd_{z} C^{(i)}_{1}).
\end{align*}
Integrating from $-l$ to $+l$ and matching $(\ref{MC2})$ applied to $\pd_{z}C^{(i)}_{1}$ leads to $(\ref{inner:interfacialsurfactantbalance})$ again.

\subsection{Asymptotic analysis for Models B and C}
The asymptotic analysis for Models B and C are similar, hence we will only sketch the analysis for Model C.  In the following, the analysis for Model B can be recovered by setting variables with index 1 to zero and replacing $c^{(2)}(q), c^{\Gamma}(q), q$ with $c, g(c), \nabla G'(c)$.  

First we express $(\ref{inst3:surfactant})$ as
\begin{align}\label{Asym:surfactantModelBC}
 \md(\xi_{1}(\varphi) c^{(1)}(q) + \xi_{2}(\varphi) c^{(2)}(q) + K \delta(\varphi, \nabla \varphi) c^{\Gamma}(q)) + \nabla \cdot \bm{J} = 0,
\end{align}
where
\begin{align*}
 \bm{J} := - \big{(} M_{c}^{(1)} \xi_{1}(\varphi) + M_{c}^{(2)} \xi_{2}(\varphi) + M_{\Gamma} K \delta(\varphi, \nabla \varphi) \big{)} \nabla q.
\end{align*}
Based on the outer and inner expansions of $\delta(\varphi, \nabla \varphi)$, we assume that $\bm{J}$ has the following outer and inner expansions: 
\begin{align*}
 \bm{J} & = \eps^{-2} \bm{J}^{\text{bulk}}_{-2} + \eps^{-1} \bm{J}^{\text{bulk}}_{-1} + \bm{J}^{\text{bulk}}_{0} + \dots, \\
 \bm{J} & = \eps^{-2} \bm{J}^{\text{int}}_{-2} + \eps^{-1} \bm{J}^{\text{int}}_{-1} + \bm{J}^{\text{int}}_{0} + \dots,
\end{align*}
where, for example, 
\begin{align*}
& \bm{J}^{\text{bulk}}_{-2} = 0, \quad \bm{J}^{\text{bulk}}_{-1} = -M_{\Gamma}(c^{\Gamma}_{0})W(\varphi_{0}) \nabla q_{0},\\
& \bm{J}^{\text{int}}_{-2} = -M_{\Gamma}(C^{\Gamma}_{0})(\tfrac{1}{2}\abs{\pd_{z}\Phi_{0}}^{2} + W(\Phi_{0}))\pd_{z}Q_{0}\bm{\nu}.
\end{align*}
The matching conditions for $\bm{J}$ are as follows (see \cite{article:GarckeStinner06}): As $z \to \pm l$,
\begin{align}
 \bm{J}^{\text{int}}_{-2}(t,s,z) & \sim 0, \quad \pd_{z} \bm{J}^{\text{int}}_{-2}(t,s,z) \sim 0, \label{Alternate:MC0} \\
 \bm{J}^{\text{int}}_{-1}(t,s,z) & \sim (\bm{J}^{\text{bulk}}_{-1})^{\pm}(t, \bm{x}) \cdot \bm{\nu}, \quad \pd_{z}\bm{J}^{\text{int}}_{-1}(t,s,z)  \sim 0, \label{Alternate:MC1} \\
 \bm{J}^{\text{int}}_{0}(t,s,z) & \sim (\bm{J}^{\text{bulk}}_{0})^{\pm}(t,\bm{x}) + \nabla (\bm{J}^{\text{bulk}}_{-1})^{\pm}(t,\bm{x}) \cdot \bm{\nu} z. \label{Alternate:MC2} 
\end{align}

\subsubsection{Outer equations and solutions}
From equation $(\ref{inst3:chem})$ we obtain to order $-1$
\begin{align*}
 0 = \tilde{\sigma}(q_{0}) W'(\varphi_{0}),
\end{align*}
from which we obtain stable solutions $\varphi_{0} = \pm 1$ and regions $\Omega^{(1)}, \Omega^{(2)}$ defined as in previous models. We also recover the usual fluid equation, incompressibility condition to zeroth order.  

With respect to the surfactant, to order $-1$ we have
\begin{align}\label{AsympBC:Flux}
 \bm{J}^{\text{bulk}}_{-1} = - M_{\Gamma} K W(\varphi_{0}) \nabla q_{0} = 0.
\end{align}
To zeroth order we recover the bulk surfactant equations from $(\ref{Asym:surfactantModelBC})$:
\begin{align*}
 \md(\xi_{1}(\varphi_{0}) c^{(1)}(q_{0}) + \xi_{2}(\varphi_{0}) c^{(2)}(q_{0})) - \nabla \cdot (M_{c}^{(1)} \xi_{1}(\varphi_{0})\nabla q_{0} + M_{c}^{(2)} \xi_{2}(\varphi_{0})\nabla q_{0}) = 0
\end{align*}
where $\xi_1(\varphi_{0}) = \xi_1(-1) = 0$ in $\Omega^{(2)}$ and $\xi_2(\varphi_{0}) = \xi_2(1) = 0$ in $\Omega^{(1)}$.

\subsubsection{Inner equations and solutions to leading and first order}
We recover $[\bm{v}_{0} \cdot \bm{\nu}]_{-}^{+} = 0$ and obtain $\pd_{z}V_{1} \cdot \bm{\nu} + \surf \cdot \bm{V}_{0} = 0$ from equation $(\ref{inst2:mass})$ to orders $-1$ and to zeroth order respectively.  From equation $(\ref{inst2:phase})$ we obtain $\pd_{z}M_{0} = 0$ and $u_{\Gamma} = \bm{v}_{0} \cdot \bm{\nu}$ to order $-2$ and to order $-1$ respectively.  To order $-2$ equation $(\ref{inst2:momentum})$ gives $[\bm{v}_{0}]_{-}^{+} = 0$.

To order $-3$, we have from $(\ref{Asym:surfactantModelBC})$
\begin{align*}
 \pd_{z} \bm{J}^{\text{int}}_{-2} \cdot \bm{\nu} = 0,
\end{align*}
where
\begin{align*}
 \bm{J}^{\text{int}}_{-2} = - M_{\Gamma} K (\tfrac{1}{2} \abs{\pd_{z}\Phi_{0}}^{2} + W(\Phi_{0})) \pd_{z} Q_{0} \bm{\nu}.
\end{align*}
This implies that $\bm{J}^{\text{int}}_{-2} \cdot \bm{\nu}$ is constant in $z$.  Furthermore, for any $\bm{\tau}$ such that $\bm{\tau} \cdot \bm{\nu} = 0$, we have $\bm{J}^{\text{int}}_{-2} \cdot \bm{\tau} = 0$.  Hence $\bm{J}^{\text{int}}_{-2} \equiv 0$ by $(\ref{Alternate:MC0})$ and this implies $\pd_{z} Q_{0} = 0$.

Equation $(\ref{inst3:chem})$ gives to order $-1$
\begin{align*}
 0 = -\pd_{z} \big{(} K \tilde{\sigma}(Q_{0})\pd_{z}\Phi_{0} \big{)} + K\tilde{\sigma}(Q_{0})W'(\Phi_{0}).
\end{align*}
Since $\pd_{z} Q_{0} = 0$ we obtain $0 = -\pd_{zz}\Phi_{0} + W'(\Phi_{0})$ again, which gives the profile for $\Phi_{0}$ and the equipartition of energy $(\ref{inner:equipartition})$.  Hence, we obtain the same solvability condition for $\Phi_{1}$ from equation $(\ref{inst3:chem})$:
\begin{align*}
 2\mu_{0} = \tilde{\sigma}(q_{0})\kappa + [(G_{i}(c^{(i)}(q_{0})) - q_{0}c^{(i)}(q_{0})))]_{1}^{2}.
\end{align*}
As previously, equation $(\ref{inst2:momentum})$ then gives to order $-1$	
\begin{align*}
[p_{0}]_{1}^{2} \bm{\nu} - 2[\eta^{(i)} D(\bm{v}_{0})]_{1}^{2} \bm{\nu} = \kappa \tilde{\sigma}(q_{0}) \bm{\nu} + \surf \tilde{\sigma}(q_{0}).
\end{align*}
To order $-2$, we have from $(\ref{Asym:surfactantModelBC})$
\begin{align*}
 \pd_{z} \bm{J}^{\text{int}}_{-1} \cdot \bm{\nu} = \pd_{z}( \bm{J}^{\text{int}}_{-1} \cdot \bm{\nu} )  = 0,
\end{align*}
where, thanks to $\pd_{z}Q_0 = 0$,
\begin{align*}
 \bm{J}^{\text{int}}_{-1} = -M_{\Gamma} (c^{\Gamma}(Q_0)) 2 K W(\Phi_{0}) (\surf Q_{0} + \pd_{z}Q_{1} \bm{\nu}).
\end{align*}
This implies that
\begin{align*}
 \pd_{z} \big{(} M_{\Gamma} 2 K W(\Phi_{0}) \pd_{z}Q_{1} \big{)} = 0.
\end{align*}
Integrating from $-l$ to $z$ and matching $(\ref{MC0})$ applied to $\Phi_{0}$ gives that 
\begin{align*}
 \pd_{z} Q_1 = 0 \mbox{ whenever } \abs{\Phi_{0}} < 1.
\end{align*}

\subsubsection{Inner equations and solutions to second order}
To order $-1$, equation $(\ref{Asym:surfactantModelBC})$ gives
\begin{multline*}
2KW(\Phi_{0}) \big{(} \nd c^{\Gamma}(Q_{0}) + \bm{V}_{0} \cdot \surf c^{\Gamma}(Q_{0}) \big{)} + \bm{V}_{1} \cdot \bm{\nu} \pd_{z}(2 K W(\Phi_{0}) c^{\Gamma}(Q_{0}))  \\
= - \surf \cdot \bm{J}^{\text{int}}_{-1} -\pd_{z} \bm{J}^{\text{int}}_{0} \cdot \bm{\nu}
\end{multline*}
where, using the already obtained results, $\bm{J}^{\text{int}}_{-1} = -M_{\Gamma}(c^{\Gamma}(Q_0)) 2 K W(\Phi_{0}) \surf Q_{0}$.

Proceeding as in Section \ref{Sec:2ndOrder}, the left hand side yields
\begin{align*}
 \md (c^{\Gamma}(q_{0})) + c^{\Gamma}(q_{0}) \surf \cdot \bm{v}_{0}.
\end{align*}
For the right hand side, the integration from $-l$ to $+l$ gives
\begin{align*}
 - \surf \cdot \left ( \int_{-l}^{+l} \bm{J}^{\text{int}}_{-1} \right ) - \left. \bm{J}^{\text{int}}_{0} \cdot \bm{\nu} \right \vert_{-l}^{+l},
\end{align*}
where
\begin{align*}
  - \surf \cdot \left ( \int_{-l}^{+l} \bm{J}^{\text{int}}_{-1} \right ) = \surf \cdot \big{(}M_{\Gamma} \surf q_{0}\big{)}
\end{align*}
and (\ref{Alternate:MC2}), (\ref{AsympBC:Flux}) give
\begin{align*}
-\left. \bm{J}^{\text{int}}_{0} \cdot \bm{\nu} \right \vert_{-l}^{+l} =  -\left. \bm{J}^{\text{bulk}}_{0} \cdot \bm{\nu} \right \vert_{-}^{+} = - \big{(} - M_{c}^{(2)} \nabla q_{0} + M_{c}^{(1)} \nabla q_{0} \big{)} \cdot \bm{\nu} = [\bm{J}_{c,0}^{(i)}]_{2}^{1} \bm{\nu}.
\end{align*}
Hence we obtain the surface surfactant equation
\begin{align*}
 \md (c^{\Gamma}(q_{0})) + c^{\Gamma}(q_{0}) \surf \cdot \bm{v}_{0} = \surf \cdot \big{(} M_{\Gamma} \surf q_{0} \big{)} + [\bm{J}_{c,0}^{(i)}]_{2}^{1} \bm{\nu}.
\end{align*}

\section{Numerical experiments}\label{numerics}
In this section we report on numerical experiments that serve to support the above asymptotic analysis and illustrate that the proposed phase field models are able to describe phenomena that can be observed in physical experiments. Since the phase field approach to two-phase flow has been intensively studied already and the extension consists of accounting for the surfactant dynamics, the numerical experiments are designed to focus on the latter one. 

\subsection{Surfactant adsorption dynamics in 1D}
We first carefully investigate the adsorption of surfactants to interfaces in a one-dimensional setting where we exclude the effects of fluid transport ($\bm{v} = 0$) and focus on the dynamics between bulk and interfacial surfactants.  We assume that the surfactant is insoluble in $\Omega^{(1)}$ and the sharp interface model is a variant of the Ward--Tordai problem defined on a bounded domain.  For the phase field models we assume that $\varphi$ is given, then the dimensionless equations of Model A simplifies down to (dropping the index $*$ and the index $2$ for the bulk phase)
\begin{align*}
\pd_{t} \big{(} \xi(\varphi)c \big{)} - \pd_{x} \Big{(} \frac{1}{\text{Pe}_{c}}\xi(\varphi)\pd_{x} c \Big{)} = \beta \delta(\varphi, \pd_{x}\varphi) \big{(} \gamma'(c^{\Gamma}) - G'(c) \big{)}, \\
\pd_{t} \big{(} K\delta(\varphi, \pd_{x} \varphi)c^{\Gamma} \big{)} - \pd_{x} \Big{(} \frac{K}{\text{Pe}_{\Gamma}} \delta(\varphi, \pd_{x} \varphi)\pd_{x} c^{\Gamma} \Big{)} =  -\beta \delta(\varphi, \pd_{x}\varphi) \big{(} \gamma'(c^{\Gamma}) - G'(c) \big{)}. 
\end{align*}
For Model B we have one equation instead,
\begin{align*}
\pd_{t} \big{(} \xi(\varphi)c + K\delta(\varphi, \pd_{x} \varphi)g(c) \big{)} - \pd_{x} \Big{(} \frac{1}{\text{Pe}_{c}}\xi(\varphi) \pd_{x} c + \frac{K}{\text{Pe}_{\Gamma}} \delta(\varphi, \pd_{x} \varphi)\pd_{x} c \Big{)} = 0,  
\end{align*}
and for Model C, we replace $c, g(c), \pd_{x}c$ by $c(q), c^{\Gamma}(q), \pd_{x}q$ in the above equation.  

To support the asymptotic analysis we test
\begin{itemize}
\item the $\eps$-convergence of the profile of $c(x,1)$;
\item the $\eps$-convergence of the profile of $c^{\Gamma}(0,t)$;
\item the $\eps$-convergence of $\abs{\gamma'(c^{\Gamma}) - G'(c)}$ at $x = 0, t = 1$.
\end{itemize}
The third test only applies to Model A when $\beta$ is scaled with $\eps^{-1}$, as the Dirichlet-type condition $\gamma'(c^{\Gamma}) = G'(c)$ for instantaneous adsorption is enforced in the limit $\eps \to 0$.  

To measure the $\eps$-convergence of the profiles, we look at the difference $\abs{c_{PF} - c_{SI}}$ and $\abs{c^{\Gamma}_{PF} - c^{\Gamma}_{SI}}$ where $c^{\Gamma}_{PF}(x,t)$ and $c_{PF}(x,t)$ are the interfacial and bulk densities of the phase field models respectively, while $c^{\Gamma}_{SI}(t)$ and $c_{SI}(x,t)$ denote the interfacial and bulk densities of the sharp interface model respectively.  We will be comparing $\{(\ref{Numerics:SIM}),(\ref{Numerics:SIMNonInst})\}$ with Model A $(\alpha > 0)$ and $\{(\ref{Numerics:SIM}),(\ref{Numerics:SIMInst})\}$ with Model A $(\alpha \to 0)$ and Model B.  The numerical methods described in this section have been implemented using the software MATLAB, Version 7.11.0 (R2010b), \cite{book:Matlab2010}.

\subsubsection{Sharp interface model}
Set $\Omega = [0, 1]$ and $\Gamma$ as the point $x=0$, the dimensionless sharp interface model is
\begin{equation}\label{Numerics:SIM}
\begin{array}{lr}
\pd_{t} c = \frac{1}{\text{Pe}_{c}}\pd_{xx} c & \quad \text{ in } (0,1], \\[6pt]
\pd_{t} c^{\Gamma} = \frac{1}{\text{Pe}_{c}}\pd_{x} c & \quad \text{ at } x = 0, 
\end{array}
\end{equation}
together with
\begin{align}\label{Numerics:SIMNonInst}
\frac{\alpha}{\text{Pe}_{c}} \pd_{x} c = -(\gamma'(c^{\Gamma}) - G'(c)) & \text{ at } x = 0 
\end{align}
for non-instantaneous adsorption or
\begin{align}\label{Numerics:SIMInst}
 c^{\Gamma}(t) = g(t) = (\gamma')^{-1}(G'(c))) & \text{ at } x = 0
\end{align}
for instantaneous adsorption.  We impose the following initial-boundary conditions 
\begin{align*}
 c(x=1,t) = 1, \quad c(x, t=0) = 1, \quad c^{\Gamma}(t=0) = c^{\Gamma}_{0}.
\end{align*}
This is a version of the famous Ward--Tordai problem on a bounded interval, see \cite{article:WardTordai46}.  We solve the problem via a finite-difference scheme:  Let $0 = x_{1} < \dots < x_{N} = 1$ be a uniform discretisation of $\Omega$ with mesh size $h = 1/N$.  Let $\Delta t = 1/N_{f}$ for integer $N_{f} \in \N$ be a time step and define $t_{n} = n \Delta t$ for $n=0, \dots, N_{f}$.  Let $\theta = \Delta t/(\text{Pe}_{c} h^2)$ and denote $c^{n}(x) = c(x,t_{n})$.  Then given $\bm{c}^{n} = (c^{n}(x_{1}), \dots, c^{n}(x_{N-1}), c^{n}(x_{N}))$, the solution at time $t_{n}$, we solved for $\bm{c}^{n+1} = (c^{n+1}(x_{1}), \dots, c^{n+1}(x_{N-1}), c^{n+1}(x_{N}))$, which for $\{(\ref{Numerics:SIM}),(\ref{Numerics:SIMNonInst})\}$ satisfies
\begin{align*}
\left ( \begin{array}{cccccc}
1+ 2\theta & -2\theta & 0 & \dots & \dots & 0 \\
-\theta & 1+ 2\theta & - \theta & 0 & \dots & 0 \\
\vdots & \ddots & \ddots & \ddots & \ddots  & \vdots \\
0 & \dots &  0 & -\theta & 1+2\theta & -\theta \\
0 & \dots &  \dots & \dots & 0 & 1 \\
\end{array} \right ) \bm{c}^{n+1} = \left ( \begin{array}{c}
c^{n}(x_{1}) + \tfrac{2h \text{Pe}_{c}\theta}{\alpha}(\gamma'(c^{\Gamma,n}) - G'(c^{n}(x_{1}))) \\
c^{n}(x_{2}) \\
\vdots \\
c^{n}(x_{N-1}) \\
1
\end{array} \right ),
\end{align*}
and then
\begin{align*}
c^{\Gamma,n+1} = c^{\Gamma,n} + \theta h (c^{n+1}(x_{2}) - c^{n+1}(x_{1})).
\end{align*}
For $\{(\ref{Numerics:SIM}),(\ref{Numerics:SIMInst})\}$, we have to solve
\begin{align*}
\left ( \begin{array}{cccccc}
\theta h & - \theta h & 0 & \dots & \dots & 0 \\
-\theta & 1+ 2\theta & - \theta & 0 & \dots & 0 \\
\vdots & \ddots & \ddots & \ddots & \ddots  & \vdots \\
0 & \dots &  0 & -\theta & 1+2\theta & -\theta \\
0 & \dots &  \dots & \dots & 0 & 1 \\
\end{array} \right ) \bm{c}^{n+1} + \left ( \begin{array}{c}
g(c^{n+1}(x_{1})) \\
0 \\
\vdots \\
0 \\
0
\end{array} \right ) = \left ( \begin{array}{c}
g(c^{n}(x_{1})) \\
c^{n}(x_{2}) \\
\vdots \\
c^{n}(x_{N-1}) \\
1
\end{array} \right ).
\end{align*}

\subsubsection{Phase field model}

We use the one-sided version for each of the above phase field models. We choose the potential $W$ to be of double-obstacle type (hence $K = \tfrac{2}{\pi}$). This has the advantage that the phase field variable $\varphi$ lies strictly in the interval $[-1,1]$ and interfacial layer has constant width equal to $\eps \pi$.  The asymptotic analysis suggests that to leading order $\varphi(x) = \sin(\tfrac{x}{\eps})$ for $\abs{x} \leq \eps \tfrac{\pi}{2}$, and thanks to equipartition of energy $\delta(\varphi, \pd_{x} \varphi)$ simplifies to
\begin{align*}
\delta(\varphi, \pd_{x} \varphi) = 
\begin{cases}
\frac{1}{\eps}\abs{\cos(\tfrac{x}{\eps})}^2, & \quad \abs{x} \leq \eps\frac{\pi}{2}, \\
0, & \quad \abs{x}> \eps \frac{\pi}{2}.
\end{cases}
\end{align*}
The cutoff function $\xi(\varphi(x))$ is chosen to be
\begin{align*}
 \xi(x) = 
\begin{cases}
1, & x \geq \eps \frac{\pi}{2}, \\
\frac{1}{2} (1+\frac{1}{2}(\frac{x}{y})(3-(\frac{x}{y})^2), & \abs{x} < \eps \frac{\pi}{2}, \\
0, & x \leq -\eps\frac{\pi}{2},
\end{cases}
\end{align*}
where $y$ is the integer part of $\eps \tfrac{\pi}{2}$.

For the discretisation we employ linear finite elements and the method of lines. Let $\Delta t = \tfrac{1}{N_{f}}$ for integer $N_{f} \in \N$ be a time step and define $t_{n} = n \Delta t$ for $n=0, \dots, N_{f}$.  Let $\mathcal{T}_{h}$ be a uniform subdivision of the interval $[-1,1]$ consisting of subintervals with size $h$.  Let $N$ be the number of vertices with coordinates denoted by $\{x_{1}, \cdots, x_{N}\}$.   Let $\mathcal{N}$ be the set of vertex indices and for an index $i \in \mathcal{N}$ let $\omega_{i}$ denote the neighbouring vertices connected to vertex $i$ (i.e. $w_{i} = \{ x_{i-1}, x_{i+1}\}$).  Furthermore, based on the functional form of $\delta$ and $\xi$, we define
\begin{align*}
 \mathcal{X}_{h} & = \{ i \in \mathcal{N} : \text{ there exists } j \in \omega_{i} \text{ such that } \xi(x_{j}) > 0 \}, \\
 \mathcal{D}_{h} & = \{ i \in \mathcal{N} : \text{ there exists } j \in \omega_{i} \text{ such that } \delta(x_{j}) > 0 \}.
\end{align*}
Let 
\begin{align*}
 \mathcal{S}^{h} := \{v_{h} \in C^0([-1,1]) \colon v_{h} \in P^1([x_{i}, x_{i+1}]), i =1, \dots, N-1\}
\end{align*}
be the discrete finite-element space.  For $\eta \in C^{0}([-1,1])$ we define the interpolation operator $\Pi^{h} : C^0([-1,1]) \to \mathcal{S}^{h}$ to be
\begin{align*}
 \Pi^{h}(\eta) := \sum_{i=1}^{N} \eta(x_{i})\chi_{i},
\end{align*}
where $\chi_{j}(x)$ denote the standard basis function such that $\chi_{j} \in C^{0}([-1,1])$ and $\chi_{j}$ is a linear polynomial on each interval $[x_{i}, x_{i+1}]$ satisfying $\chi_{j}(x_{i}) = \delta_{ji}$ for all $i, j = 1, \dots, N$. 
Using the method of \cite{article:ElliottStinnerStylesWelford11}, we can find the finite-element function $c^{\Gamma,n+1}_{h}(x) = c^{\Gamma}_{h}(x, t_{n+1}) \in \mathcal{S}^{h}$ such that $c^{\Gamma,n+1}_{h}(x_{j}) = 0$ if $j \notin \mathcal{D}_{h}$ and satisfying
\begin{align*}
&\frac{K}{\Delta t}  \left ( \int_{-1}^{1} \Pi^{h}(\delta c^{\Gamma,n+1}_{h}\chi_{j}) - \Pi^{h}( \delta c^{\Gamma,n}_{h}\chi_{j}) \right )  + \int_{-1}^{1} \frac{K}{\text{Pe}_{\Gamma}}\Pi^{h}(\delta) \pd_{x} c^{\Gamma,n+1}_{h} \pd_{x} \chi_{j}\\
& \quad = - \int_{-1}^{1} \Pi^{h}(\beta \delta (\gamma'(c^{\Gamma,n}_{h}) - G'(c^{n}_{h})) \chi_{j}), \quad \forall j \in \mathcal{D}_{h}.
\end{align*}
The method for $c^{n+1}_{h}(x) = c_{h}(x, t_{n+1}) \in \mathcal{S}^{h}$ is analogous, whereby $c^{n+1}_{h}(x_{j}) = 0$ if $j \notin \mathcal{X}_{h}$ and satisfies
\begin{align*}
& \frac{1}{\Delta t} \left ( \int_{-1}^{1} \Pi^{h}(\xi c^{n+1}_{h}\chi_{j}) - \Pi^{h}(\xi c^{n}_{h}\chi_{j}) \right ) + \int_{-1}^{1} \frac{1}{\text{Pe}_{c}}\Pi^{h}(\xi) \pd_{x} c^{n+1}_{h} \pd_{x} \chi_{j} \\
& \quad = \int_{-1}^{1} \Pi^{h}(\beta \delta (\gamma'(c^{\Gamma,n}_{h}) - G'(c^{n}_{h})) \chi_{j}), \quad \forall j \in \mathcal{X}_{h}.
\end{align*}
For Model B, we seek $c^{n+1}_{h} \in \mathcal{S}^{h}$ such that $c^{n+1}_{h}(x_{j}) = 0$ if $j \notin \mathcal{X}_{h} \cup \mathcal{D}_{h}$ and satisfying
\begin{multline*}
 \frac{1}{\Delta t} \left ( \int_{-1}^{1} \Pi^{h}((\xi c^{n+1}_{h} + K \delta g(c^{n+1}_{h}))\chi_{j}) \right. - \left. \int_{-1}^{1} \Pi^{h}((\xi c^{n}_{h}+ K \delta g(c^{n}_{h}))\chi_{j}) \right ) \\
 + \int_{-1}^{1} \Pi^{h} \left ( \frac{\xi}{\text{Pe}_{c}} + \frac{K \delta}{\text{Pe}_{\Gamma}} \right ) \pd_{x} c^{n+1}_{h} \pd_{x} \chi_{j} = 0, \quad \forall j \in \mathcal{X}_{h} \cup \mathcal{D}_{h}. 
\end{multline*}
We remark that the scheme for Model C in this setting is structurally similar to the scheme of Model B.  Hence in the subsequent one-dimensional experiments we will only implement the schemes for Models A and B, while Model C will be the subject of investigation in the two-dimensional experiments due to its two-sided nature.

\subsubsection{Numerics for Model A}

\begin{figure}[p]
\centering
\subfigure[]{\includegraphics[width=0.48\textwidth]{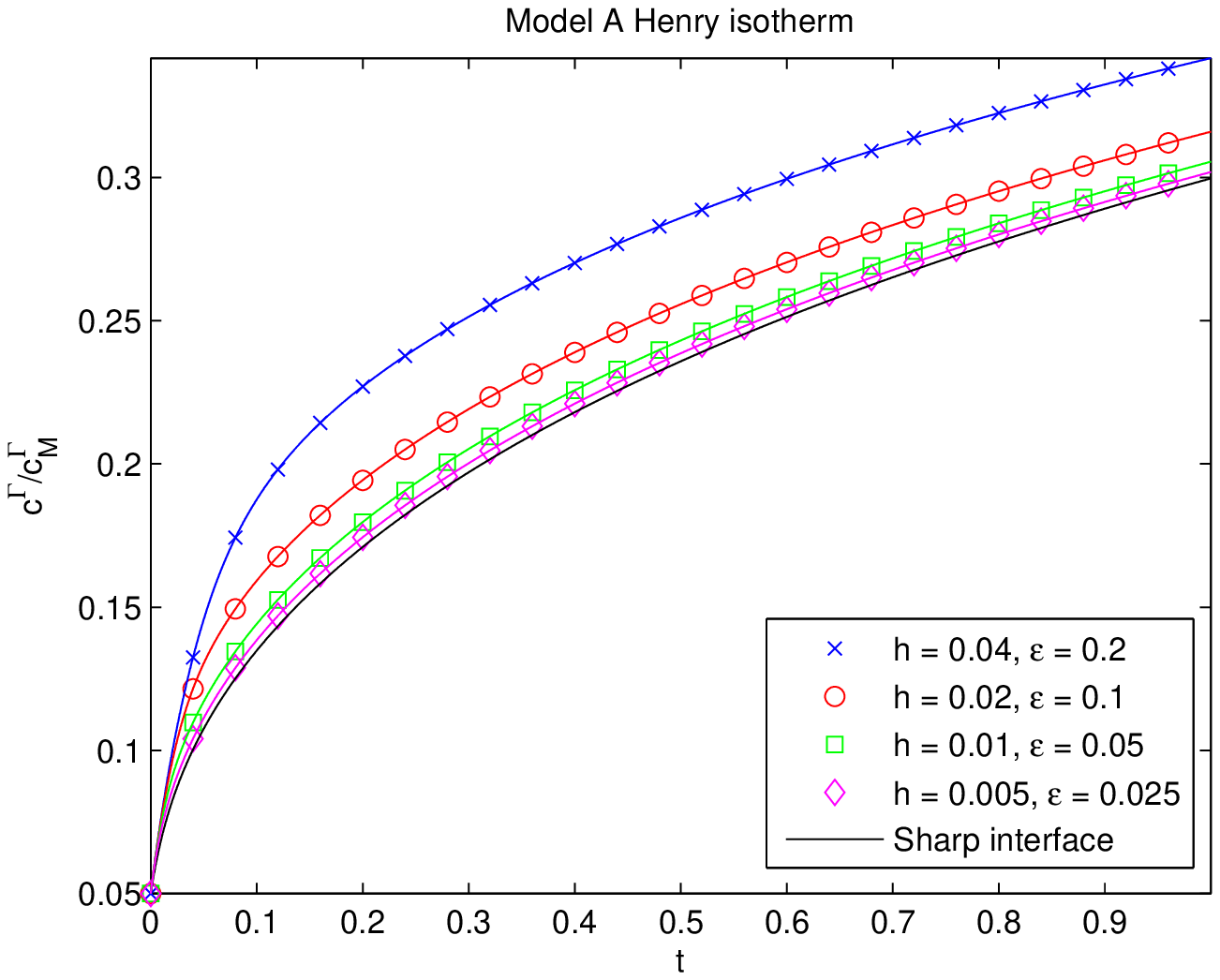}}  
\subfigure[]{\includegraphics[width=0.48\textwidth]{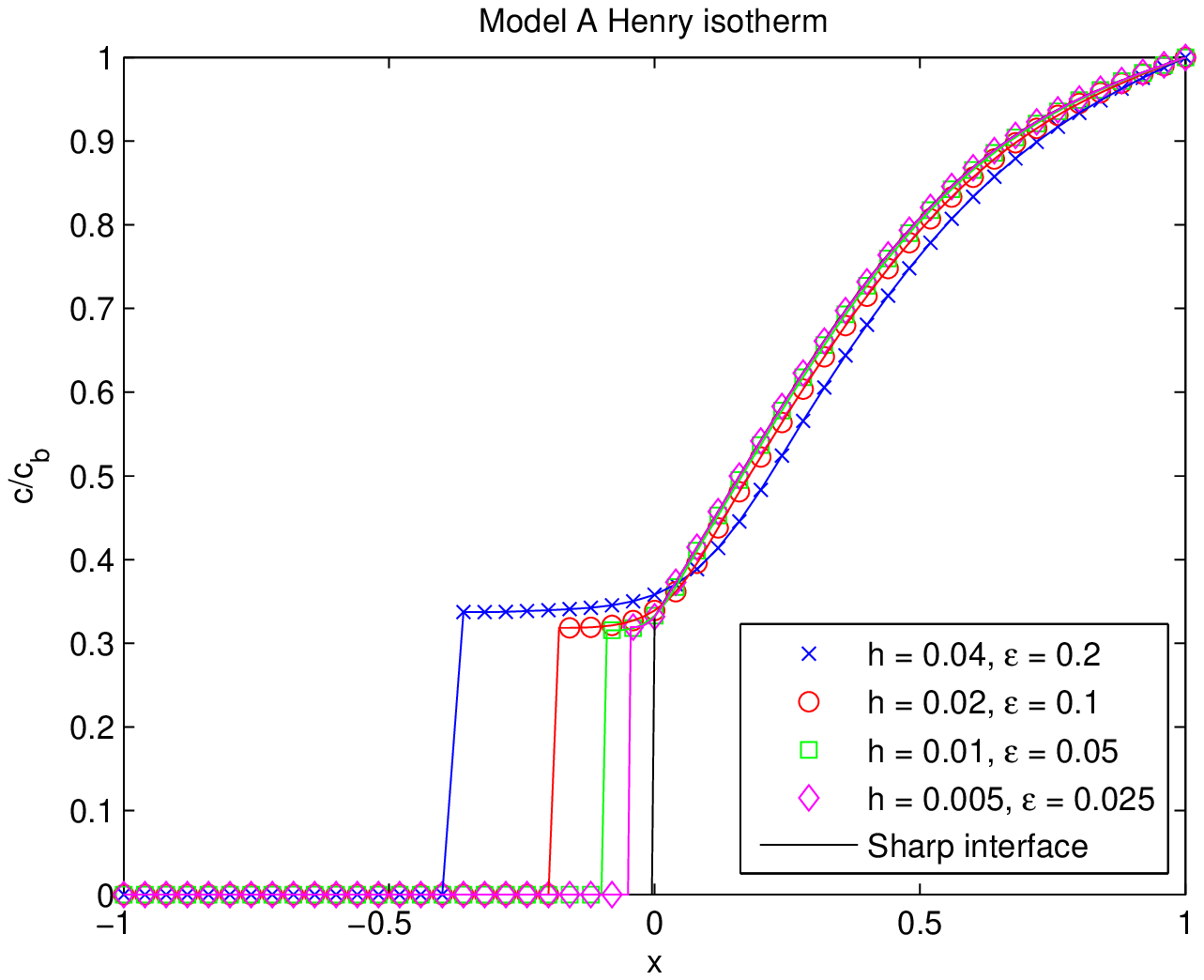}} \\
\subfigure[]{\includegraphics[width=0.48\textwidth]{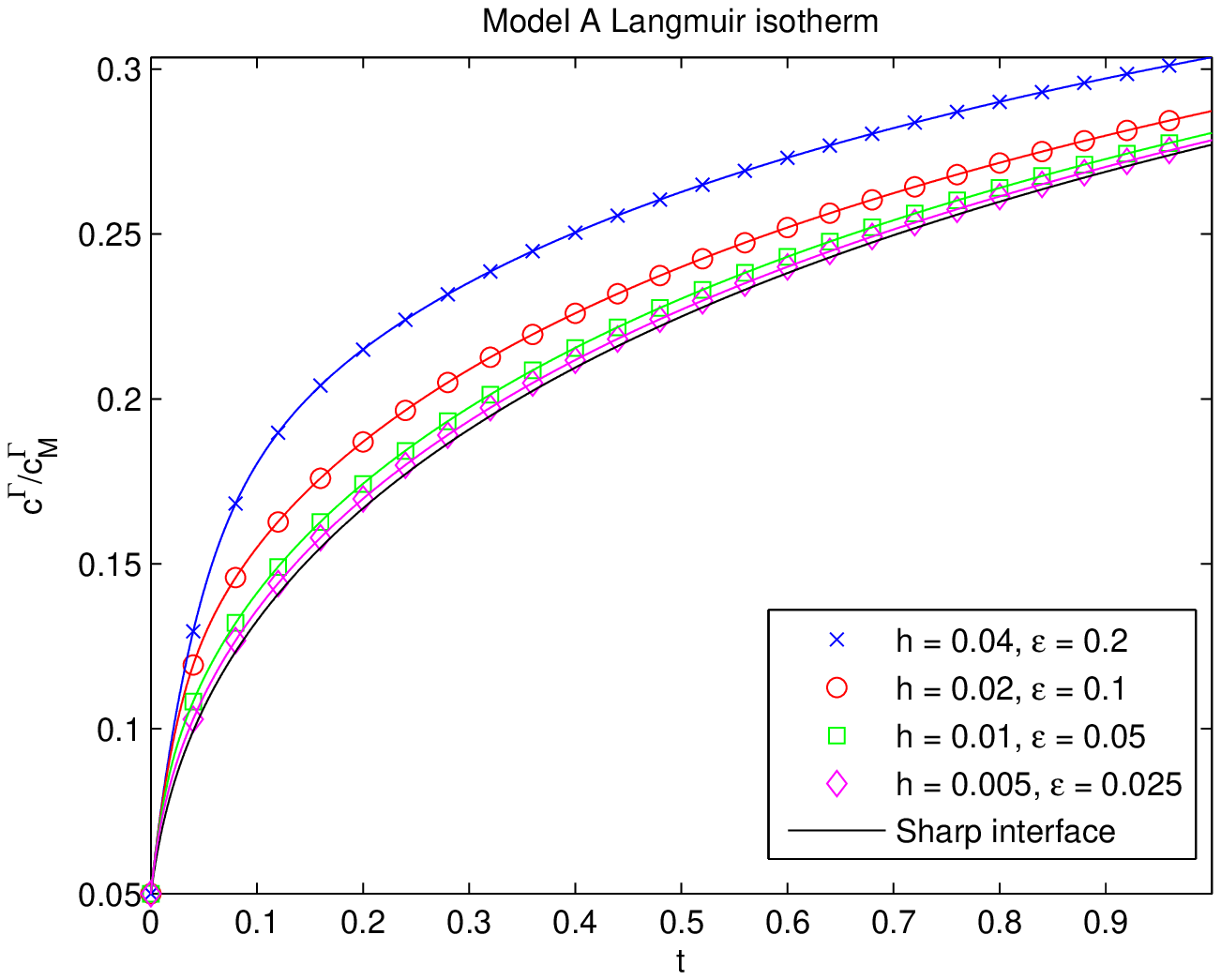}}  
\subfigure[]{\includegraphics[width=0.48\textwidth]{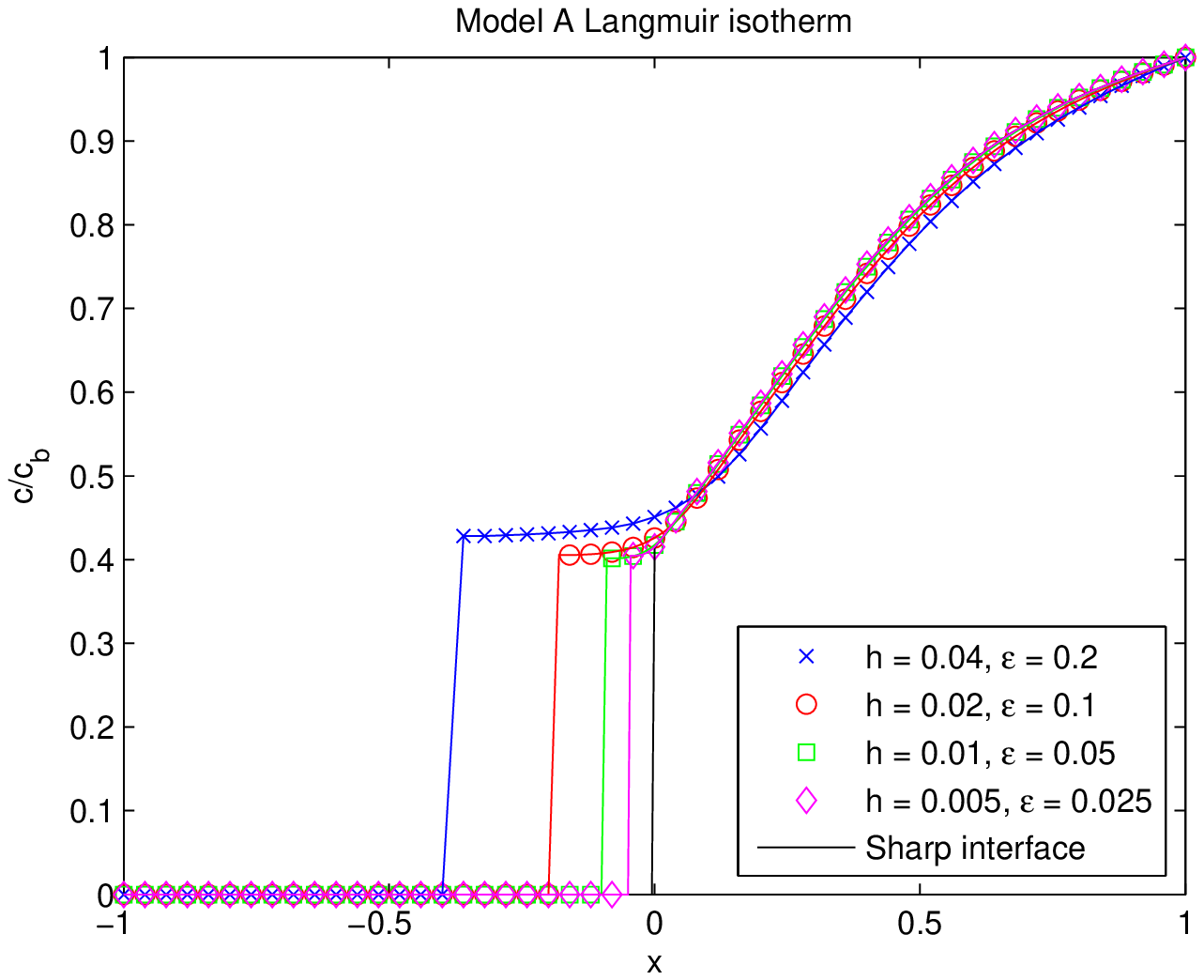}} \\
\caption{Model A $\eps$-convergence for (a) the profile of $c^{\Gamma}(x=0,t)$ and (b) the profile of $c(x,t=1)$ with the Henry isotherm, (c) the profile of $c^{\Gamma}(x=0,t)$ and (d) the profile of $c(x,t=1)$ with the Langmuir isotherm.  The parameter $\alpha$ is chosen to be 1.}
\label{fig:ModelANonInstConvergence}
\end{figure}

\begin{figure}[p]
\centering
\subfigure[]{\includegraphics[width=0.48\textwidth]{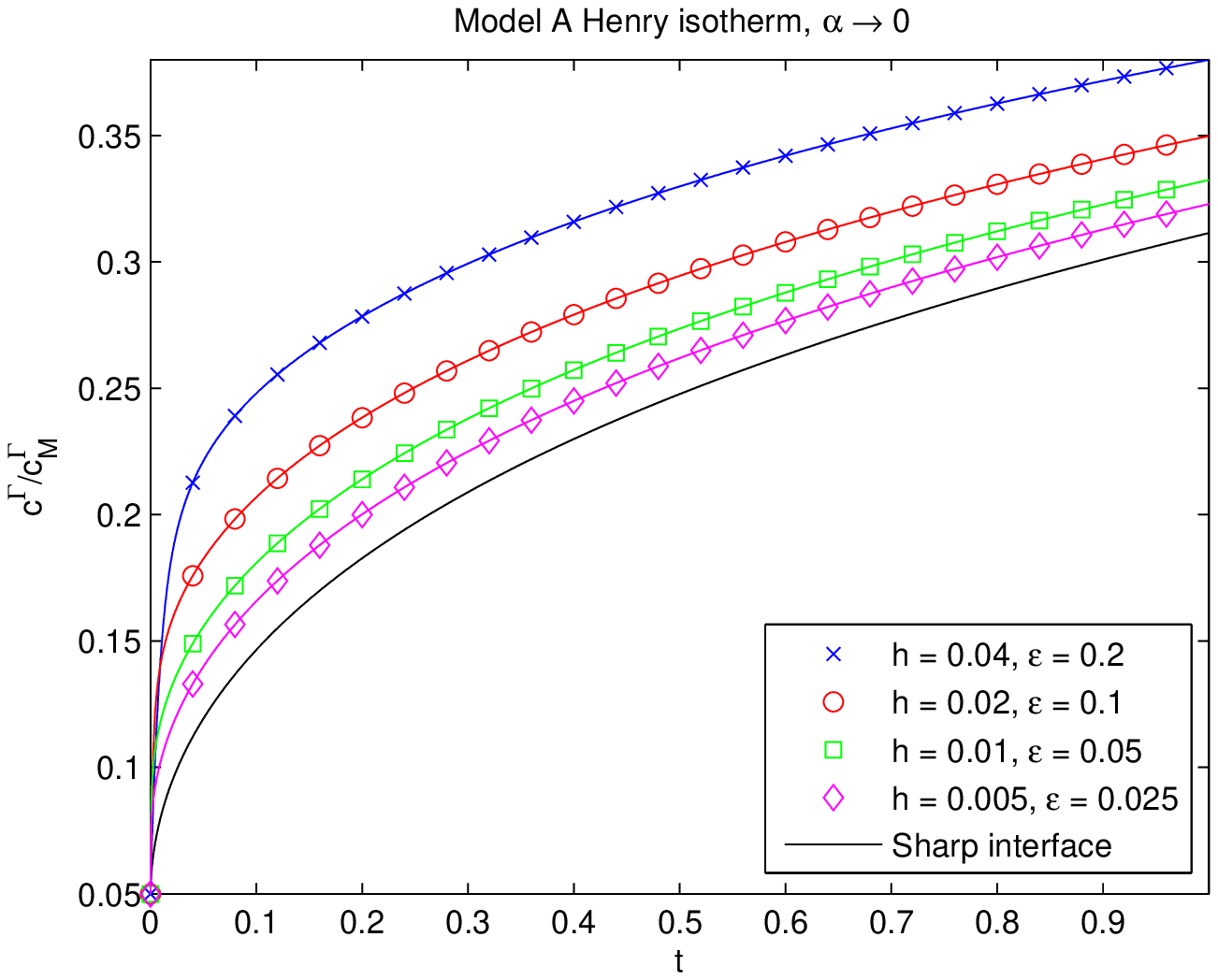}}  
\subfigure[]{\includegraphics[width=0.48\textwidth]{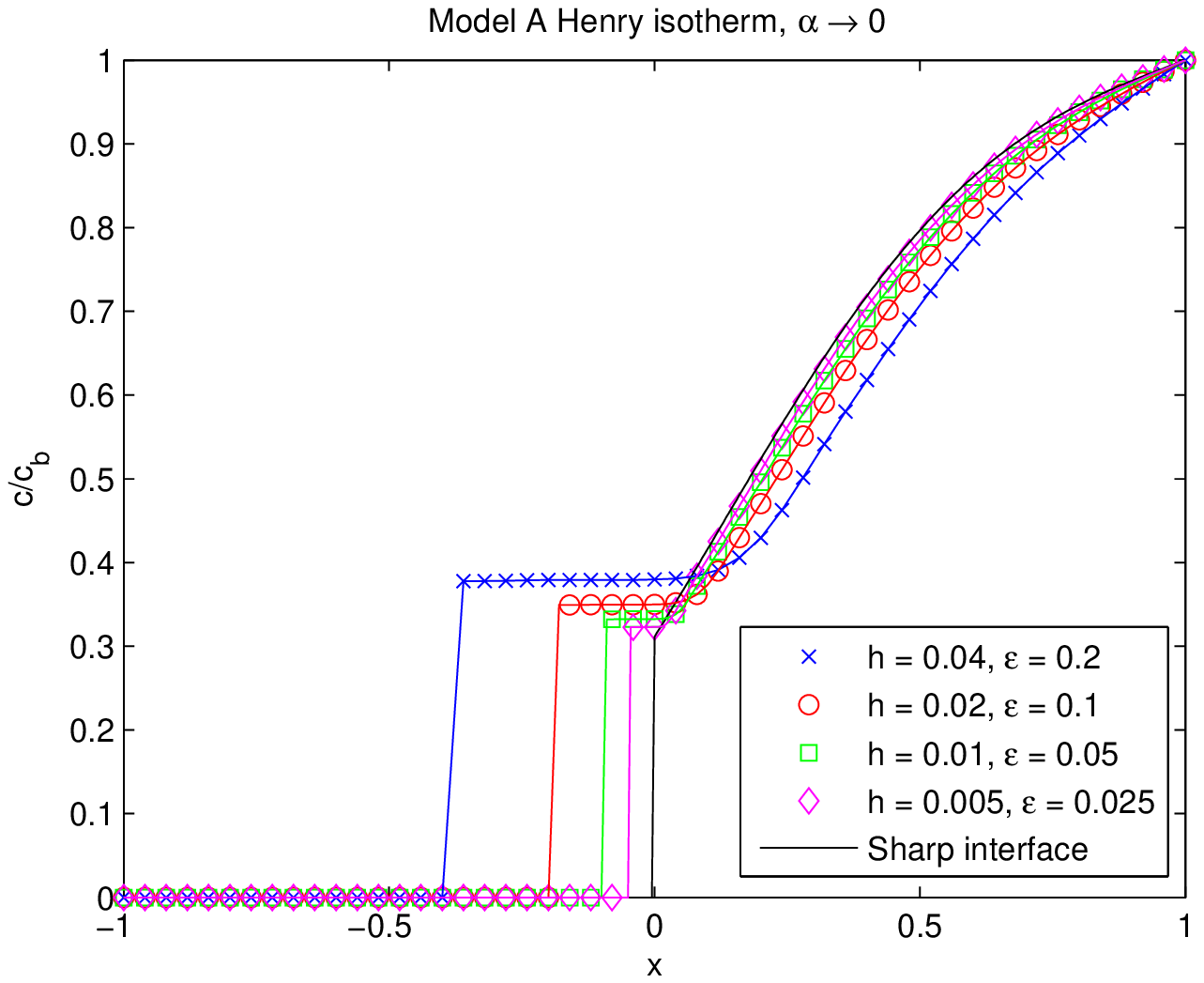}} \\
\subfigure[]{\includegraphics[width=0.48\textwidth]{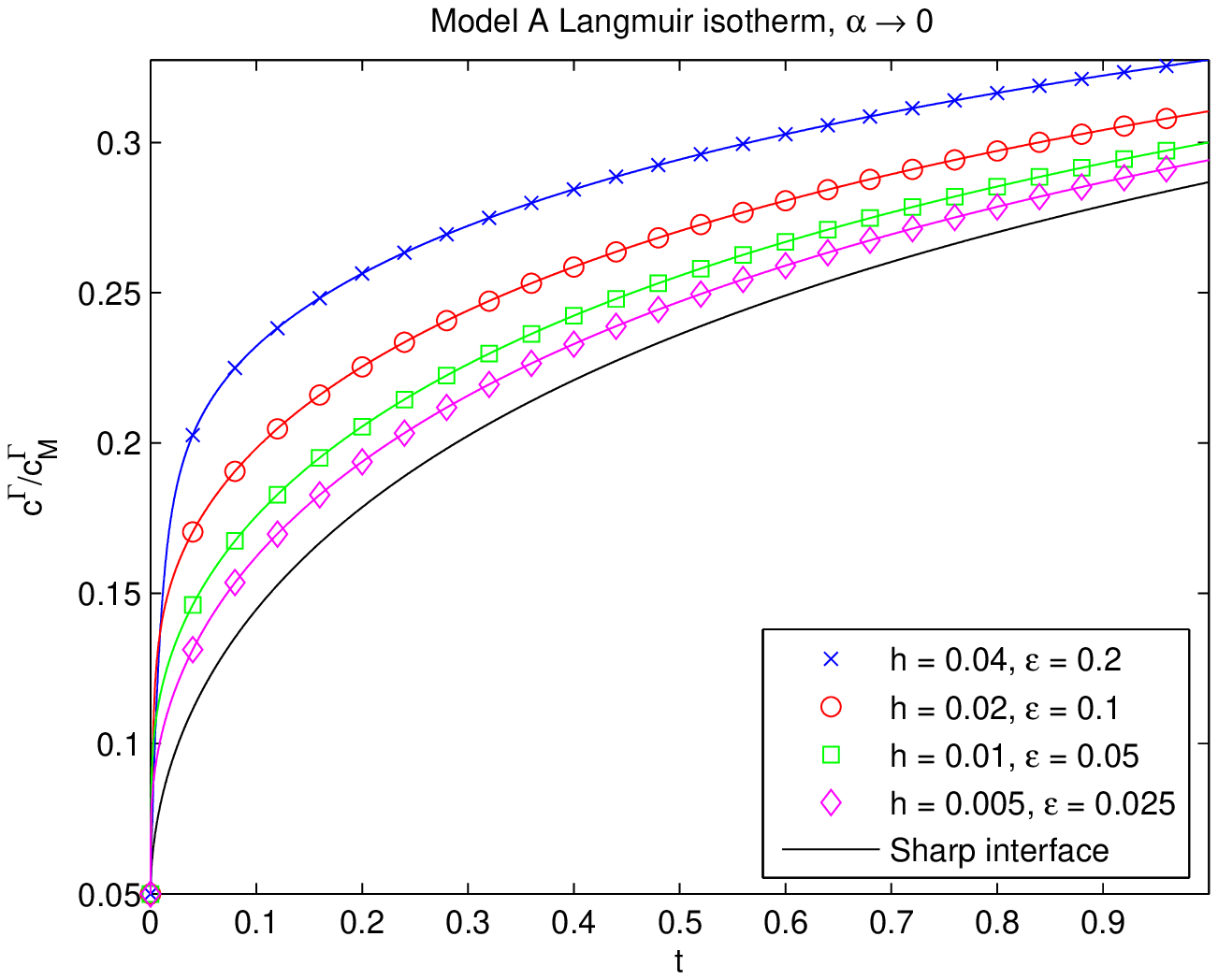}}  
\subfigure[]{\includegraphics[width=0.48\textwidth]{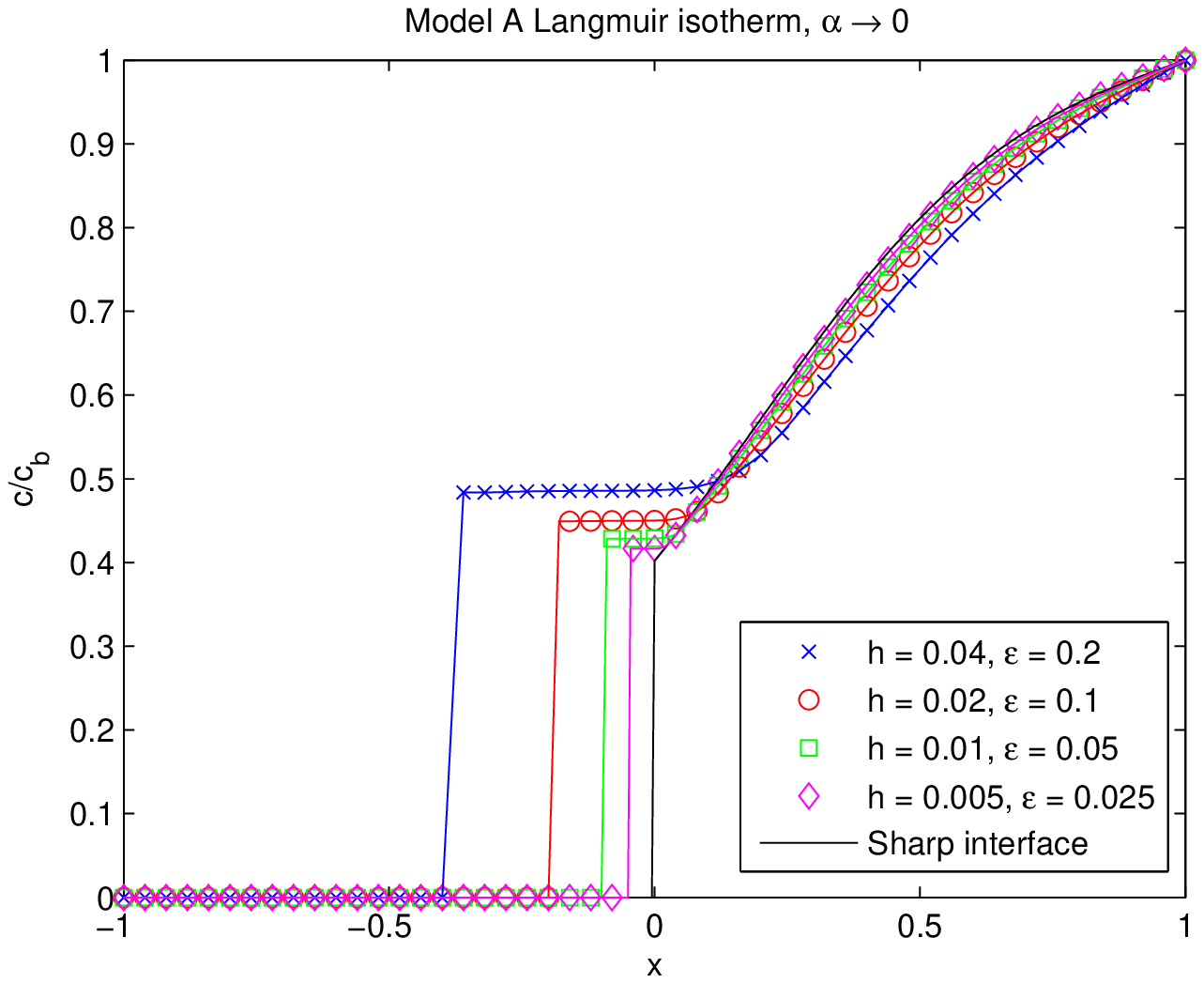}} \\
\caption{Model A, $\eps$-convergence for (a) the profile of $c^{\Gamma}(x=0,t)$ and (b) the profile of $c(x,t=1)$ with the Henry isotherm, (c) the profile of $c^{\Gamma}(x=0,t)$ and (d) the profile of $c(x,t=1)$ with the Langmuir isotherm.  The parameter $\alpha$ is chosen to be $\eps$.}
\label{fig:ModelAInstConvergence}
\end{figure}

We observed the following regarding the choice of model parameters:
\begin{itemize}
\item \textit{Interfacial Peclet number $\text{Pe}_{\Gamma}$}: Fixing $\alpha = 0.2$ and $\text{Pe}_{c} = 1$, we explored the effects of varying $\text{Pe}_{\Gamma}$.  For $\text{Pe}_{\Gamma} = 1$ we observed that the profile for $c^{\Gamma}$ across the interfacial layer is linear when $\eps = 0.2$ or $0.1$, but decreasing $\eps$ to $0.05$ or $0.025$ give a more uniform profile across the interface.  Moreover, we can achieve a constant profile for larger values of $\eps$, i.e. $\eps = 0.2$ or $0.1$, by decreasing $\text{Pe}_{\Gamma}$ to $0.01$.

\item \textit{Bulk Peclet number $\text{Pe}_{c}$}: Fixing $\text{Pe}_{\Gamma} = 0.01$, $\alpha = 0.2$, we observe that the profile of $c$ across the interface is linear for $\text{Pe}_{c} = 0.1$ when $\eps = 0.2, 0.1, 0.05, 0.025$.  When $\text{Pe}_{c}$ is increased to $10$, we observe a constant profile in $(-\eps \frac{\pi}{2},0)$ and a linear profile in $(0,\eps \frac{\pi}{2})$.  The size of these regions seems to be invariant for fixed $\text{Pe}_{c}$ as we reduced $\eps$ from $0.2$ to $0.025$.
\end{itemize}

These initial experiments with model parameters motivate the following choice for the convergence tests:  We choose $\alpha = 1$, $\beta = \tfrac{2}{\pi}$, $\text{Pe}_{\Gamma} = 0.01$ and $\text{Pe}_{c} = 10$.  The other parameters of the model are $c^{\Gamma}_{M} = 1, c(x,0) = 1, c^{\Gamma}(x,0) = 0.05$.  The mesh size $h$ is taken from $\{0.08, 0.04, 0.02, 0.01, 0.005\}$ and the corresponding value of $\eps$ is chosen from $\{0.4, 0.2, 0.1, 0.05, 0.025\}$.  To ensure that the numerical scheme is stable, for each test we choose a time step $\Delta t \leq h^2$.  

In the case of fixed $\alpha > 0$ we refer to Table $\ref{tbl:NonInstModelA}$ for the $\eps$-convergence in the difference in $c^{\Gamma}(0,1)$ and $c(0,1)$ between the phase field model and the sharp interface model and Figure $\ref{fig:ModelANonInstConvergence}$ for the $\eps$-convergence of the profiles.

\begin{table}[ht]
 \centering
\begin{tabular}{|c|c|c|c|}
\hline
$h$ & $\eps$ & \gape{$\abs{c^{\Gamma}_{PF}(0,1) - c^{\Gamma}_{SI}(1)}$} & $\abs{c_{PF}(0,1) - c_{SI}(0,1)}$ \\
\hline
0.08 & 0.4 & 0.0974417 & 0.0732749 \\
0.04 & 0.2 & 0.0419969  & 0.0265120  \\
0.02 & 0.1 & 0.0163026  & 0.0076752  \\
0.01 & 0.05& 0.0058420  & 0.0015298  \\
0.005 & 0.025 & 0.0022358  & 0.0002207  \\
\hline
\end{tabular}
\begin{tabular}{|c|c|c|c|}
\hline
$h$ & $\eps$ & \gape{$\abs{c^{\Gamma}_{PF}(0,1) - c^{\Gamma}_{SI}(1)}$} & $\abs{c_{PF}(0,1) - c_{SI}(0,1)}$ \\
\hline
0.08 & 0.4 & 0.0596860 & 0.0963854  \\
0.04 & 0.2 & 0.0265857 & 0.0364079 \\
0.02 & 0.1 & 0.0102234 & 0.0115916 \\
0.01 & 0.05& 0.0035830 & 0.0030918  \\
0.005 & 0.025 & 0.0013697 & 0.0009629 \\
%0.0025 & 0.0125 & 0.0007884 & 0.0007920 \\
\hline
\end{tabular}
\caption{Convergence table for Model A, non-instantaneous adsorption ($\alpha  = 1$), Henry isotherm (top) and Langmuir isotherm (bottom).}
\label{tbl:NonInstModelA}
\end{table}

We also considered the scaling $\alpha = \eps$ (or $\beta = \eps^{-1}$) and from Figure $\ref{fig:ModelAInstConvergence}$  and Tables \ref{tbl:ModelAInstHen} and \ref{tbl:ModelAInstLangmuir} we observed the $\eps$-convergence in the difference in $c^{\Gamma}(0,1)$ and $c(0,1)$ between the phase field model and the sharp interface model. Furthermore, we note that the maximum and mean difference of $\abs{\gamma'(c^{\Gamma}) - G'(c)}$ in the interfacial layer decreases linearly as $\eps \to 0$.
\begin{table}[p]
 \centering
\begin{tabular}{|c|c|c|c|}
\hline
$h$ & $\eps$ & \gape{$\abs{c^{\Gamma}_{PF}(0,1) - c^{\Gamma}_{SI}(1)}$} & $\abs{c_{PF}(0,1) - c_{SI}(0,1)}$ \\
\hline
0.08 & 0.4 & 0.1191555 & 0.1175129 \\
0.04 & 0.2 & 0.0685148 & 0.0682569  \\
0.02 & 0.1 & 0.0383807 & 0.0384228  \\
0.01 & 0.05& 0.0209969 & 0.0210621  \\
0.005 & 0.025 & 0.0114668 & 0.0115106 \\
\hline
\hline
$h$ & $\eps$ & $\max \abs{\gamma' - G'}$ & $\text{ave} \abs{\gamma' - G'}$ \\
\hline
0.08 & 0.4 & 0.5882511 & 0.1085532  \\
0.04 & 0.2 & 0.3540145 & 0.0572062 \\
0.02 & 0.1 & 0.2061245 & 0.0316161  \\
0.01 & 0.05& 0.1128733 & 0.0168467  \\
0.005 & 0.025 & 0.0594562  & 0.0087458 \\
\hline
\end{tabular}
\caption{Convergence table for Model A, instantaneous adsorption ($\alpha = \eps$), Henry isotherm.}
\label{tbl:ModelAInstHen}
\end{table}

\begin{table}[p]
 \centering
\begin{tabular}{|c|c|c|c|}
\hline
$h$ & $\eps$ & \gape{$\abs{c^{\Gamma}_{PF}(0,1) - c^{\Gamma}_{SI}(1)}$} & $\abs{c_{PF}(0,1) - c_{SI}(0,1)}$  \\
\hline
0.08 & 0.4 & 0.0687143 & 0.1452171 \\
0.04 & 0.2 & 0.0420765 & 0.1452171 \\
0.02 & 0.1 & 0.0249919 & 0.0506682 \\
0.01 & 0.05& 0.0146093  & 0.0292756  \\
0.005 & 0.025 & 0.0087232 & 0.0173523 \\
\hline
\hline
$h$ & $\eps$ & $\max \abs{\gamma' - G'}$  & $\text{ave} \abs{\gamma' - G'}$ \\
\hline
0.08 & 0.4 & 0.4014189  & 0.0759004  \\
0.04 & 0.2 & 0.2347884  & 0.0389953  \\
0.02 & 0.1 & 0.1326851  & 0.0210856   \\
0.01 & 0.05& 0.0711437  & 0.0110897  \\
0.005 & 0.025 & 0.0370265  & 0.0057192  \\
\hline
\end{tabular}
\caption{Convergence table for Model A, instantaneous adsorption ($\alpha = \eps$), Langmuir isotherm.}
\label{tbl:ModelAInstLangmuir}
\end{table}

\subsubsection{Numerics for Model B}

\begin{figure}[p]
\centering
\subfigure[]{\includegraphics[width=0.48\textwidth]{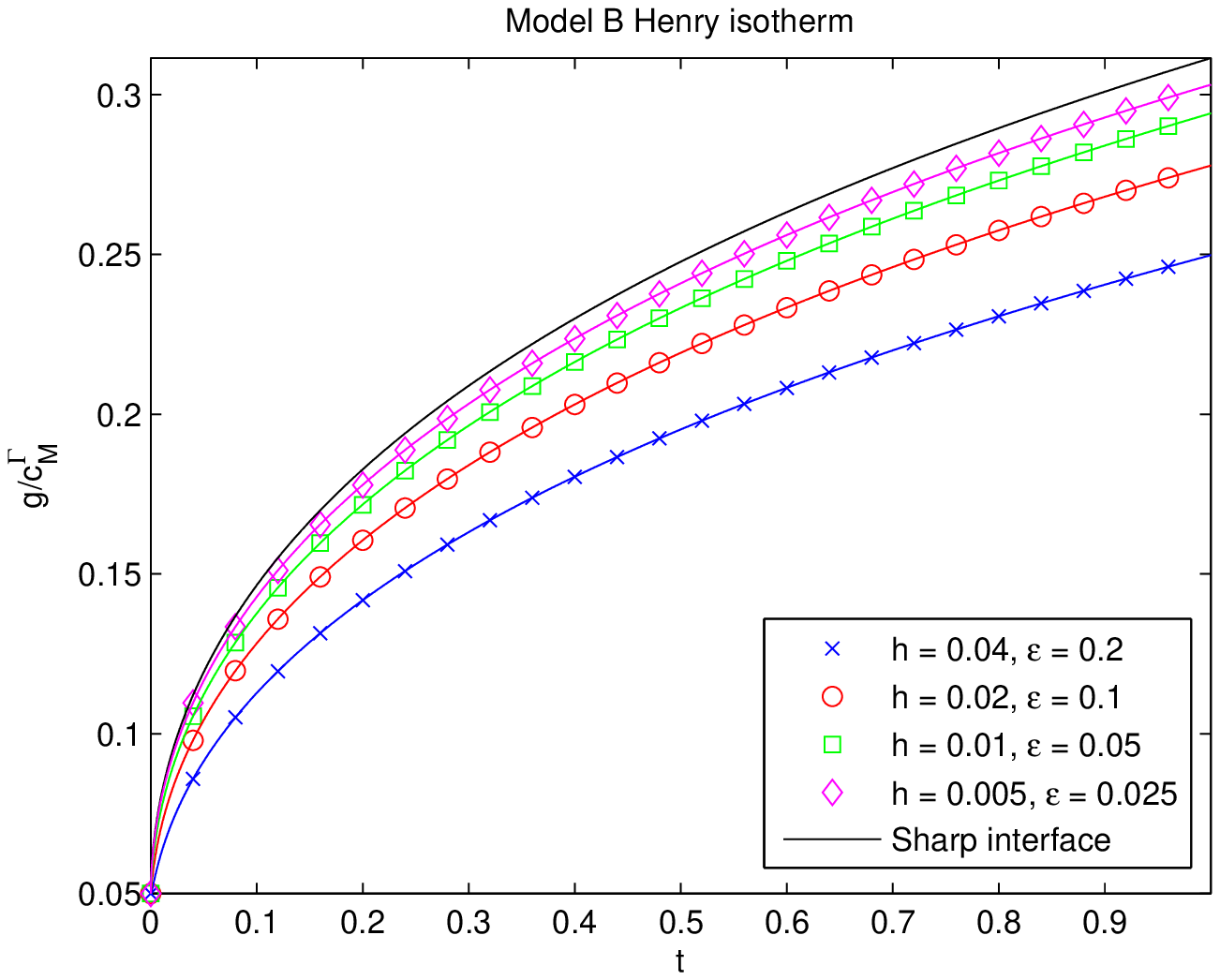}}  
\subfigure[]{\includegraphics[width=0.48\textwidth]{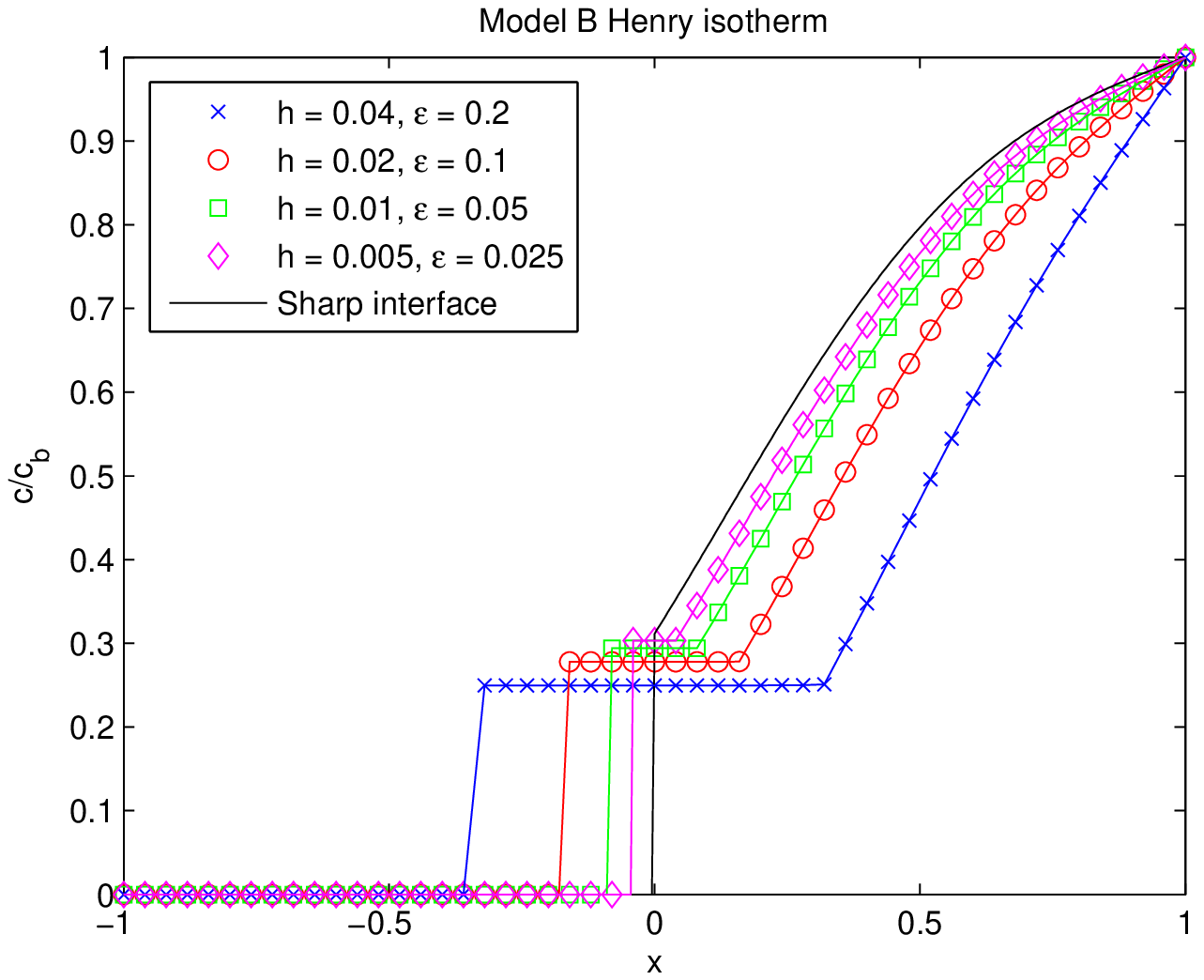}} \\
\subfigure[]{\includegraphics[width=0.48\textwidth]{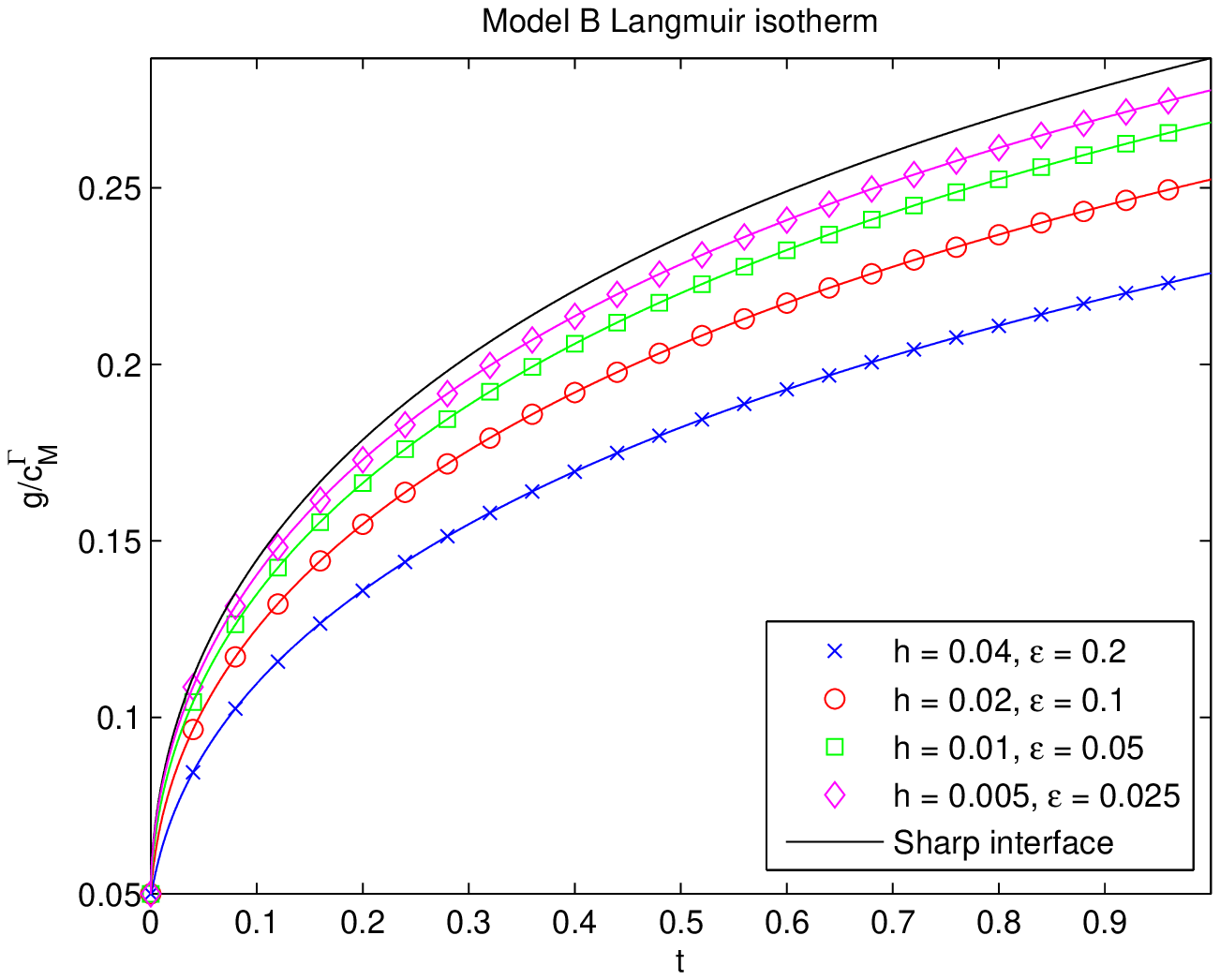}}  
\subfigure[]{\includegraphics[width=0.48\textwidth]{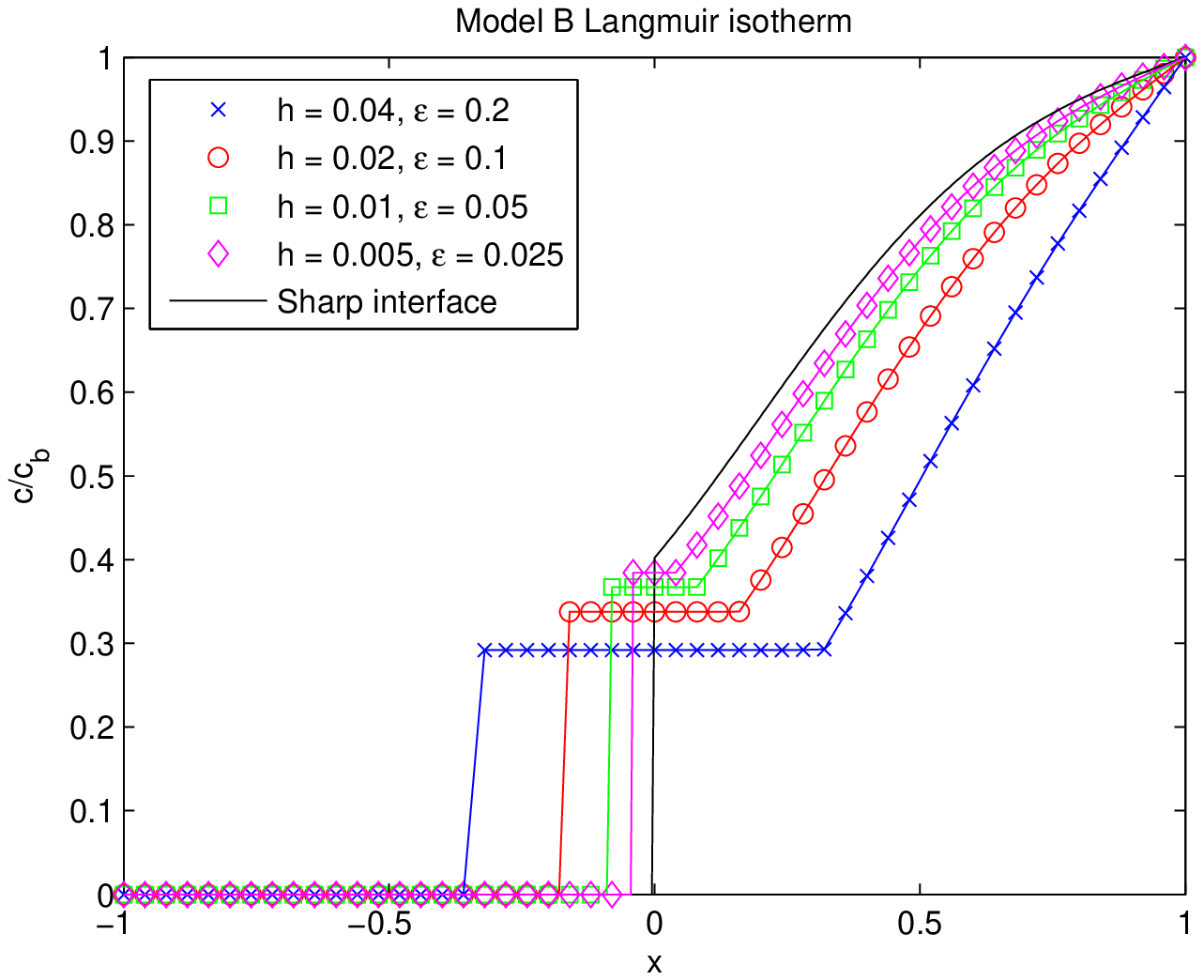}} \\
\caption{Model B $\eps$-convergence for (a) the profile of $g(x=0,t)$ and (b) the profile of $c(x,t=1)$ with the Henry isotherm, (c) the profile of $g(x=0,t)$ and (d) the profile of $c(x,t=1)$ with the Langmuir isotherm.}
\label{fig:ModelBConvergence}
\end{figure}

For Model B, since we have instantaneous adsorption, we can infer the difference of $\abs{c_{PF}(0,1) - c_{SI}(0,1)}$ from $\abs{c^{\Gamma}_{PF}(0,1) - c^{\Gamma}_{SI}(1)}$ via the adsorption isotherms.  Hence Table \ref{tbl:ModelBInst} displays only the difference $\abs{c^{\Gamma}_{PF}(0,1) - c^{\Gamma}_{SI}(1)}$ for the Henry and Langmuir isotherms, in which we observe $\eps$-convergence along with Figure $\ref{fig:ModelBConvergence}$.  The model parameters are chosen to be the same as in Model A.

\begin{table}[p]
 \centering
\begin{tabular}{|c|c|c|c|}
\hline
$h$ & $\eps$ & Henry  & Langmuir  \\
\hline
0.08 & 0.4 & 0.0938706 &  0.0895642 \\
0.04 & 0.2 & 0.0616441 &  0.0593439  \\
0.02 & 0.1 & 0.0336103 &  0.0330060  \\
0.01 & 0.05& 0.0172770 &  0.0168309  \\
0.005 & 0.025 & 0.0083055 & 0.0076996 \\
\hline
\end{tabular}
\caption{Convergence table for Model B}
\label{tbl:ModelBInst}
\end{table}

\subsection{2D Simulations}

In this section we present some results of numerical simulations in two spatial dimensions in order to qualitatively illustrate the effectivity of our approach. In a first setting we expose a droplet of a fluid suspended in another fluid to a shear flow. Under moderate shear rates the droplet's shape attains a steady state. This shape changes in the presence of the surfactant. Of particular interest to us is the dependence of the shape on the isotherm. In a second setting we start with a droplet at rest (in particular, in equilibrium with respect to the surfactant). Then we supply surfactant on one of the sides of the simulation box and investigate how far the droplet is sucked towards this side due to the Marangoni effect. As we are mainly interested in the effect of the surfactant on a qualitative basis we make convenient assumptions with respect to the two-phase flow, namely, that the fluids have the same mass densities and viscosities and that a Dirichlet boundary condition holds for the velocity. Also, the surfactant related parameters and data do not correspond to any specific species or systems.

Both dynamic adsorption (Model A) and instantaneous adsorption (Model C) have been considered. In both cases, the Navier-Stokes-Cahn-Hilliard system was solved following the lines of \mathcite{article:KayStyWel08} but we employed the double-obstacle potential for $W(\varphi)$. The saddle point problem arising from \mathref{phaseNonD:mass} and \mathref{phaseNonD:AlternateMomentum} has been solved with a preconditioned GMRES \mathcite{article:SilElmKayWat01}. For the phase field equation \mathref{phaseNonD:phase} together with \mathref{phaseNonD:chem} in form of a variational inequality we have employed a Gauss-Seidel type iteration as described in \mathcite{article:BarNurSty04}. 

For Model A, we always considered Fickian diffusion by setting $M^{(i)}_{c,*}(c_{*}) = 1 / (G''_{*}(c_{*}) \text{Pe}_{c,i})$ and $M_{\Gamma,*}(c^{\Gamma}_{*}) = 1 / (\gamma''_{*}(c^{\Gamma}_{*}) \text{Pe}_{\Gamma,i})$. We also replaced $\delta_{*}(\varphi, \nabla_{*} \varphi)$ by $2 W(\varphi) / \eps_{*}$ in the surfactant equation \mathref{phaseNonD:interface} which effects the validity of the energy inequality but doesn't change the result of the asymptotic analysis. The reason is that the method developed in \mathcite{article:ElliottStinnerStylesWelford11} can directly be applied. We leave a careful study of the impact of the gradient term for future investigations. In analogy to \mathcite{article:ElliottStinnerStylesWelford11} a method for the degenerate bulk surfactant equations \mathref{phaseNonD:bulk} has been developed. The methods have been implemented using the software ALBERTA, Version 2.0.1, \mathcite{book:SchSie2005}.

In the surfactant equation \mathref{phaseNonDinst3:surfactant} for Model C we assumed constant mobilities, $M^{(i)}_{c,*}(c_{*}(q_{*})) = 1/ \text{Pe}_{c,i}$ and $M_{\Gamma,*}(c_{*}^{\Gamma}(q_{*})) = 1/ \text{Pe}_{\Gamma}$, and we also replaced $\delta_{*}(\varphi, \nabla_{*} \varphi)$ by $2 W(\varphi) / \eps_{*}$ for not having to deal with $\nabla_{*} \varphi$ in the diffusion term. Whenever no closed formula for $c^{\Gamma}_{*}$, $c_{*}^{(1)}$, or $c_{*}^{(2)}$ as a function of $q_{*}$ was available we employed a Newton method. In the same way we also dealt with the nonlinear system of equations emerging from the finite element discretisation of the surfactant equation. 

With regards to parameters and functions appearing in non-dimensional equations of the phase field models we have in both settings: $K = 2 / \pi$, $\lambda_\rho = 1$, $\lambda_\eta = 1$, $\rrr{Ca} = 0.1$, 
\[
\xi_{1}(\varphi) = 
\begin{cases}
1, & \quad 1 \leq \varphi, \\
\frac{1}{2} (\varphi + 1), & \quad -1 < \varphi < 1, \\
0, & \quad \varphi \leq -1,
\end{cases}
\]
and $\xi_{2}(\varphi) = 1 - \xi_{1}(\varphi)$ where we set $\xi_{i}'(\varphi) = 0$ if $| \varphi | \geq 1$.

\subsubsection{Droplet in shear flow}

\begin{figure}
 \begin{center}
  \includegraphics[width=0.49\textwidth]{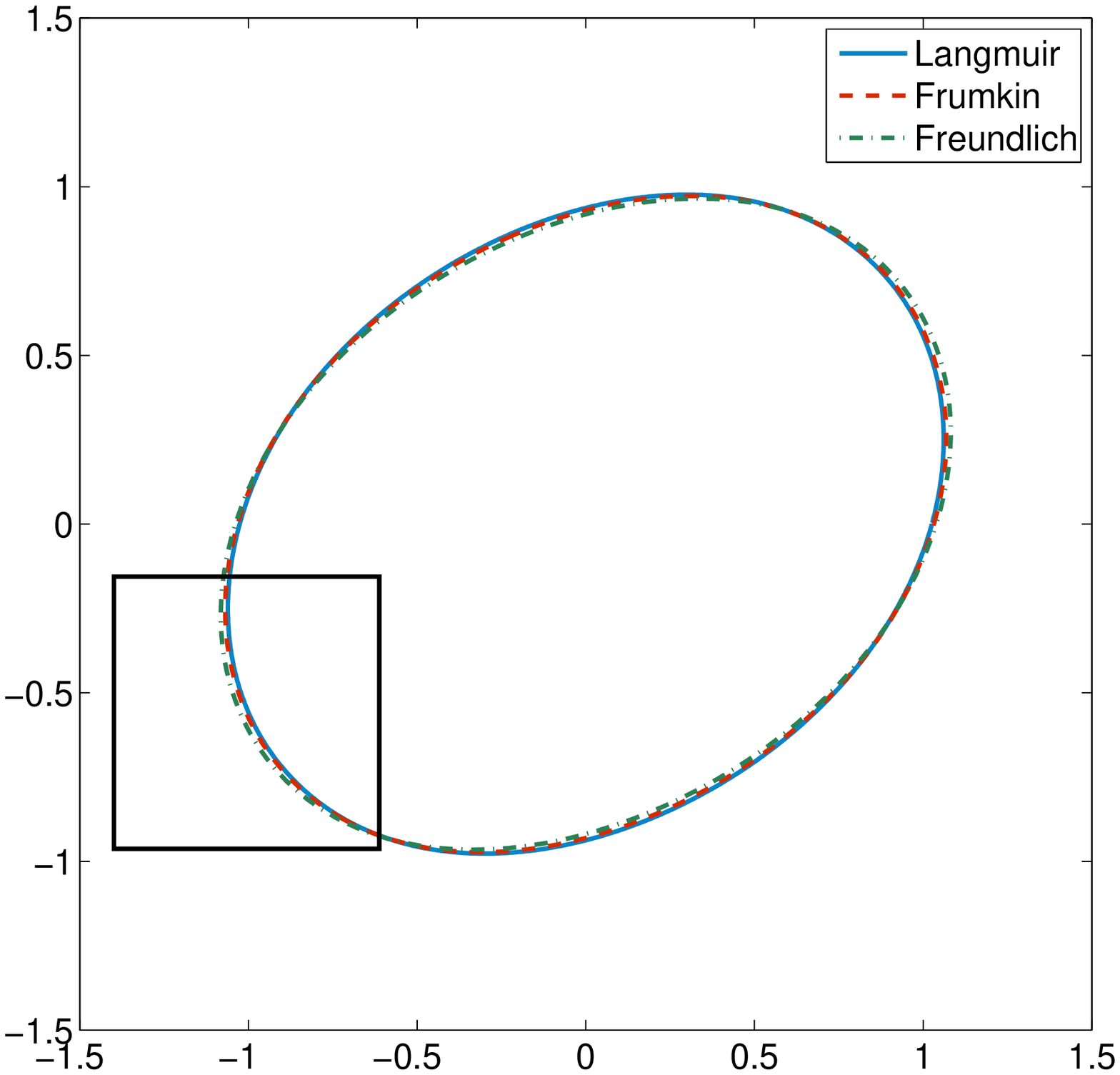} \hfill \includegraphics[width=0.49\textwidth]{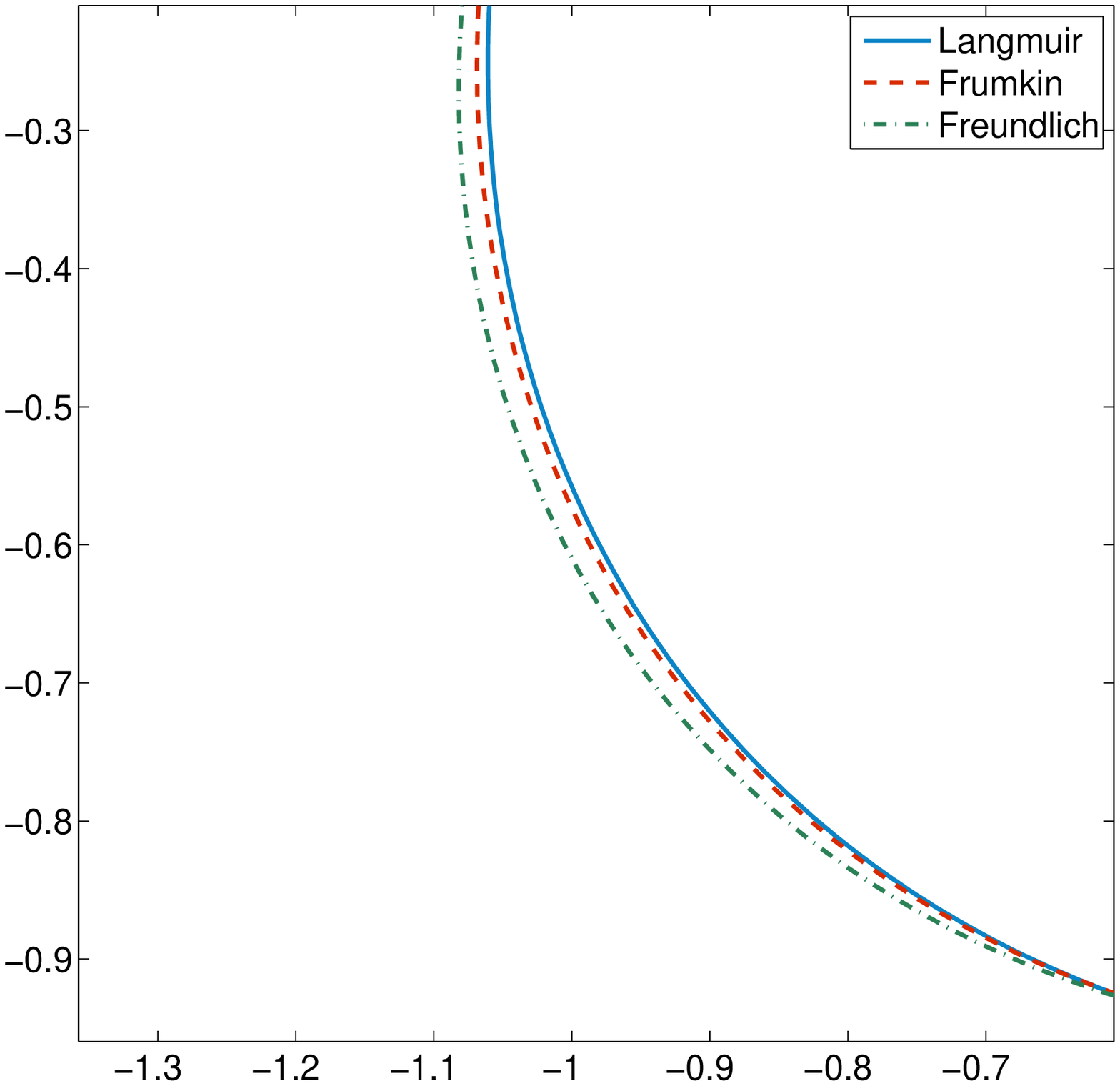}
 \end{center}
 \caption{Droplet in shear flow: Zero level sets of $\varphi$ for several isotherms, $\eps = 0.0565685425 \approx 0.08 / \sqrt{2}$, $t = 10$. The right graph displays a zoom into the square indicated on the left graph.}
 \label{fig:2D_caseA_shapes}
\end{figure}

\begin{figure}
 \begin{center}
  \includegraphics[width=0.49\textwidth]{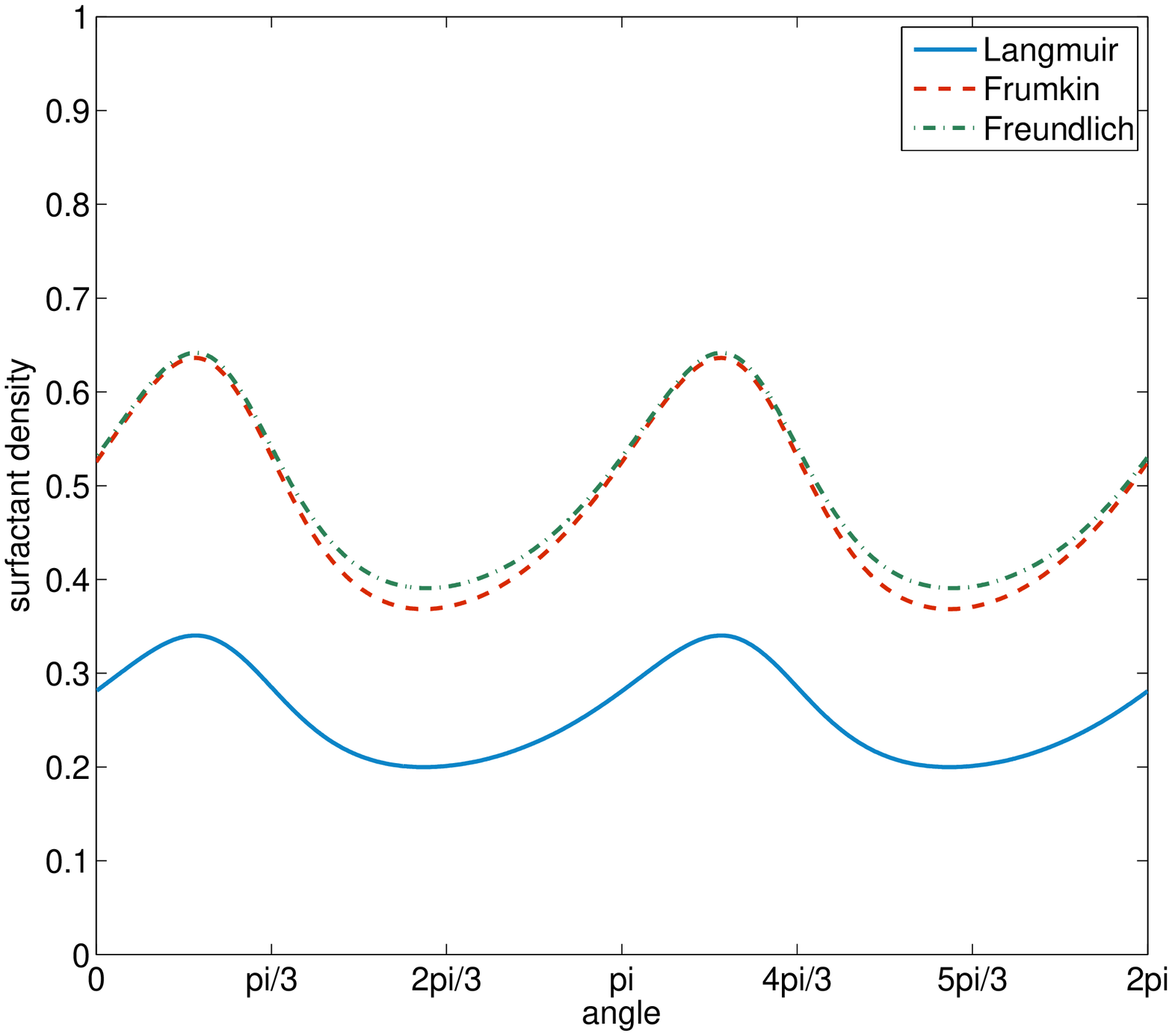} \hfill \includegraphics[width=0.49\textwidth]{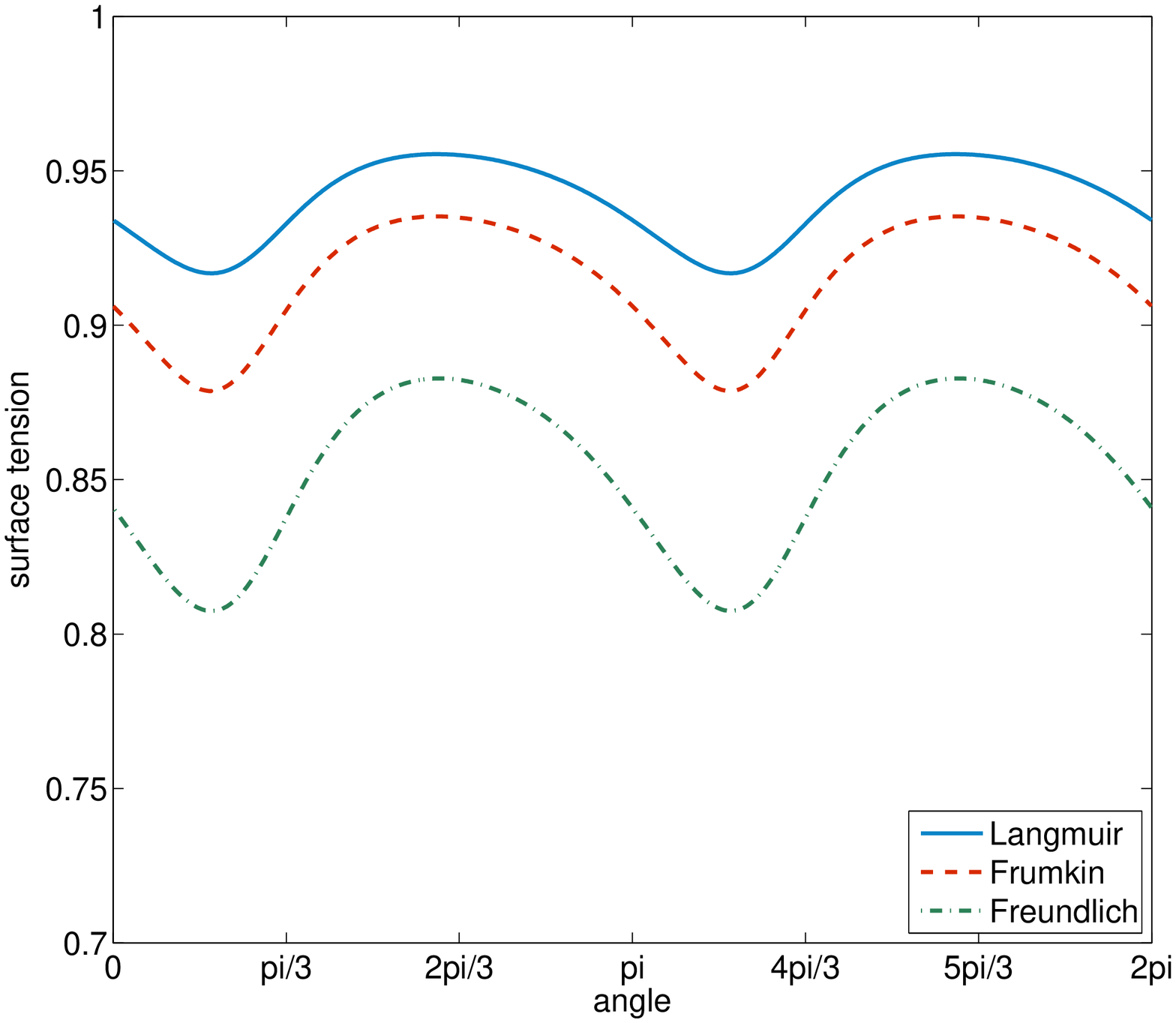}
 \end{center}
 \caption{Droplet in shear flow: Interface surfactant density $c^\Gamma_{*}$ (left) and surface tension $\sigma_{*}(c_{*}^{\Gamma})$ (right) plotted over the angle formed by the line from the centre to a boundary point and the x-axis for several isotherms, $\eps = 0.0565685425 \approx 0.08 / \sqrt{2}$, $t = 10$.}
 \label{fig:2D_caseA_graphs}
\end{figure}

\begin{figure}
 \begin{center}
  \includegraphics[width=0.49\textwidth]{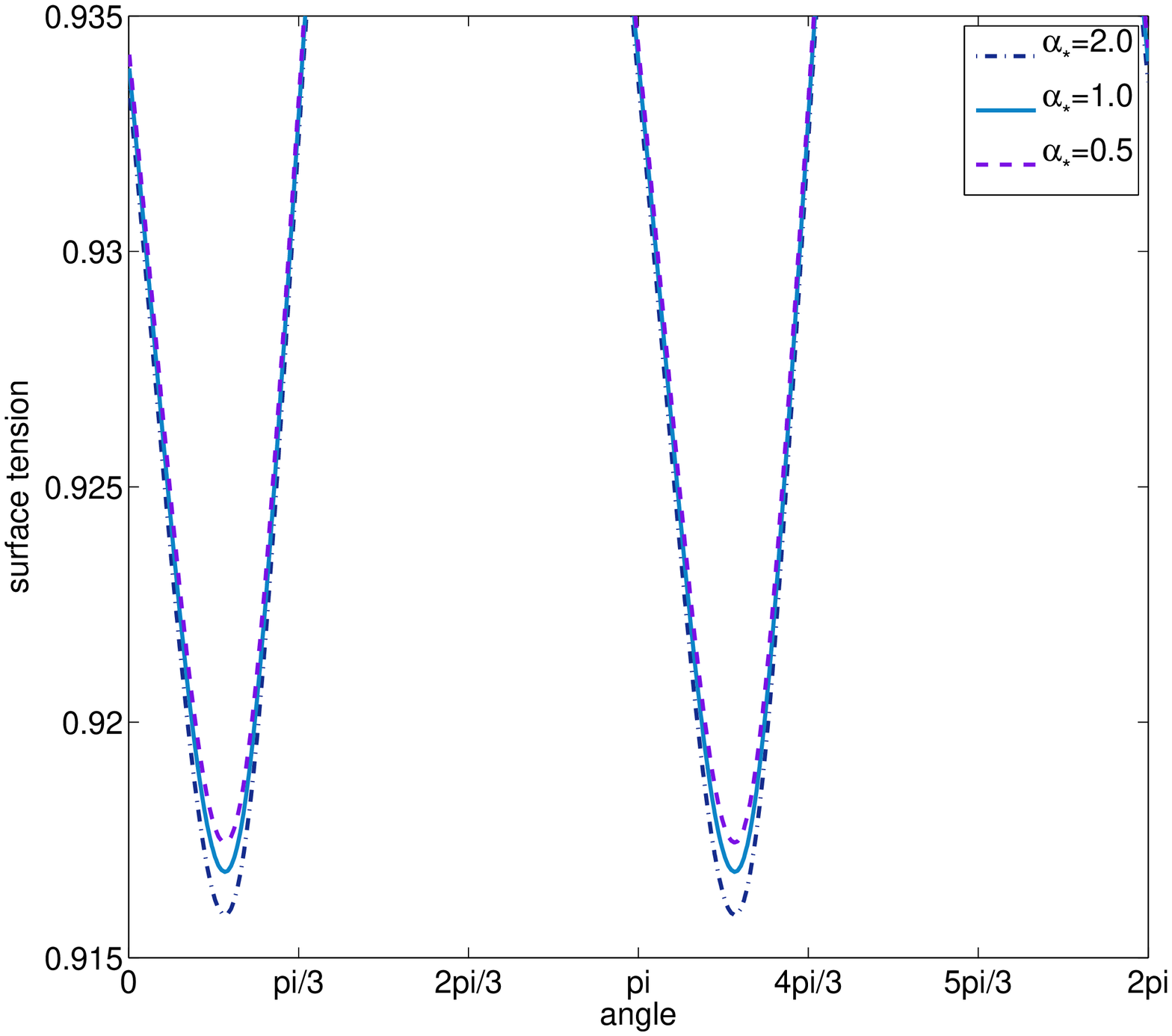} \hfill \includegraphics[width=0.49\textwidth]{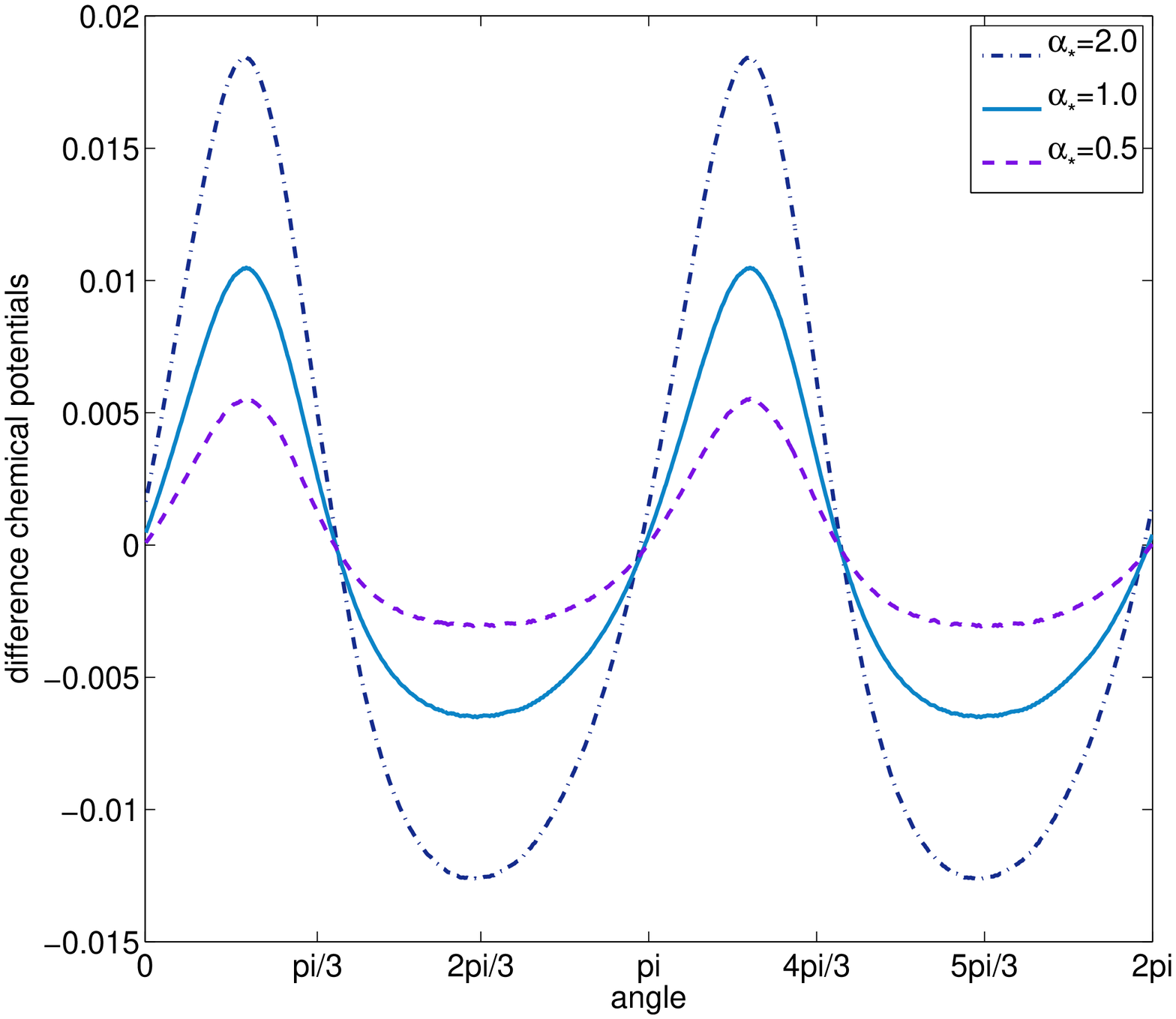}
 \end{center}
 \caption{Droplet in shear flow: Surface tension $\sigma_{*}(c^{\Gamma}_{*})$ at the tips of the droplet (left) and difference of surface and bulk chemical potentials $\gamma_{*}'(c^{\Gamma}_{*}) - G_{*}'(c_{*}^{(2)})$ (right) plotted over the angle formed by the line from the centre to a boundary point and the x-axis for several values of $\alpha_{*}$, $\eps = 0.0565685425 \approx 0.08 / \sqrt{2}$, $t = 10$.}
 \label{fig:2D_caseA_chempot}
\end{figure}

On the domain $\Omega=[-5,5] \times[-2,2] \subset \R^2$ the velocity was initialised with $\bbb{v}(x_1,x_2,0) = 0$. On the upper and lower boundary $\{ x_2 = 2 \}$ and $\{ x_2 = -2 \}$ we then increased the velocity linearly in time to $\bbb{v}(x_1,x_2,t) = (x_2/2, 0)$, $t \geq 0.1$. On the two sides $\{ x_1 = -5 \}$ and $\{ x_1 = 5 \}$ we imposed the condition $\bbb{v}(x_1,x_2,t) = 0$. The phase field was initialised with $\varphi(\bbb{x},0) = \psi((\| \bbb{x} \|_2  - 1)/ \eps)$ where 
\begin{equation} \label{eq:psi_init}
\psi(z) = 
\begin{cases}
+1, & \text{ for } z \geq \frac{\pi}{2}, \\
\sin(z), & \text{ for } \abs{z} < \frac{\pi}{2}, \\
-1, & \text{ for } z \leq -\frac{\pi}{2},
\end{cases}
\end{equation}
which yields a circular diffuse interface of radius one and centre $\bbb{m} = (0,0)$. Furthermore, we set $\rrr{Re} = 0.1$ and $m_{*}(\varphi) = \frac{1}{2} (1 - \varphi^{2})_{+}$.

We investigated Model A with $\rrr{Pe}_\Gamma = 2.5$, $\rrr{Pe}_{c,i} = 2.5$, and $\alpha^{(i)}_{*} = 1$ for $i=1,2$ for the following isotherms, see Table \ref{tbl:Isotherms} (assuming the same free energies in the two bulk phases, thus dropping the index):
\begin{itemize}
 \item Langmuir ($B=0.2$, $\sigma_{0}=1$, $K=10$);
 \item Frumkin ($B=0.2$, $\sigma_{0}=1$, $K=10$, $A=0.4$);
 \item Freundlich ($B=0.2$, $\sigma_{0}=1$, $K=10$, $N=1.5$, $A_c=1.0$).
\end{itemize}
The initial bulk surfactant density was $c_{*}^{(1)} = c_{*}^{(2)} = 1/(10e) \approx 0.03679$, and the interfacial surfactant density $c^{\Gamma}_{*}$ was the equilibrium value (thus, depending on the isotherm).

At time $t=10$ the droplets seemed to have attained stationary shapes. These are displayed in Figure \ref{fig:2D_caseA_shapes} for several isotherms. For our parameters we found that the Langmuir isotherm leads to the least deformed shape while the shape for the Freundlich isotherm is most deformed when comparing with the initial circular shape. A common measure for the deformation is the Taylor deformation parameter $D_{Tay} = (L-B) / (L+B)$ where $L$ and $B$ are the maximum and the minimum distance to the centre, respectively. We obtained the following values:
%\begin{table}[p]
%\centering
\begin{center}
\begin{tabular}{|l|l|l|l|}
\hline
isotherm & Langmuir & Frumkin & Freundlich \\
\hline
$D_{Tay}$ & 0.143298 & 0.148370 & 0.160821 \\
\hline
\end{tabular}
\end{center}
In Figure \ref{fig:2D_caseA_graphs} we display the surface surfactant density and the surface tension along the interface between the two fluids which qualitatively reveal the usual distribution, for instance, compare with \mathcite{article:LaiTsengHuang08}.

We also investigated a change in the adsorption parameter $\alpha_{*}^{(i)}$ (both always equal for the two phases, whence we drop the upper index). The impact on the shape is small in comparison with the isotherm. For the Langmuir isotherm, we obtained the deformation parameters
\begin{center}
\begin{tabular}{|l|l|l|l|}
\hline
adsorption parameter & $\alpha_{*} = 2.0$ & $\alpha_{*} = 1.0$ & $\alpha_{*} = 0.5$ \\
\hline
$D_{Tay}$ & 0.143395 & 0.143298 & 0.143241 \\
\hline
\end{tabular}
\end{center}
In Figure \ref{fig:2D_caseA_chempot} the difference of the chemical potentials at the interface is displayed, revealing the expected convergence to zero when the adsorption parameter $\alpha_{*}$ decreases.

\subsubsection{Marangoni effect}
\label{sec:num2D_caseB}

\begin{figure}
 \begin{center}
  \includegraphics[width=0.49\textwidth]{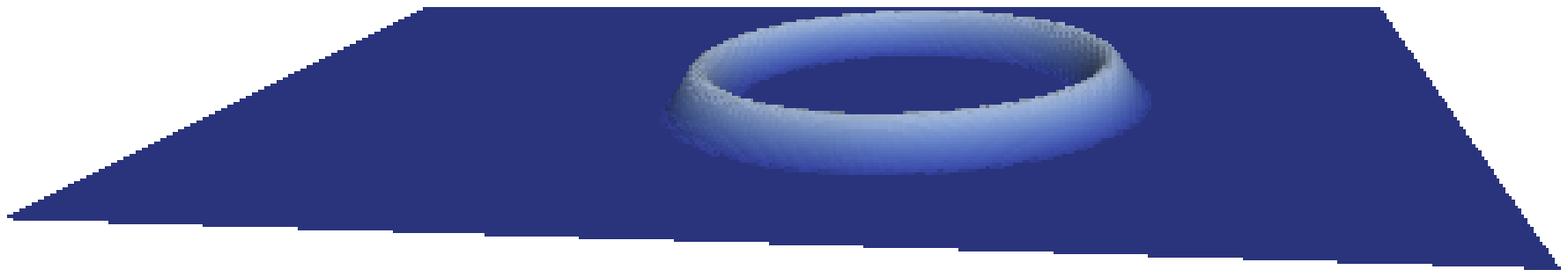} \hfill \includegraphics[width=0.49\textwidth]{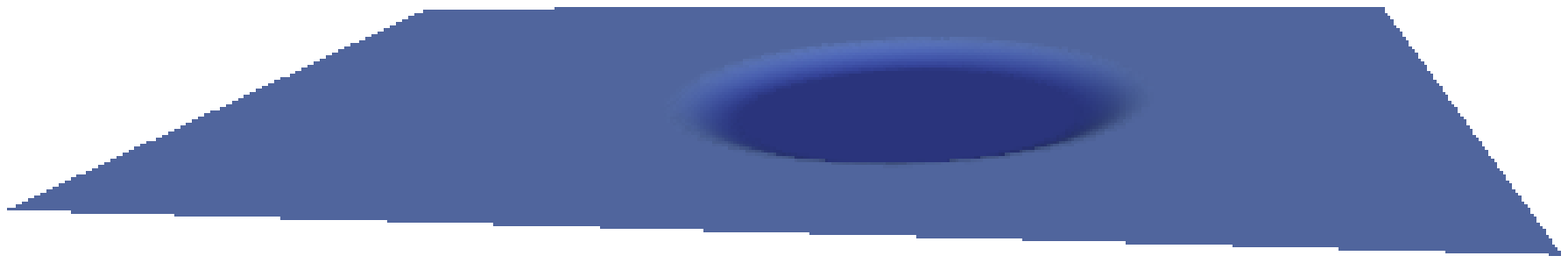} \\
  \includegraphics[width=0.49\textwidth]{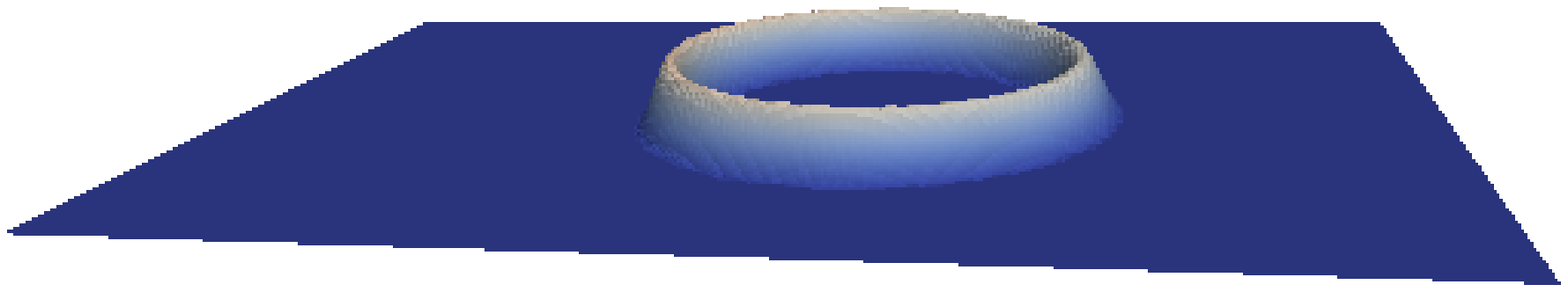} \hfill \includegraphics[width=0.49\textwidth]{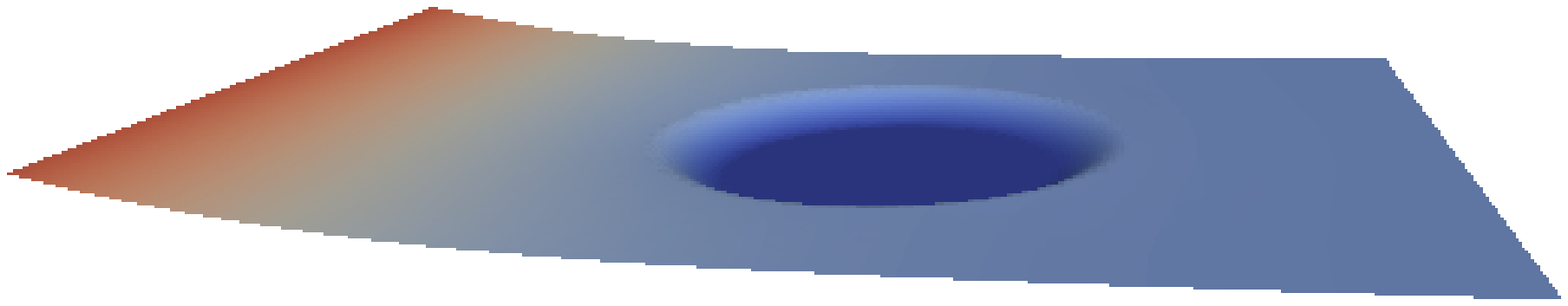} \\ 
  \includegraphics[width=0.49\textwidth]{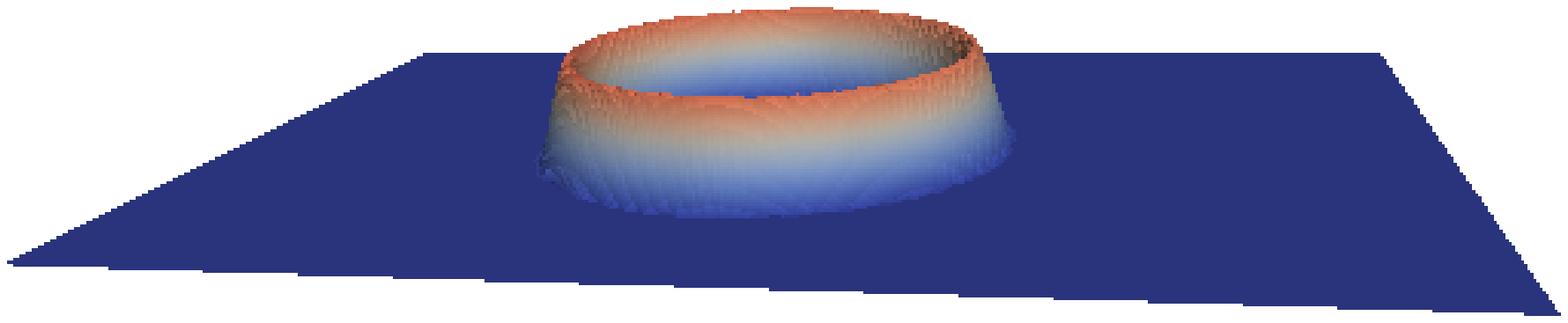} \hfill \includegraphics[width=0.49\textwidth]{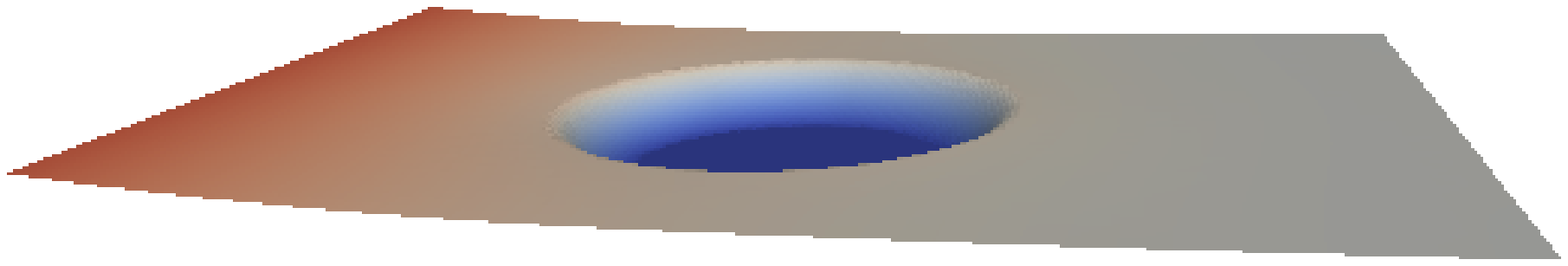} \\
  \includegraphics[width=0.49\textwidth]{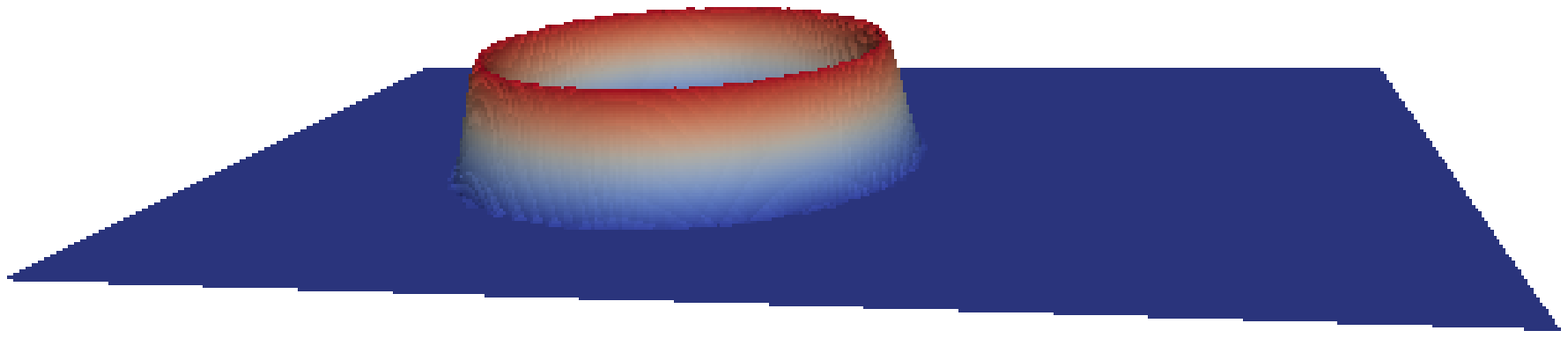} \hfill \includegraphics[width=0.49\textwidth]{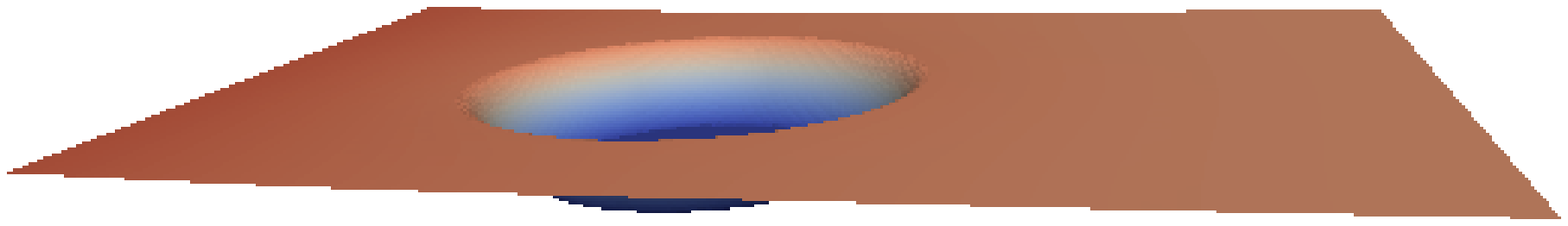} \\
 \end{center}
 \caption{Marangoni effect on a surfactant laden droplet due to the provision of surfactant at the boundary. Computed fields $2W(\varphi) c_{*}^\Gamma(q_{*})$ (left) and $\xi_1(\varphi) c_{*}^{(1)}(q_{*})$ (right) are plotted over the domain $\Omega = [-3,3] \times [-2,2]$ ($x$-axis from left to right, $y$-axis from front to rear, $z$-axis or height indicates the value of the field) at times $t=0, 10, 40, 100$ (top down) for a simulation performed with the Frumkin isotherm data (see Section \ref{sec:num2D_caseB}) and $\eps = 0.12$. The data range is between 0.0 (blue) and about 0.585 (red).}
 \label{fig:2D_caseB_movement}
\end{figure}

We now consider the domain $\Omega = [-3,3]\times [-2,2]$. Both velocity and pressure are initialised with $0$, and this is also the Dirichlet boundary condition for the velocity. For the phase field we set $\varphi(\bbb{x},0) = \psi((\| \bbb{x} - \bbb{m} \|_2 - 1)/ \eps)$ with $\psi$ given as in \mathref{eq:psi_init} and $\bbb{m} = (0.5,0)$ which corresponds to a circular diffuse interface of radius one around $\bbb{m}$. The Reynolds number is $\rrr{Re} = 10$ and we chose $m_{*}(\varphi) = (1 - \varphi^{2})_{+}$. 

Simulations were performed with Model C where we set $\rrr{Pe}_{c,i} = \rrr{Pe}_\Gamma = 10.0$, $i=1,2$ and used the following free energies (again, the free energies in the two bulk phases are assumed to be the same so that the index is dropped):
\begin{itemize}
 \item Langmuir ($B = 1$, $\sigma_{0} = 2$, $K = 2.5$);
 \item Frumkin ($B = 1$, $\sigma_{0} = 2$, $K = 2.5$, $A=0.4$);
 \item Freundlich ($B = 1$, $\sigma_{0} = 2$, $K = 1$, $N=1.5$, $A_c=0.6$).
\end{itemize}
The field $q_{*}$ was initialised such that $c_{*}^{(1)}(q_{*}) = c_{*}^{(2)}(q_{*}) = 0.1$ . During the time interval $[0,0.1]$ we linearly increased $q_{*}$ on the boundary $\{ x_1 = -3 \}$ such that, at $t=0.1$, $c_{*}^{(1)}(q_{*}) = 0.5$.

As a consequence, the droplet moved in $-x_1$ direction towards the source of the surfactant as exemplary illustrated in Figure \ref{fig:2D_caseB_movement} for the Frumkin isotherm data. Initially at rest, the supply of surfactant on the boundary leads to a surfactant gradient at the interface of the droplet. Since $\sigma_{*}$ is decreasing in $c^{\Gamma}_{*}$ the related Marangoni force $\nabla_\Gamma \sigma_{*}(c^\Gamma_{*})$ points into the opposite direction and, thus, leads to a drift towards the source of the surfactant. In the long term, the system reaches a steady state again with spatially homogeneous distributions of the surfactant in both phases and on the interface, which is fairly achieved at time $t=100.0$. For our choice of parameters the Freundlich isotherm lead to the most significant displacement $d_{x_1}$ along the $x_1$ axis while the Langmuir isotherm lead to the least significant displacement:
\begin{center}
\begin{tabular}{|l|l|l|l|}
\hline
 & Langmuir & Frumkin & Freundlich \\
\hline
$d_{x_1}$ & -1.055512 & -1.087783 & -1.114869 \\
\hline
\end{tabular}
\end{center}

\section{Appendix}\label{appendix}
We use the following result from Alt \cite{article:Alt09} to reformulate the strong form of the surfactant equations $(\ref{SIM:eq3}),(\ref{SIM:eq6}),(\ref{SIM:eq7})$ into an equivalent distributional form.  Let $\mathcal{D}'(\Omega)$ denote the space of distributions on $\Omega$.
\begin{thm}[Alt \cite{article:Alt09} Section 2.7 \& Theorem 2.8]\label{Alt:thm}
Given an open set $D \subset \R \times \R^{d}$ consisting of two open sets $\Omega^{(1)}$ and $\Omega^{(2)}$ separated by a smooth evolving hypersurface $\Gamma$, in particular, $\Gamma \subset D$ has no boundary within $D$.  For $(t,x) \in \Gamma$ we let $\bm{\nu}_{i}(t,x) \in (T_{x}(\Gamma(t)))^{\perp} \subset \R^{d}$ be the external unit normal of $\Omega^{(i)}(t)$.  Then $\bm{\nu}_{1} + \bm{\nu}_{2} = 0$.  Denote by $\chi_{\Omega^{(1)}}, \chi_{\Omega^{(2)}}, \delta_{\Gamma}$ the following distributions:
\begin{align*}
 \int_{D} f d \chi_{\Omega^{(i)}} = \int_{\R} \int_{\Omega^{(i)}(t)} f(t,x), \quad \int_{D} f d \delta_{\Gamma} = \int_{\R} \int_{\Gamma(t)} f(t,x).
\end{align*}
Then a single balance law is an equality of the form
\begin{align}\label{Alt:distributionallaw}
 \pd_{t} E + \nabla \cdot \bm{Q} = F \text{ in } \mathcal{D}'(D)
\end{align}
with distributions given by
\begin{align*}
 E = \sum_{i=1,2} e^{(i)} \chi_{\Omega^{(i)}} + e^{\Gamma} \delta_{\Gamma}, \; \bm{Q} = \sum_{i=1,2} \bm{q}^{(i)} \chi_{\Omega^{(i)}} + \bm{q}^{\Gamma} \delta_{\Gamma}, \; F = \sum_{i=1,2} f^{(i)} \chi_{\Omega^{(i)}} + f^{\Gamma} \delta_{\Gamma}, 
\end{align*}
where $e^{(i)}, q^{(i)}_{j}, f^{(i)} : \overline{\Omega^{(i)}} \to \R$ and $e^{\Gamma}, q^{\Gamma}_{j}, f^{\Gamma} : \Gamma \to \R$ are smooth functions. Then the distributional law $(\ref{Alt:distributionallaw})$ is equivalent to the following:
\begin{enumerate}
 \item For $i = 1,2$ in $\Omega^{(i)}$:
\begin{align*}
 \pd_{t} e^{(i)} + \nabla \cdot \bm{q}^{(i)} = f^{(i)}.
\end{align*}
\item For all $(t,x) \in \Gamma$:
\begin{align*}
 (\bm{q}^{\Gamma} - e^{\Gamma}\bm{u}_{\Gamma})(t,x) \in T_{x}(\Gamma(t)).
\end{align*}
\item On $\Gamma$:
\begin{align*}
 \pd_{t} e^{\Gamma} + \bm{u}_{\Gamma} \cdot \nabla e^{\Gamma} - e^{\Gamma} \bm{\kappa}_{\Gamma} \cdot \bm{u}_{\Gamma} + \surf \cdot (\bm{q}^{\Gamma} - e^{\Gamma}\bm{u}_{\Gamma}) = f^{\Gamma} + \sum_{i=1,2}(\bm{q}^{(i)} - e^{(i)} \bm{u}_{\Gamma}) \cdot \bm{\nu}_{i},
\end{align*}
\end{enumerate}
where $\bm{u}_{\Gamma}$ is the unique velocity vector such that 
\begin{align*}
 T_{(t,x)}\Gamma = \text{span} \{ (1, \bm{u}_{\Gamma}(t,x)) \} \oplus (\{0\} \times T_{x} \Gamma(t)),
\end{align*}
and $\bm{\kappa}_{\Gamma}$ is the curvature vector defined by 
\begin{align*}
 \surf \cdot \bm{n} = - \bm{\kappa}_{\Gamma} \cdot \bm{n},
\end{align*}
for spatial normal vector fields $\bm{n}(t,x) \in (T_{x}\Gamma(t))^{\perp}$.
\end{thm}

For the reformulation, we assume as in \cite{article:TeigenLiLowengrubWangVoigt09} that $c^{\Gamma}$ is extended off $\Gamma$ constant in the normal direction, hence $\surf c^{\Gamma} = \nabla c^{\Gamma}$.  Define 
\begin{align*}
 j_{1} = \frac{1}{\alpha^{(1)}}(\gamma'(c^{\Gamma}) - G_{1}'(c^{(1)})), \quad j_{2} = \frac{1}{\alpha^{(2)}}(\gamma'(c^{\Gamma}) - G_{2}'(c^{(2)})),
\end{align*}
then by the definition of $\md(\cdot)$, the divergence-free property of $\bm{v}$ and that $\nabla \gamma'(c^{\Gamma}) = \gamma''(c^{\Gamma}) \nabla c = \gamma''(c^{\Gamma}) \surf c^{\Gamma} = \surf \gamma'(c^{\Gamma})$, equation $(\ref{SIM:eq6})$ can be written as
\begin{align*}
 \pd_{t} c^{\Gamma} + \surf \cdot (c^{\Gamma} \bm{v} - M_{\Gamma} \nabla \gamma'(c^{\Gamma})) = -(j_{1} + j_{2}).
\end{align*}
Choosing $e^{(i)} = \bm{q}^{(i)}_{j} = f^{(i)} = 0$ for $i = 1,2$, $1 \leq j \leq d$ and $e^{\Gamma} = c^{\Gamma}, \bm{q}^{\Gamma} = c^{\Gamma} \bm{v} - M_{\Gamma} \nabla \gamma'(c^{\Gamma})$, $f^{\Gamma} = -(j_{1} + j_{2})$.  Theorem \ref{Alt:thm} implies that the distributional form
\begin{align}\label{Appendix:interfacedistributional}
 \pd_{t}(\delta_{\Gamma} c^{\Gamma}) + \nabla \cdot (\delta_{\Gamma} c^{\Gamma} \bm{v} - M_{\Gamma} \delta_{\Gamma} \nabla \gamma'(c^{\Gamma})) = -\delta_{\Gamma} (j_{1} + j_{2})
\end{align}
is equivalent to 
\begin{align*}
 \pd_{t} c^{\Gamma} + \bm{u}_{\Gamma} \cdot \nabla c^{\Gamma} - c^{\Gamma} \bm{\kappa}_{\Gamma} \cdot \bm{u}_{\Gamma} + \surf \cdot (c^{\Gamma}\bm{v} - M_{\Gamma}\surf \gamma'(c^{\Gamma}) - c^{\Gamma}\bm{u}_{\Gamma}) = -(j_{1} + j_{2}) \text{ on } \Gamma.
\end{align*}
We have $\surf \cdot (c^{\Gamma} \bm{u}_{\Gamma}) = - c^{\Gamma} \bm{\kappa}_{\Gamma} \cdot \bm{u}_{\Gamma}$ and $\bm{u}_{\Gamma} = (\bm{v} \cdot \bm{\nu}_{1})\bm{\nu}_{1}$ implies $\bm{v} = \bm{u}_{\Gamma} + \bm{v}_{\tau}$.  Furthermore, $\surf \cdot (c^{\Gamma} \bm{v}) = \surf c^{\Gamma} \cdot \bm{v}_{\tau} + c^{\Gamma} \surf \cdot \bm{v}$.  Hence equation $(\ref{eq:interfacedistributional})$ is equivalent to $(\ref{SIM:eq6})$.  For $i = 1$, choose $e^{(2)} = q^{(2)}_{j} = f^{(1)} = f^{(2)} = e^{\Gamma} = q^{\Gamma}_{j} = 0$ for $1 \leq j \leq d$ and $e^{(1)} = c^{(1)}, f^{\Gamma} = j_{1}, \bm{q}^{(1)} = c^{(1)} \bm{v} - M_{c}^{(1)} \nabla G_{1}'(c^{(1)})$.  Then the distributional form
\begin{align}\label{Appendix:bulk1distributional}
 \pd_{t}(\chi_{\Omega^{(1)}}c^{(1)}) + \nabla \cdot (\chi_{\Omega^{(1)}} c^{(1)} \bm{v} - \chi_{\Omega^{(1)}}M_{c}^{(1)} \nabla G_{1}'(c^{(1)})) = \delta_{\Gamma} j_{1}
\end{align}
is equivalent to
\begin{align*}
 \pd_{t}(c^{(1)}) + \nabla \cdot (c^{(1)} \bm{v} - M_{c}^{(1)}\nabla G_{1}'(c^{(1)})) & = 0, \text{ in } \Omega^{(1)}, \\
 M_{c}^{(1)} \nabla G_{1}'(c^{(1)})\cdot \bm{\nu}_{1} & = j_{1}, \text{ on } \Gamma.
\end{align*}
Similarly, choosing $e^{(1)} = q^{(1)}_{j} = f^{(1)} = f^{(2)} = e^{\Gamma} = q^{\Gamma}_{j} = 0$ for $1 \leq j \leq d$ and $e^{(2)} = c^{(2)}$, $f^{\Gamma} = j_{2}$, $\bm{q}^{(2)} = c^{(2)} \bm{v} - M_{c}^{(2)} \nabla G_{2}'(c^{(2)})$.  Then the distributional form
\begin{align}\label{Appendix:bulk2distributional}
 \pd_{t}(\chi_{\Omega^{(2)}}c^{(2)}) + \nabla \cdot (\chi_{\Omega^{(2)}} c^{(1)} \bm{v} - \chi_{\Omega^{(2)}}M_{c}^{(2)} \nabla G_{2}'(c^{(2)})) = \delta_{\Gamma} j_{2}
\end{align}
is equivalent to
\begin{align*}
 \pd_{t}(c^{(2)}) + \nabla \cdot (c^{(2)} \bm{v} - M_{c}^{(2)}\nabla G_{2}'(c^{(2)})) & = 0, \text{ in } \Omega^{(2)}, \\
 -M_{c}^{(2)} \nabla G_{2}'(c^{(2)})\cdot \bm{\nu}_{1} & = j_{2}, \text{ on } \Gamma
\end{align*}
as $\bm{\nu}_{2} = - \bm{\nu}_{1}$.  Thus the bulk and interfacial surfactant equations can be reformulated in the distributional forms $ (\ref{Appendix:interfacedistributional})-(\ref{Appendix:bulk2distributional})$.

\medskip

{\bf Acknowledgement.} This research has been supported by the British Engineering and Physical Sciences Research Council (EPSRC), Grant EP/H023364/1 and by the SPP 1506 ``Transport Processes at Fluidic Interfaces'' of the German Science Foundation (DFG) through the grant GA 695/6-1.

\medskip

\bibliographystyle{plain}
\bibliography{GLS_SolSurf}
% % % % 
% % % % \begin{thebibliography}{10}
% % % % 
% % % %           %
% % % % 
% % % %           % and use \bibitem to create references.
% % % % 
% % % %           %
% % % % 
% % % % \bibitem{A}Author, {\em title of paper}, Journal Name
% % % % Volume,  page numbers, year.
% % % % 
% % % %           % Format for Journal Reference. For example
% % % % 
% % % % \bibitem{T1}C. Taubes, {\em The Seiberg-Witten invariants
% % % %  and symplectic forms}, Math. Res. Letters, 1, 809--822, 1994.
% % % % \end{thebibliography}
\end{document}